%% file: cluster_main.tex
\documentclass[fleqn,usenatbib]{mnras}

\usepackage{newtxtext,newtxmath}

\usepackage[T1]{fontenc}
\usepackage{orcidlink}
\usepackage{ifthen}

\newcommand{\orcid}[1]{\ifthenelse{\equal{#1}{}}{\relax}{\,\orcidlink{#1}}}

\DeclareRobustCommand{\VAN}[3]{#2}
\let\VANthebibliography\thebibliography
\def\thebibliography{\DeclareRobustCommand{\VAN}[3]{##3}\VANthebibliography}

\usepackage{graphicx}	
\usepackage{amsmath}	
\usepackage{subcaption}

\title[ASPIRE Correlation Function]{Clustering of $z\sim6.6$ Quasars and [O III] Emitters Constrains Host Halo Masses and Duty Cycles in 25 ASPIRE Fields
}

\input{authors_list}

\date{Accepted XXX. Received YYY; in original form ZZZ}

\pubyear{2024}

\begin{document}

\font\sevenrm=cmr7
\def\oiii{[O\,\textsc{iii}]}
\def\FeII{Fe~{\sevenrm II}}
\def\FeIIf{[Fe~{\sevenrm II}]}
\def\SIII{[S~{\sevenrm III}]}
\def\HeI{He~{\sevenrm I}}
\def\HeII{He~{\sevenrm II}}
\def\NeV{[Ne~{\sevenrm V}]}
\def\OIV{[O~{\sevenrm IV}]}

\def\mpch{~\mathrm{cMpc}~h^{-1}}
\def\rp{r_{\rm p}}
\def\msun{{\rm {M_{\odot}}}}
\def\mminq{M^{\rm QSO}_{h, \rm min}}
\def\mming{M^{\rm \oiii}_{h, \rm min}}

\label{firstpage}
\pagerange{\pageref{firstpage}--\pageref{lastpage}}
\maketitle


\begin{abstract}
We use data from the JWST ASPIRE Wide Field Slitless Spectroscopy (WFSS) program to measure the auto-correlation function of \oiii-emitters at $5.3<z<7.0$ and the quasar--\oiii-emitter cross-correlation function around 25 ASPIRE quasars ($6.51 < z < 6.82$; $\langle z\rangle = 6.6$). 
We use synthetic source injection to calibrate the selection function, which we combine with the large-volume cosmological simulation FLAMINGO-10k (2.8 cGpc box) to construct realistic mock observations. Our simulation-based approach offers two key advantages: (1) the clustering models capture nonlinear structure growth and scale-dependent bias on small scales beyond analytic prescriptions, and (2) covariance matrices from mock realizations account for cosmic variance and correlated uncertainties across radial bins.
Our clustering measurements yield a correlation length of 
$r_0^{\rm GG} = 4.7^{+0.5}_{-0.6}~\mpch$ for the \oiii-emitter auto-correlation with a fixed power-law slope $\gamma_{\rm GG}=1.8$, 
and $r_0^{\rm QG} = 8.9^{+0.9}_{-1.0}~\mpch$ for the quasar--\oiii-emitter cross-correlation with $\gamma_{\rm QG}=2.0$. 
We jointly estimate the halo masses by assuming a step-function halo occupation distribution (HOD) with mass-dependent covariance matrices. Our analysis yields 
$\log (M_{h, \mathrm{min}}^{\mathrm{[OIII]}}/\msun) = 10.55^{+0.11}_{-0.12}$ 
for \oiii-emitter host halos and 
$\log (M_{h, \mathrm{min}}^{\mathrm{QSO}}/\msun) = 12.13^{+0.31}_{-0.38}$ 
for quasar host halos. Based on these minimum halo masses, we estimate duty cycles of 
$2.5^{+1.3}_{-0.8}\%$ for \oiii-emitters, 
and $0.3^{+4.8}_{-0.3}\%$ for quasars. The low quasar duty cycle implies UV-bright lifetimes of only $t_{\rm Q} =2.64^{+39.15}_{-2.61}~\rm Myr$, representing $\lesssim 10\%$ of a Salpeter $e$-folding time. This indicates that the observed UV-luminous phase contributes minimally to total SMBH mass assembly, placing tight constraints on early black hole growth models.
\end{abstract}



\section{Introduction}
\label{sec:introduction}

The detection of quasars at high redshift hosting supermassive black holes (SMBHs) with masses around $\sim 10^9 ~ \rm M_\odot$ \citep{Mortlock2011, Matsuoka2019, Yang2020, Wang2021, Fan2023}, as well as recent findings of active galactic nuclei (AGNs) candidates at $z \gtrsim 8$ with black holes masses of $\gtrsim 10^6 ~\rm M_\odot$ \citep{Bunker2023, Bogdan2024}, 
poses significant challenges to our current understanding of black hole formation and growth in the early Universe.
In particular, the existence of SMBHs with $\sim 10^9 - 10^{10} \, \rm M_\odot$ in the centers of luminous quasars at $z \gtrsim 6$, when the universe was less than a billion years old, raises questions about the efficiency and mechanisms of black hole accretion \citep{WMC2008, Shankar2010,Conroy2013, Mazzucchelli2017, Banados2018, Davies2019, Yang2020, Wang2021}. 

To observationally constrain the growth scenarios of early SMBH, 
it is essential to determine not only how massive these early SMBHs are when active,
but also what fraction of their growth occurs during the UV-bright phase that we observe as quasars. Quasar clustering provides a direct way to quantify this through measurements of the quasar duty cycle, defined as the ratio between the number density of observed quasars ($n_{\rm quasar}$) and the number density of their host dark matter halos ($n_{\rm halo}$): $f_{\rm duty} \equiv n_{\rm quasar}/n_{\rm halo}$. This ratio reflects the fraction of cosmic time during which a typical halo hosts a luminous quasar. Under the assumption that each halo undergoes one quasar phase of duration $t_{\rm Q}$, the duty cycle equals the quasar lifetime as a fraction of the Hubble time: $f_{\rm duty} \sim t_{\rm Q}/t_{\rm H}$, where $t_{\rm H}$ is the Hubble time at the redshift of observation \citep{Efstathiou1988, Cole1989, HaimanHui2001, Martini2001}. 
Under the assumption that each halo undergoes one quasar phase of duration $t_{\rm Q}$, the duty cycle directly reflects the fraction of cosmic time during which a typical halo hosts a luminous quasar. Therefore, combining clustering-based duty cycle estimates with SMBH mass measurements provides insight into how much of the total SMBH mass assembly occurs during these luminous phases, while still allowing for the possibility that a substantial fraction of growth may take place during radiatively inefficient or obscured accretion modes that contribute little to the observed quasar luminosity \citep{Baados2025}.

To find $n_{\rm halo}$ 
a key step is to find the characteristic halo mass that can host a quasar.
As predicted by the $\Lambda$CDM cosmology, more massive dark matter halos exhibit stronger clustering due to their higher bias, which enhances the correlation length \citep{Kaiser1984, MoWhite2002}. By assuming that quasars and galaxies represent a subsample of the halo population, we can use their spatial distribution to model the distribution of their host halos. This allows us to statistically infer the typical halo mass from their observed clustering strength \citep{Martini2001}.

SMBH growth relies on sustained, Eddington-limited accretion to achieve the extreme SMBH masses observed at high redshifts \citep{Salpeter1964}. Achieving $\sim 10^{9}$–$10^{10}~\rm M_\odot$ by $z \gtrsim 6$ requires black holes to gain mass over multiple Salpeter times, where the Salpeter time is the characteristic $e$-folding timescale for black hole growth under Eddington limited accretion. This requires that a significant fraction of SMBH growth occurs while accreting efficiently enough to build their mass within the limited cosmic time available \citep[e.g.][]{Tanaka2009}. Because luminous quasars at $z \gtrsim 6$ are extremely rare in terms of space density ($\sim 1~\rm cGpc^{-3}$ at $z\sim6$, see e.g., \citealt{Wang2019, Schindler2023}), a large duty cycle for the UV-bright phase would imply that these sources inhabit comparably rare dark matter halos, i.e., the most massive collapsed structures at early times. Conversely, if quasars resided in more common, lower-mass halos, their observed scarcity would require a correspondingly small duty cycle.

Clustering analysis through the two-point correlation function (2PCF) has become a powerful tool to probe the environments and host halo masses of high-redshift quasars \citep{Osmer1981, Shen2007}. While quasar clustering measurements at redshifts $1 \lesssim z \lesssim 3$ indicate host halo masses on the order of $10^{12}~\msun$, increasing to $\sim 10^{13}~\msun$ at $z \sim 4$ \citep{Porciani2004, Croom2005, Coil2007, Pizzati2024a, GinerMascarell2025}, high-redshift measurements of quasar auto-correlation functions become challenging due to the low spatial density of luminous quasars at $z \gtrsim 6$. As a result, cross-correlation measurements between quasars and galaxies offer a complementary approach for determining host halo masses at high redshift \citep{GarciaVergara2017, GarciaVergara2019, Garcia-Vergara2022}.

Before JWST, empirical studies of quasar environments at $z>5$ yielded mixed results: narrow-band searches for Ly$\alpha$ emitters (LAEs) and Lyman break galaxies (LBGs) as tracers sometimes reported galaxy excesses around high-$z$ quasars \citep[e.g.,][]{Zheng2006, Kim2009, Morselli2014, Wang2023, Pudoka2024}, whereas others found no significant overdensities \citep{Willott2005, Banados2013, Simpson2014, Lambert2024}. These mixed conclusions likely reflected observational limitations—including small fields of view, heterogeneous depth, and redshift offsets between narrow-band filters and quasar systemic redshifts—which complicate comparisons across surveys \citep{Overzier2016}. Furthermore, only a few spectroscopic observations have shown that the quasars are
living in megaparsec-scale (comoving) galaxy overdensities at $z\sim6$ \citep{Bosman2020, Mignoli2020, Meyer2022}.

Recent observations with \textit{JWST}/NIRCam using the Wide Field Slitless Spectroscopy (WFSS) mode have opened new opportunities to probe quasar environments with higher sensitivity and larger spatial coverage. Several ongoing programs are now using WFSS to search for line-emitting galaxies, including \oiii\ and H$\alpha$ emitters, in quasar fields. These studies have revealed an abundance of \oiii-emitting galaxies around some high-redshift quasars with $z\gtrsim6$ \citep{Wang2023, Eilers2024, Kashino2023, Champagne2025}, suggesting that luminous quasars may reside in significant overdensities. However, recent measurements of the quasar–\oiii emitter cross correlation for two quasars at $z\simeq7.3$ find no such excess \citep{Schindler2025b}, indicating that quasar environments may vary from object to object and providing tentative evidence for a non monotonic redshift evolution in quasar clustering properties \citep{Schindler2025b}. The growing sample of emission-line galaxy tracers has enabled more precise clustering analyses to study quasar environments and the growth of supermassive black holes.

However, a major uncertainty for the halo mass measurements using the correlation functions arises since the field-to-field variance is large: 
when the number of observed fields is limited, statistical fluctuations in the large-scale structure from one field to another, known as field-to-field or cosmic variance, introduce noise into the measurement of the correlation function. Each field samples only a small 
portion of the underlying matter distribution, so the measured clustering amplitude from a small number of fields represents a noisy realization of the cosmic mean. Consequently, the uncertainty scales roughly as the intrinsic variance of the density field divided by the square root of the number of fields ($\propto \sigma / \sqrt{N}$). When only a few independent fields are available, this variance may not average down effectively and can dominate over the Poisson errors from pair counting, becoming the main source of uncertainty in the inferred clustering strength and halo mass.

More importantly, the presence of positive off-diagonal elements in the correlation matrix is expected in addition to the cosmic variance. When galaxy clustering is estimated from pair counts, any large-scale fluctuation in the galaxy density (for example, a locally overdense region) boosts the number of pairs across a wide range of separations. Consequently, measurements in different radial bins rise or fall together, producing positive correlations between bins. Put differently, because the same galaxies contribute to pairs at multiple separations, the errors in the clustering measurements are inherently correlated across scale bins.
Estimating the halo mass from limited number of sample can lead to inconclusive results \citep{Yang2005, Robertson2010}. In addition, the field-to-field variance itself may encode valuable information about the underlying halo mass. A comprehensive analysis that models the full probability density function of galaxy counts will be presented in future work. 
Meanwhile, using a fainter yet more abundant quasar population (\( M_{1450} \approx -25 \)), the Subaru High-z Exploration of Low-luminosity Quasars (SHELLQs) Collaboration \citep{Arita2023} has recently carried out the first measurement of the quasar autocorrelation function at \( z \sim 6 \). 
Their results suggest that these quasars reside in dark matter haloes with masses comparable to, or slightly higher than, those inferred at lower redshifts, although the associated uncertainties remain considerable.

The combination of \oiii-emitters auto-correlation and quasar–\oiii-emitters cross-correlation measurements using \textit{JWST} WFSS data has now enabled the first constraints on the dark matter halo masses hosting galaxies and quasars at $\langle z\rangle = 6.25$ \citep{Eilers2024, Pizzati2024b}. While the use of a large sample of \oiii-emitters improves the statistical precision of the galaxy clustering signal, the quasar–\oiii-emitters cross-correlation function is being measured for the first time in these studies. In particular, the EIGER clustering analysis is based on only four quasar fields (albeit with wider mosaics per field compared to single pointing), making it especially susceptible to field-to-field variance. Still, it is important to keep in mind that field-to-field variance remains a dominant contributor to the total uncertainty and can outweigh the statistical errors in the correlation function measurements (e.g.,\citealt{Yang2005}).

This work extends previous analyses in three specific aspects. 
First, we target quasars at higher redshift ($\langle z\rangle = 6.6$ versus $\langle z\rangle = 6.25$) and include fainter quasars, with ASPIRE magnitudes $-25.25 > M_{1450} > -27.38$ compared to EIGER’s brighter range $-26.63 > M_{1450} > -29.14$ \citep{Eilers2024}. Second, we analyze a larger number of independent sightlines: 25 ASPIRE quasar fields with single pointings each. Although the total numbers of \oiii-emitters are comparable between EIGER and ASPIRE (and thus Poisson counting errors are similar), the greater number of fields in ASPIRE more effectively suppresses cosmic (field-to-field) variance than the EIGER sample which contains four quasar fields. 
Third, we correctly propagate measurement uncertainties by jointly fitting the \oiii auto correlation and the quasar–\oiii\ cross correlation with a full covariance model.

In particular, we jointly infer the host halo masses from the \oiii\ auto-correlation and the quasar–\oiii\ cross-correlation, explicitly modeling their covariance. The dominant covariance arise from (i) coherent large-scale structure within each field (e.g., a filament) that boosts correlated counts across separations, and (ii) shared counting statistics in the pair counts—for example, in the \oiii\ auto-correlation the same galaxies contribute to multiple separation bins (e.g., under the Landy–Szalay estimator; \citealt{LZ1993}). Therefore, off-diagonal terms in the covariance matrices for the correlation functions are non-negligible. 
However, estimating the full covariance is challenging and requires large-volume simulations and forward-modeled mocks that capture within-field large-scale structure and non-linear clustering, which is more relevant at high redshift.  While \citet{Eilers2024} presented the first quasar–\oiii\ cross-correlation measurement, their halo-mass inference did not propagate field-to-field variance; here we include such variance in a unified likelihood, yielding more reliable constraints of the estimate of the quasar and \oiii-emitters host halo masses.

Moreover, the covariance changed with both the tracer and the quasar host halo mass. As halos get more massive, the tracer bias $b(M)$ rises and galaxies cluster more strongly, which boosts the diagonal variances and also the off-diagonal correlations (see, e.g., \citealt{Findlay2025}). To capture this, it is necessary to generate mocks based on the minimum halo masses that can host the quasar or \oiii-emitters, rather than adopting a fixed, mass-independent covariance.

To that end, we employ a dark-matter-only cosmological simulation, \texttt{FLAMINGO-10k} (Schaller et al., in prep.; see also \citealt{Pizzati2024a, Pizzati2024b}) from the FLAMINGO suite of simulations \citep{Schaye2023, Kugel2023}, to generate quasar and galaxy mock catalogs within a 2.8 cGpc volume. This enables us to account for field-to-field variance, critical given the limitations of observational sample sizes and clustering fluctuations within small fields.
To achieve realistic clustering estimates, our mock catalogs include selection functions that use observational selection specific to the \textit{JWST}/WFSS fields. The selection function incorporates spatial coverage and sensitivity, enabling simulations of quasar and galaxy fields that include realistic observational selection effects. If unaccounted for, these effects will bias clustering measurements. In order to estimate the dark matter halo mass hosting quasars and galaxies and infer their duty cycle, we compare the ASPIRE result with the halo correlation function generated, where the inference procedure includes the mass-dependent covariance matrix. By incorporating field-to-field variance and observational selection effects in our mock catalogs, we can more robustly link observed clustering signals to the underlying halo populations, thereby placing tighter bounds on quasar lifetimes and the growth of supermassive black holes at $z\sim6.6$. 

The paper is structured as follows. 
Section \ref{sec:data_red} briefly describes the WFSS data reduction and \oiii-emitter catalog construction. Section \ref{sec:rand_and_maps} describes the ASPIRE selection function as well as the random catalog generation based on a Monte Carlo injection with spectral coverage and sensitivity map. We measure the correlation function based on the observed \oiii-emitter 
catalog and the Monte Carlo injected random catalog in Section \ref{sec:measuments}. In Section \ref{sec:cov} we discuss the mock generation based on \texttt{FLAMINGO-10k} simulation
and the construction of a mass-dependent covariance matrix. In Section \ref{sec:inference}, we describe the inference procedure for the quasar and \oiii-emitters' minimum halo mass and their results based on the covariance matrices. Section \ref{sec:duty_cycle} provides the duty cycle measurements for quasar and \oiii-emitter. The discussion for the results on the inferred minimum halo mass and the implications on the measured duty cycle is presented in Section \ref{sec:discussion}. Section \ref{sec:conclusion} provides the conclusion.
\begin{figure*}
\centering
	\begin{tabular}{@{}cccccc@{}}
	\includegraphics[width=0.89\columnwidth]{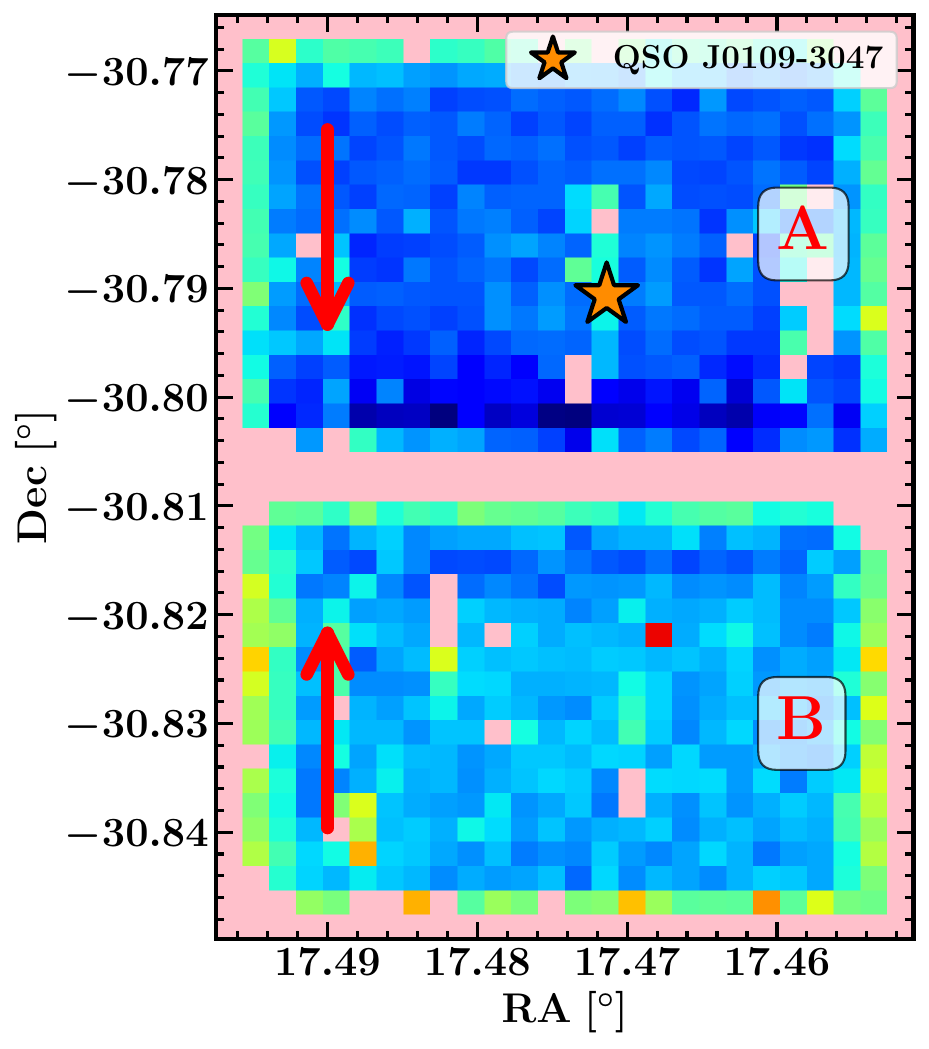}
	\includegraphics[width=\columnwidth]{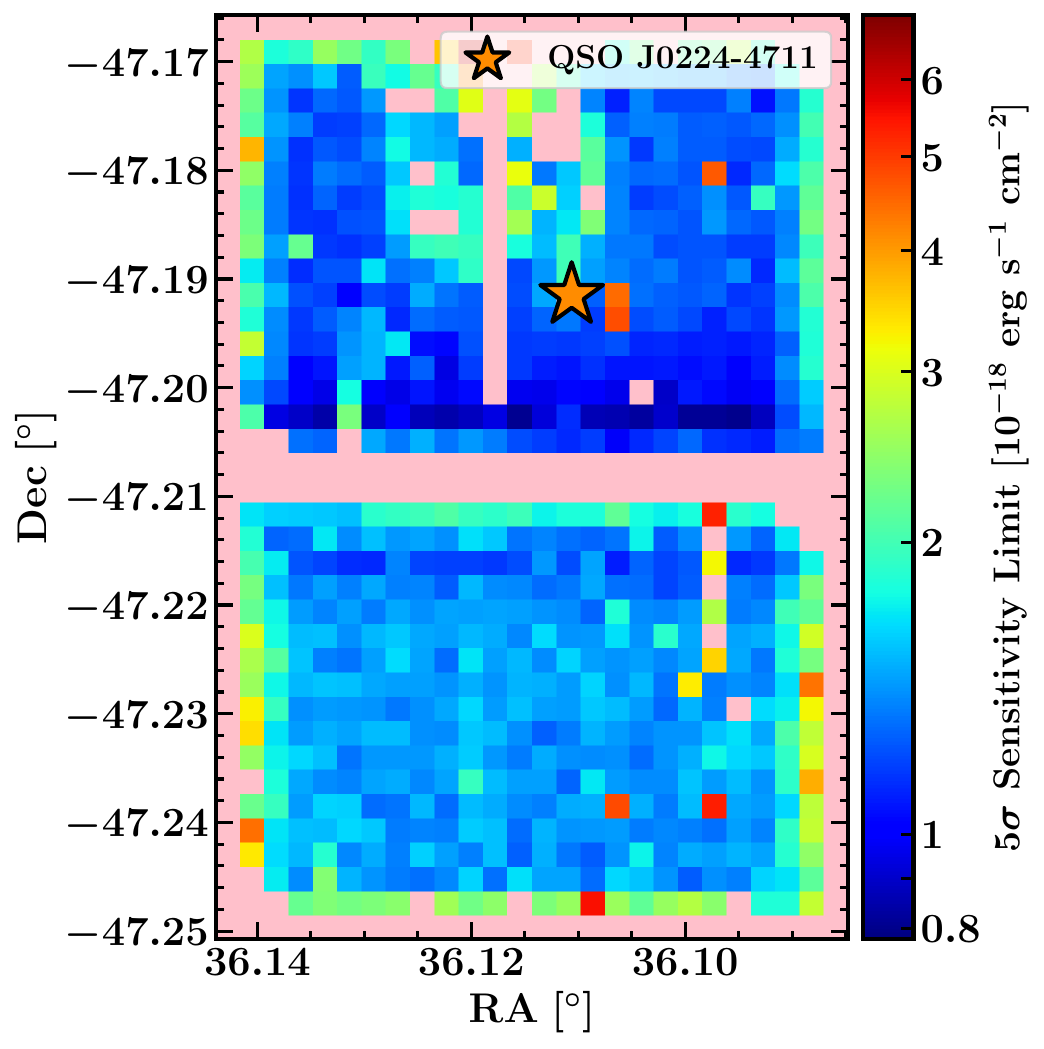}
	\end{tabular}
    \caption{Two examples of the sensitivity map for quasar field J0109-3047 and J0224-4711, the orange stars mark the location of the quasars. Each pixel value (\texttt{Flim}) on the sensitivity map is computed from the S/N of the injected source. The inject mock grid has size 40 rows $\times$ 25 columns, and the direction of the columns follows the direction of the dispersion. Grid points with no detection (i.e., 
    the redshift of the mock source is misidentified with $|z-z_{\rm true}|>0.01$), where $z_{\rm true}=6.5$ for the injected mocks, are shown in pink. 
    In the left panel, we mark the dispersion direction of Grism R (dispersion across detector rows) as red arrows for modules A and B. The color bars correspond to the $5\sigma$ flux limit, where deeper bluer color means more sensitive to the faint sources. The vertical stripes of non-detection within the FOV correspond to the contamination due to the dispersed foreground bright sources.}
    \label{fig:sensitivity}
\end{figure*}

\section{[OIII]-Emitting Galaxy Catalog}
\label{sec:data_red}
The ASPIRE JWST program targets 25 quasars using the F356W filter for WFSS observation in the long-wavelength (LW) channel, which provides a spectral wavelength coverage of $\sim 3 - 4\mu$m. Simultaneously, F200W imaging is obtained in the short-wavelength (SW) channel. A detailed discussion of the \oiii-emitter catalog construction is presented in Wang et al. 
in preparation; 
a brief summary of the selection process is as follows.

The catalog construction utilizes \textit{JWST}/NIRCam WFSS observations to search for \oiii-emitters without relying on a prior photometric selection, combining analysis of both 1D and 2D continuum-subtracted spectra. First, a median-filtered continuum model (with a window size of 51 pixels; 10 \AA/pixel) is generated and subtracted from the optimally extracted 1D spectra. The resulting signal-to-noise (S/N) spectrum is then convolved with the instrumental resolution ($\sim2.3$ pixels, \citealt{Wang2023}). 

Emission-line candidates are identified by searching for peaks with $\mathrm{S/N}>1.2$ in the smoothed spectrum. To remove artifacts from the analysis of the 1D continuum-subtracted spectra, each peak is required to have a total integrated $\mathrm{S/N}>2$ within $\pm2$ original pixels of the peak center. Independently from the 1D continuum-subtracted spectra analysis, a search for bright blobs is performed in the continuum-subtracted 2D spectra. \texttt{Photutils} \citep{photutils} is used to detect blobs with at least three connected pixels having $\mathrm{S/N}>0.8$ and an integrated flux significance $>2\sigma$. Potential \oiii-emitters must have $\mathrm{S/N}>5$ for \oiii~$\lambda5008$ and $\mathrm{S/N}>2$ for either \oiii~$\lambda4960$ or H$\beta$, where the total S/N was measured by integrating $\pm2$ original pixels from the center of the peak. 
Only objects confirmed as \oiii-emitter candidates by both the 1D and 2D methods are retained. Each source is then visually inspected.
In this paper, we use only high-confidence emitters with \texttt{Score\_VI=3} for the clustering analysis. This selection yields a sample of 487 \oiii-emitters across the 25 quasar fields, covering a redshift range of $5.34 < z < 6.98$. The sample spans $41.80 < \log(L_{\rm [O\,III]}/\rm{erg\,s^{-1}}) < 43.71$ with a mean luminosity of $\langle \log L_{\rm [O\,III]} \rangle = 42.53$.

\section{Completeness and Selection Function}
\label{sec:rand_and_maps}

To estimate the correlation functions, we use the Landy-Szalay estimator \citep{LZ1993, GarciaVergara2017}, which requires generating a mock catalog of unclustered galaxies. These mock galaxies are randomly distributed in spatial position and redshift, within the field of view and redshift coverage of each ASPIRE quasar field. However, to perform an unbiased clustering analysis, it is essential to account for the spatial and spectral selection effects introduced by the NIRCam WFSS configuration, which affect whether a source would be detectable in the survey. In practice, detector gaps in the NIRCam WFSS modules A and B, spectral-overlap contamination when the dispersed trace of a bright source intersects the \oiii\ emission, and position-dependent sensitivity across the field all affect the selection. The effective survey footprint is also redshift dependent, since the dispersed locations of the \oiii\ lines shift with redshift in a non-trivial way.

To characterize the survey completeness, we construct two key products: (1) a spectral coverage map, which determines whether an \oiii-emitter
at a given sky position and redshift has the emission lines (e.g., \oiii$\lambda\lambda$4960, 5008) falling within the WFSS field of view, accounting for WFSS dispersion and detector boundaries; and (2) a sensitivity map, which provides the flux detection limit across the field, derived from mock source injection and spectral extraction, and reflects variations in exposure depth, background level, and instrumental sensitivity (e.g., between modules A and B). Since the detector sensitivity is only a function of the spacial location on the detector (not a function of the \oiii-emitters' redshift), we only generate the sensitivity map at a fixed redshift and interpolate the sensitivity map with the coverage map to determine the redshift dependence of completeness for \oiii-emitters.

These two maps are essential for constructing realistic mock and random catalogs. A source might lie at a redshift where its dispersed \oiii\ emission lines fall outside the detector, or onto the detector gap. Such sources will be excluded using the coverage map. Even if a source is spatially covered, it may still fall below the local detection threshold due to varying sensitivity, which is modeled by the sensitivity map. To determine which mock sources would be observable, we first filter them through the redshift-dependent coverage map, and then apply a Monte Carlo detection step using the local flux limit from the sensitivity map. This procedure ensures that our mocks faithfully reflect the ASPIRE selection function and enables accurate measurement of the galaxy and quasar correlation functions. 

\subsection{Mock \oiii-emitters Generation}
\label{sec:mock_generation}




\subsubsection{Redshift-dependent spatial coverage}
\label{sec:coverage}
The spatial coverage of the WFSS depends on both the sky location and redshift. In order to take into account the redshift-dependent sky coverage, we generate a grid of points based on the F356W direct imaging. The pixel location in the direct image is mapped to the sky coordinate, which is eventually mapped to the ($x, y$) pixel coordinate of each WFSS frame using the WCS solution. For each of the grid points located in the direct imaging mapped onto the WFSS frame, we apply the dispersion function from \cite{Sun2024jwst} to determine where the dispersed mock emission line is located on the detector. Depending on the location and the wavelength of the redshifted line, there is a unique value ($\Delta x(x,y,\lambda_{\rm obs}), \Delta y(x,y,\lambda_{\rm obs})$). In other words, given a source position on the sky and its redshift, we can calculate where its emission line will appear on the dispersed image relative to its location on the detector in the direct image (i.e., without dispersion).

The displaced position $(x+\Delta x, y+\Delta y)$ of each dispersed \oiii\ line is first checked against the boundary of each WFSS frame. For each dithered exposure, a line is flagged as covered if its dispersed position lies within the image boundary. The coverage masks for the two \oiii\ lines, $\lambda4960$ and $\lambda5008$, are then combined using a logical AND, so that an \oiii-emitter is considered covered only if both lines fall within the boundary in a given exposure. This procedure is repeated for all dithered exposures, and the per-exposure results are combined using a logical OR, meaning that an emitter is included in the final coverage map if the two lines are jointly covered in at least one of the dithered frames.

\begin{figure}
    \centering
    \includegraphics[width=0.47\textwidth]{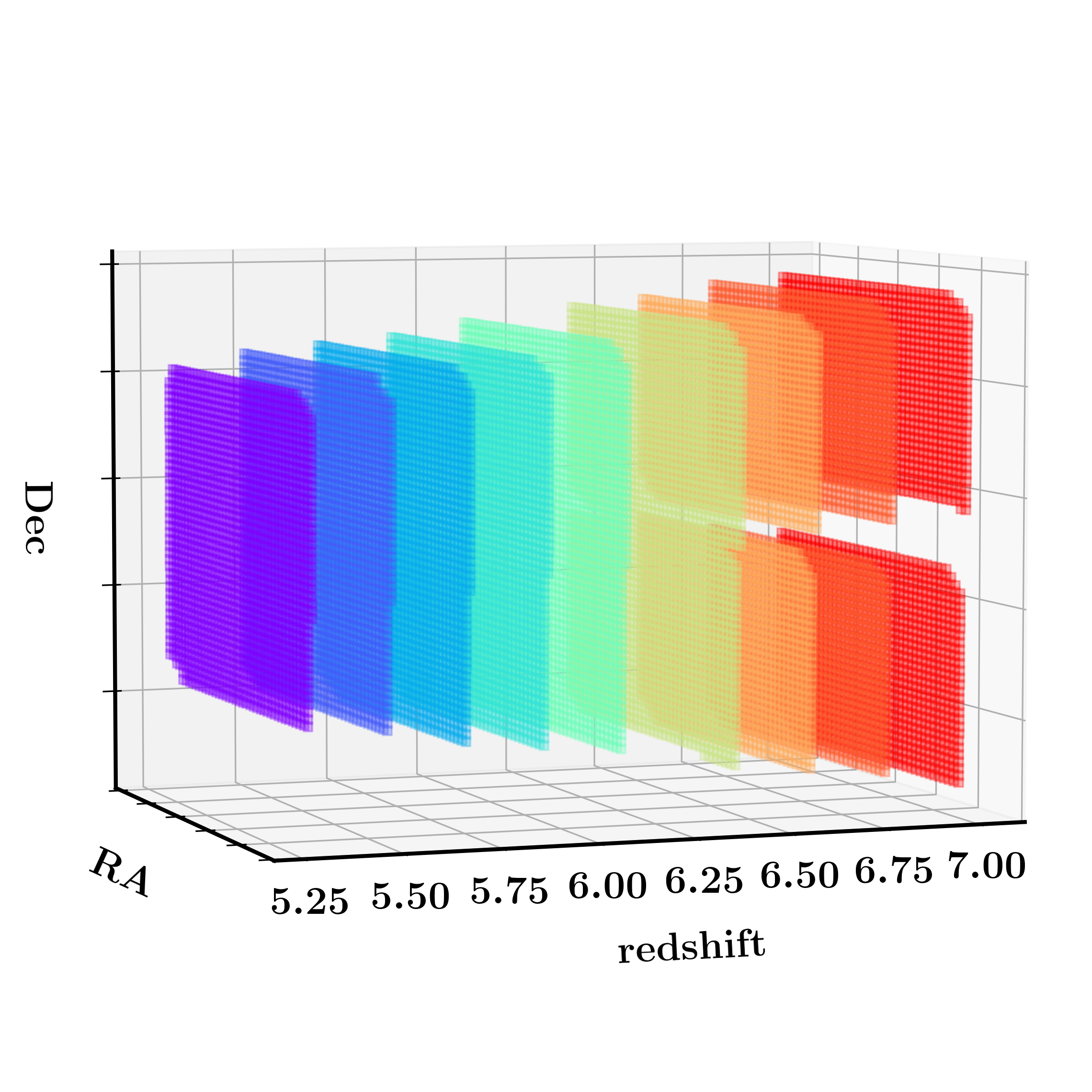}
    \caption{The spatial coverage map of quasar field J0109-3047 ($z_{\rm QSO}=6.79$) for 9 example redshifts between $z=5.3$ (blue) to $7.0$ (red). The colored grid points show the region where \texttt{covered=True}.}
    \label{fig:spatial_coverage}
\end{figure}

Finally, the coverage map is looped over $z=5.3$ to $7.0$ (which corresponds to the redshift range of \oiii-emitters that can be covered with F356W filter) to take into account of the redshift-dependent spatial coverage. In summary, this produces a stack of images in the direct-imaging coordinate frame with a boolean mask \texttt{covered} (\texttt{covered=True} where both \oiii\ lines are covered, \texttt{covered=False} otherwise), where the third dimension indexes the redshift. Figure \ref{fig:spatial_coverage} illustrates the spatial coverage maps for nine example redshifts spanning $z=5.3$ (blue) to $z=7.0$ (red). The colored grid points mark regions where \texttt{covered=True}. At $z<6.3$, the dispersed locations of the two \textit{JWST} NIRCam modules overlap, producing a contiguous combined coverage. At higher redshifts ($z>6.3$), the dispersion shifts the lines such that the module A and module B footprints separate, and gaps begin to appear between the two coverage regions. We then repeat the procedure to generate coverage maps for all 25 quasar fields.
\subsubsection{Spatially dependent sensitivity limit}
\label{sec:sensitivity}
To take into account the spatial sensitivity variations due to variations in exposure time, background level for the grism data, and the sensitivity difference between module A and module B, we injected a grid of mock sources into the observed WFSS data. We generate a grid of mock sources with fixed redshift and flux, where the spatial location of the grid points are given from the extent of the direct imaging mosaic generated from \texttt{unfold\_jwst} 
\citep[Wang et al. in preparation]{Wang2023}, a Python package for NIRCam/WFSS data reduction and emission line searching. The detailed description will be presented in Wang et al. in preparation. 

The grid is constructed such that its horizontal axis matches the dispersion direction (Fig. \ref{fig:sensitivity}), 
rather than the RA–DEC axes on the sky. This choice improves efficiency when the field of view is rotated with respect to the RA–DEC frame: if we choose the grid that aligns with RA and DEC, many grid cells would fall outside the footprint when the dispersion direction is not parallel to the sky coordinates, leading to a large number of empty grid points.
To avoid redshift confusion (i.e., multiple injected sources along the grism dispersion direction so there are more than one possible redshift solutions), the mock sources are generated on a row-by-row basis.

After generating the grid of mock emission line sources along a row, we extract the spectrum of these sources using \texttt{unfold\_jwst} \citep[Wang et al. in preparation]{Wang2023}. We show two example mock sources with $z_{\rm true}=6.5$ extracted with \texttt{unfold\_jwst} in Fig. \ref{fig:appendix_mock_o3}. To characterize the sensitivity, we inject mock sources with a known flux $f_{\rm mock}$ and measure the extracted signal-to-noise ratio $(S/N)_{\rm mock}$. Since the signal-to-noise ratio scales linearly with flux, we can determine the $5\sigma$ detection threshold at each grid point as:
\begin{equation}
F_{\rm lim} \equiv \frac{5 \, f_{\rm mock}}{(S/N)_{\rm mock}}.
\end{equation}
This gives the minimum flux a source would need at that position to be detected at $S/N = 5$.

To illustrate the grid of injected mock \oiii-emitters, we show an example using $f_{\rm mock} = 10^{-17}~\rm erg\,s^{-1}\,cm^{-2}$, which is about 5 times brighter than the typical reported sensitivity limit for the WFSS mode \citep[Wang et al. in preparation]{Wang2023}. For consistency, for the two lines, the FWHM of both \oiii$\lambda4959$ and $\lambda5007$ emission lines are set to be 200 $\rm km\,s^{-1}$. This value is picked such that it is close to the FWHM corresponding to the spectral resolution ($\sim 25$\AA), or $202.6 \rm ~km~s^{-1}$ at $3.7~\rm \mu m$.
We use a grid size of $40\times 25$, with 40 grid points along the dispersion direction (which happen to align with the Dec direction for
the example pointings shown in Fig. \ref{fig:sensitivity}), and 25 grid points perpendicular to the dispersion direction. The resulting sensitivity map for $z=6.5$ is shown in Fig. \ref{fig:sensitivity}. 

We inject the mock sources at a fixed redshift ($z = 6.5$) because the flux sensitivity does not depend on the redshift of the line, and only depends on its spatial location. Since the wavelength coverage of the \oiii\ lines is redshift dependent, we interpolate the redshift-dependent coverage map onto the sensitivity map at this single reference redshift. This allows us to evaluate the spatial variation of detectability without introducing sensitivity variations that are driven by redshift rather than by position on the detector.
For the two example sensitivity maps, the direction of dispersion is approximately along the y (Dec) axis, as indicated by the red arrows in the left panel of Fig. \ref{fig:sensitivity}.
The darker blue color corresponds to a smaller $5\sigma$ flux limit (higher sensitivity). The horizontal gap (pink shaded region) in the middle of the map corresponds to the gap between NIRCam module A and module B, and the vertical gaps and holes mark the region where the injected \oiii-emitters spectra collides with the dispersed spectra of a foreground bright source, so that they cannot correctly recover the injected source ($\Delta z=z_{\rm recover}-z_{\rm inject}>0.01$), which is twice the spectral resolution ($\Delta z = 0.005$) at the wavelength of the \oiii~$\lambda5007$ line.
We repeat this procedure for all 25 quasar fields, generating a corresponding sensitivity map for each using the outputs of the \texttt{unfold\_jwst} outputs.

\subsubsection{Selecting random sources to build the random catalog}
\label{sec:select_random}
Computing the correlation function with the Landy-Szalay estimator \citep{LZ1993, GarciaVergara2017} requires generating a mock catalog of unclustered galaxies. Therefore, we construct the random catalog using a Monte Carlo sampling approach, after generating the redshift-dependent spatial coverage and sensitivity maps. Rather than directly injecting millions of mock sources into the WFSS spectral images and extracting them with \texttt{unfold\_jwst} \citep[Wang et al. in preparation]{Wang2023}, we adopt a more efficient strategy. Direct injection would be computationally prohibitive for each ASPIRE field and would also cause spectral collisions, where the dispersed \oiii\ emission lines from different mock
sources overlap on the detector, complicating the extraction process. Instead, we use the coverage map to determine whether a source at a given sky position and redshift is geometrically observable, and the sensitivity map to assess whether it would be detectable above the local flux limit. This method allows us to generate a statistically representative random catalog that incorporates the ASPIRE selection function without the need for a full spectral simulation.

Our goal in this step is to determine the expected number of \oiii–emitters, $N_{\rm gal}$, after applying the ASPIRE selection function, so that the random catalogs used in the correlation–function estimator have the correct normalization. Once $N_{\rm gal}$ is known, we generate a random catalog with the same sky footprint and redshift-dependent selection as the data. To suppress Poisson noise in the random pair counts, we oversample the random catalog by generating $N_{\rm mock} = 5\times10^6$ random sources per field. This oversampling corresponds to multiplying the luminosity function by a fixed scale factor before drawing the random sources, and the same factor is divided out analytically when computing the correlation functions so that the estimator remains correctly normalized. The full correlation–function measurement procedure is described in \S\ref{sec:measuments}.
To generate the random catalog, we inject $5\times10^6$ mock sources uniformly in the RA and DEC domain covered by the F356W direct imaging. 
Then, for each random source we assign both a redshift (with redshift boundary $5.3<z<7.0$) and an \oiii\ luminosity by sampling the \oiii\ luminosity function measured in the EIGER program, which covers the same redshift range as ASPIRE using NIRCam/WFSS F356W \citep{Matthee2023}. Sampling the \oiii\ luminosity function is necessary because the flux limit and the spatial coverage vary across the field, so the detectability of a source depends on its intrinsic luminosity together with its redshift. Sampling the luminosity function provides a realistic intrinsic distribution of $(z, L_{\rm \oiii})$ for the random population, which can then be propagated through the spatially varying sensitivity map.


The expected number of \oiii-emitters with the selection function, $N_{\rm gal}$, is given by:
\begin{equation}
N_{\rm gal} = \int_{\mathcal{A}} d\Omega ~ \int_{z_{\rm min}}^{z_{\rm max}}  \int_{L_{\rm min}(z)}^{L_{\rm max}} dL~\phi(L, z) S(L,z;\hat{\boldsymbol{n}})
\frac{dV}{dz d\Omega} dz, 
\label{eq:sample_lf_luminosity}
\end{equation}
where $S(L,z;\hat{\boldsymbol{n}})$ is the selection function, which is a function of \oiii\ luminosity, redshift, and location on the sky. The domain on the sky is denoted as $\mathcal{A}$. As discussed in Sec. \ref{sec:mock_generation}, the redshift of the source determines the spatial coverage of WFSS imaging (see Fig. \ref{fig:spatial_coverage}), and the luminosity plus the redshift of the source determines whether it can be observed given the sensitivity limit (which is determined with the sensitivity limit map in Fig. \ref{fig:sensitivity}).  

Here, $L$ is the \oiii\ luminosity, which is given by:
\begin{equation}
L = f 4\pi d^2_L(z),
\label{eq:lum_to_flux}
\end{equation}
where $f$ is the flux and $d_L(z)$ is the luminosity distance of the injected \oiii-emitters, such that the lower bound for integration is $L_{\rm min}(z) = f_{\rm limit}\times 4\pi d_L(z)^2$.

Since the EIGER \oiii\ luminosity function is computed in a single redshift bin from $z=5.33$ to 6.96 \citep{Matthee2023}, we can simplify Eq. \ref{eq:sample_lf_luminosity} under the assumption that the \oiii\ luminosity function does not evolve with redshift in the redshift range of ASPIRE \oiii-emitters, so $\phi(L, z) = \phi(L)$. Combining this with Eq. \ref{eq:lum_to_flux}, we can change the integration variable to the \oiii\ flux, $f$:
\begin{align}
\label{eq:sample_lf_flux}
N_{\rm gal} &= \int_{\mathcal{A}} d\Omega ~  \int_{z_{\rm min}}^{z_{\rm max}} \int_{f_{\rm limit}}^{f_{\rm max}} df d^2_L(z) \phi(f, z) S(f,z,\hat{\boldsymbol{n}})\frac{dV}{dz d\Omega} dz \\
\label{eq:monte_carlo_pdf}
&= \int_{\mathcal{A}} d\Omega ~  \int_{z_{\rm min}}^{z_{\rm max}} \int_{f_{\rm limit}}^{f_{\rm max}}  dz df~U(f,z)~S(f,z,\hat{\boldsymbol{n}}) \\
&= \int_{\mathcal{A}} d\Omega ~  \int_{z_{\rm min}}^{z_{\rm max}} \int_{f_{\rm limit}}^{f_{\rm max}}  dz df~\mathcal N p(f,z)~S(f,z,\hat{\boldsymbol{n}}).
\label{eq:monte_carlo_normalized}
\end{align}
So the integration limits for the integration over flux are not a function of redshift. We can fix the flux limit $f_{\rm limit}$, which is an observed quantity given by ASPIRE observation. In addition, in Eq. \ref{eq:monte_carlo_pdf} we have defined the (unnormalized) differential number density:
\begin{equation}
U(f,z) = d_{L}^{2}(z)\phi(f,z)
\frac{dV}{dzd\Omega} ,
\end{equation}
and we normalize $U(f,z)$ to get the probability distribution $p(f,z)$ in the Monte Carlo integration
\begin{equation}
p(f,z) = \frac{U(f,z)}{\mathcal N}
\end{equation}
where the normalization $\mathcal N = \int d\Omega dz df \, U(f,z) \equiv N^{\rm perfect}_{\rm gal}$ recovers the total number of injected galaxies without selection function, $N_{\rm gal}^{\rm perfect}$.


Due to the complex nature of the WFSS selection function, we map out the integration in Eq. \ref{eq:monte_carlo_normalized} using a Monte Carlo method. For all injected mock \oiii-emitters with a random distribution in redshift and spatial location (where the random spatial location is generated in a rectangular region on the sky based on the F356W direct imaging mosaic), we apply the spatial coverage map for each of the redshift bins with $\Delta z=0.1$. Each mock source is assigned to the coverage map of the nearest redshift bin with $\Delta z=0.1$, and the corresponding map is used to decide whether both \oiii\ lines fall within the WFSS footprint at its sky position.
This procedure can be simplified as, for each random source with ($f_i, z_i, {\rm RA}_i, {\rm Dec}_i$), we first use the Boolean coverage map with redshift closest to $z_i$, then based on (${\rm RA}_i, {\rm Dec}_i$), we use a nearest grid point (NGP) algorithm to find the closest location on the map. If \texttt{Coverage=1}, we keep the injected random source, while we reject the source if  \texttt{Coverage=0}. Then, we filter the coverage-selected source based on the $5\sigma$ sensitivity map with a similar logic as the coverage map selection: we use NGP to find the closest location on the sensitivity map, then if $f_{i}$>\texttt{Flim}, the injected random source will be kept, and vice versa.

Unlike the completeness correction in the \oiii-emitters luminosity function (see, e.g., \citealt{Matthee2023, Lin2025_lf}) we use a step function for the selection function along the flux dimension. This means that we do not model the dependence of the completeness of flux at a fixed location in the Monte Carlo mock injection. In reality, the source with $f_i$=\texttt{Flim} is not guaranteed to be a detection due to Poisson fluctuation, and the case where $f_i$<\texttt{Flim} can also yield a detection, although the detection rate is much smaller than brighter sources. In our selection function model, a random source with a given luminosity is accepted only if its flux $f_i$ exceeds the local detection limit (\texttt{Flim}) at its position in the sensitivity map; otherwise, it is rejected.

Now, with the Monte Carlo injection, Eq. \ref{eq:monte_carlo_normalized} becomes: 
\begin{align}
\label{eq:monte_carlo_selection}
N_{\rm gal}
& \approx \left(\frac{dN}{d\Omega}\right) A_{\rm field}
\left[ \frac{1}{N_{\rm mock}}
\sum_{i=1}^{N_{\rm mock}} S(f_i,z_i,\hat{\mathbf n}_i) \right] \\
&= \left(\frac{dN}{d\Omega}\right)A_{\rm field}~\langle S\rangle ,
\end{align}
with $\langle S\rangle$ the effective completeness, defined as:
\begin{equation}
\langle S\rangle \equiv \frac{1}{N_{\rm mock}}
\sum_{i=1}^{N_{\rm mock}} S(f_i,z_i,\hat{\mathbf n}_i),
\end{equation}
such that the number of \oiii-emitters after the selection function is $N_{\rm gal} \approx N^{\rm perfect}_{\rm gal} \langle S\rangle $. The area of the spatial domain of the Monte Carlo injection is denoted as $A_{\rm field}$, and $N_{\rm mock}$ is the total number of injected mock \oiii-emitters.

In practice, we generate the random catalog by sampling the redshift distribution $z_i$ using MCMC draws from $p(z,f)$ based on the EIGER luminosity function, ensuring that the redshift distribution of the randoms reflects the flux limit of ASPIRE rather than a uniform distribution in $z$. In the MCMC sampling of flux and redshift, we use the minimum flux $f_{\rm limit}=5\times10^{-19} \rm erg~s^{-1}~cm^{-2}$, which is lower than the best flux limit in the sensitivity map shown in Fig. \ref{fig:sensitivity}. The sky coordinates are assigned independently using the Monte–Carlo injection procedure described above, in which sources are initially placed uniformly within the $A_{\rm field} = 12.2~\rm arcmin^2$ rectangular region enclosing the full NIRCam F356W mosaic direct imaging coverage, then kept/reject based on the coverage and sensitivity maps, for each quasar field. We note that the effective completeness $\langle S \rangle$ depends on the choice of minimum injected flux and redshift range, as well as the spatial domain for the Monte Carlo injection. In other words, a lower value for $f_{\rm limit}$ will decrease the effective completeness, as more injected source will be rejected as they are below the sensitivity limit given by the sensitivity map. Similarly, expanding the spatial domain of the injected mocks will include more mock points that are rejected by the spatial coverage map, lowering the effective completeness. For the value of redshift range and ($f_{\rm limit}, A_{\rm field}$) described above, we get an effective completeness of $\langle S \rangle=0.54$.






\section{Correlation Function Measurements}
\label{sec:measuments}

The volume-averaged correlation function $ \chi(r_{\rm p,\min}, r_{\rm p,\max}) $ is defined as (see \cite{Hennawi2006, GarciaVergara2017}):
\begin{equation}
\chi(r_{\rm p,\min}, r_{\rm p,\max}) = \frac{2}{V} \int_{r_{\rm p,\min}}^{r_{\rm p,\max}} \int_{0}^{\pi_{\max}} \xi(\rp, \pi) \, 2 \pi \rp \, d\rp \, d\pi,
\label{eq:volavg_corr}
\end{equation}
where $ \pi_{\max} $ is maximum the comoving distance for the cylindrical region, and $\pi$ is comoving distance along the line-of-sight. The volume of the cylindrical shell is $ V = \pi_{\rm geom.}(r_{\rm p,\max}^2 - r_{\rm p,\min}^2) \pi_{\max} $ is the volume of the cylindrical region considered, where $\pi_{\mathrm{geom}}$ refers to the mathematical constant. Here $\xi(\rp, \pi)$ is the 2-d correlation function. 
We choose the outermost bin edge $r_{\rm p,\mathrm{out}} = 5.6 \, \mpch$ and $\pi_{\max} = 7 \mpch$, which corresponds to a line-of-sight velocity of ${\rm d}v = 1037 \, \text{km s}^{-1}$ at $z=6.5$. 
Instead of fixing the maximum velocity for computing the correlation function counts (e.g., \citealt{Eilers2024}),
we choose to fix $\pi_{\max}$. The integration limit of $\pi_{\max} = 7\,h^{-1}\,\mathrm{Mpc}$ is chosen to include the physical extent of \oiii-emitters clustering around quasars, while integrating over the redshift-space distortions, which are more challenging to model. This
value is comparable to the ${\rm d}v = 1000\;\mathrm{km\,s^{-1}}$ adopted by \citet{Eilers2024}, so results should not differ significantly due to this choice.
Now, if we integrate the volume-averaged correlation function over $\rp$ up to the maximum projected radius ($r_{\rm p,\mathrm{out}}$), we recover the overdensity in the cylindrical volume ($\delta_{\rm out} + 1$) with radius $r_{\rm p,\mathrm{out}}$,
where $ r_{\rm p,\mathrm{out}} $ is the cylindrical radius and $ \pi_{\max} $ is the depth of the cylinder we consider.





Using the \texttt{Corrfunc} \citep{Sinha2020} python package, we compute the pair counts of \oiii-emitters within cylindrical bins defined by $r_{\rm p, min} < r_p < r_{\rm p, max}$ and $-\pi_{\rm max} < \pi < \pi_{\rm max}$. Specifically, we measure the \oiii-emitter--\oiii-emitter
pair counts $\langle D_{\rm G} D_{\rm G} \rangle$, 
\oiii-emitter--random pair counts $\langle D_{\rm G} R_{\rm G} \rangle$, and random--random pair counts $\langle R_{\rm G} R_{\rm G} \rangle$, all computed within these cylindrical volumes. 
These pair counts enter the Landy–Szalay estimator \citep{LZ1993}, which we use to compute the 2D \oiii emitter auto correlation function.
We integrate the 2D pair counts along the line of sight ($\pi$) direction to obtain the pair counts at different transverse separation ($r_{\rm p}$) bins.
We exclude the quasar environment ($|\pi|\leq\pi_{\rm max}=7\mpch$) when we compute auto-correlation pair counts in each of the 25 ASPIRE quasar fields. Since the pair counts are computed over cylindrical volumes, the resulting correlation function represents a volume-averaged measurement within each bin:
\begin{equation}
\chi_{\text{GG}}(\rp^{\rm min}, \rp^{\rm max}, \pi^{\rm max}) = \frac{\langle D_{\rm G} D_{\rm G} \rangle - 2 \langle D_{\rm G} R_{\rm G} \rangle+ \langle R_{\rm G} R_{\rm G} \rangle}{\langle R_{\rm G} R_{\rm G}\rangle}.
\label{eq:auto}
\end{equation}

For the quasar--\oiii-emitter cross correlation function, we use the estimator in \cite{Davis1983} instead:
\begin{equation}
\chi_{\text{QG}}(\rp^{\rm min}, \rp^{\rm max}, \pi^{\rm max}) = \frac{\langle D_{\rm Q} D_{\rm G} \rangle} {\langle D_{\rm Q} R_{\rm G} \rangle} -1,
\label{eq:cross}
\end{equation}
where $\langle D_{\rm Q} R_{\rm G} \rangle$ is the quasar--\oiii-emitter pair count. Since each ASPIRE pointing contains only a single quasar at a fixed location, we do not generate a random catalog for quasars ($R_Q$). Table~\ref{table:correlation_measurements} presents the measurements of the 
\oiii-emitter auto-correlation function and the quasar--\oiii-emitter cross-correlation function, computed out to $r_p = 5.6\,h^{-1}\mathrm{cMpc}$. The auto-correlation pair counts are listed as $\langle D_{\rm G} D_{\rm G} \rangle$, $\langle D_{\rm G} R_{\rm G} \rangle$, and $\langle R_{\rm G} R_{\rm G} \rangle$, while the cross-correlation counts are listed as $\langle D_{\rm Q} D_{\rm G} \rangle$ and $\langle D_{\rm Q} R_{\rm G} \rangle$. All pair counts are computed in cylindrical bins with projected separation $r_p \in (\rp^{\rm min}, \rp^{\rm max})$ and line-of-sight extent $\pi_{\rm max} = 7\,h^{-1}\mathrm{cMpc}$.

The values of $\langle D_{\rm G}R_{\rm G}\rangle$, $\langle R_{\rm G}R_{\rm G}\rangle$, and $\langle D_{\rm Q}R_{\rm G} \rangle$ must be normalized using the expected number density of \oiii–emitters after applying the ASPIRE selection function (see \S\ref{sec:select_random}). Because the survey is incomplete and flux–limited, this normalization cannot be obtained by a simple volume integral; instead it is determined through the Monte–Carlo evaluation of the selection function described in the previous section. In practice, we oversample the random catalog by injecting $5\times10^6$ random \oiii–emitters per field in order to suppress Poisson noise in the pair counts. However, the Landy–Szalay estimator requires the effective number of randoms corresponding to the true expected number of galaxies, not the oversampled catalog. Therefore, the normalization is based on the downscaled effective number of randoms $N_{R_{\rm G}}$, which is the effective number of random galaxies expected in a blank field after applying the ASPIRE selection function. The normalized pair counts are given by:
\begin{align}
\langle R_{\rm G} R_{\rm G} \rangle &= \frac{R_{\rm G} R_{\rm G}}{N_{R_G}(N_{R_G} - 1)/2}, \nonumber \\
\langle D_{\rm G} R_{\rm G} \rangle &= \frac{D_{\rm G} R_{\rm G}}{N_{D_G} N_{R_G}}, \nonumber \\
\langle D_{\rm Q} R_{\rm G} \rangle &= \frac{D_{\rm Q} R_{\rm G}}{N_{D_Q} N_{R_G}},
\end{align}
where $R_GR_G$, $D_GR_G$, and $D_QR_G$ 
are the raw pair counts in each bin computed from \texttt{Corrfunc} \citep{Sinha2020}, and $N_{D_{\rm G}}$, $N_{D_{\rm Q}}$, and $N_{R_{\rm G}}$ are the total number of 
galaxies, quasars, and random galaxies, respectively. 

\begin{table*}
\begin{center}
\caption{Pair counts for the \oiii-emitter auto-correlation and quasar--\oiii-emitter cross-correlation measurements for 25 ASPIRE fields. For the auto-correlation, we exclude the quasar environment and only select the \oiii-emitters with ($|\pi|\leq\pi_{\rm max}=7\mpch$). For the cross-correlation, we only keep \oiii-emitters in the quasar environment ($\left|\Delta v_{\rm QSO}\right|<1000\, \rm km~h^{-1}$) to compute the quasar-\oiii\ pair counts. Errors ($\chi_{\rm GG^{Pois.}_{\rm err}}$ and $\chi_{\rm QG^{Pois.}_{\rm err}}$) are computed based on $\sqrt{\langle D_{\rm G}D_{\rm G} \rangle}$ and $\sqrt{\langle D_{\rm Q}D_{\rm G} \rangle}$, respectively. In comparison, covariance-based errors ($\chi_{\rm GG^{cov}_{err}}$ and $\chi_{\rm QG^{cov}_{err}}$) are computed from the diagonal elements in the covariance matrices (See Fig. \ref{fig:covar_sim}) for a fiducial minimum halo mass model with $\log (\mming / \msun, \log (\mminq / \msun) = (10.6, 12.2)$.}
\label{table:correlation_measurements}
\resizebox{\textwidth}{!}{%
\begin{tabular}{||c | c | c | c | c | c | c | c || c | c | c | c | c | c ||} 
\hline
$\rp^{\rm min}$ & $\rp^{\rm max}$ & $\langle D_{\rm G}D_{\rm G} \rangle$ & $\langle D_{\rm G}D_{\rm R} \rangle$ & $\langle D_RD_R \rangle$ & $\chi_{\rm GG}$ & $\chi_{\rm GG^{cov}_{err}}$ & $\chi_{\rm GG^{Pois.}_{err}}$
& $\langle D_{\rm Q}D_{\rm G} \rangle$ & $\langle D_{\rm Q}D_{\rm R} \rangle$ & $\chi_{\rm QG}$ & $\chi_{\rm QG^{cov}_{err}}$ & $\chi_{\rm QG^{Pois.}_{err}}$\\
$[h^{-1}{\rm cMpc}]$ & $[h^{-1}{\rm cMpc}]$ & & & & & & & & & & \\
\hline\hline
0.06 & 0.10 & 10 & 0.142 & 0.210 & 47.32 & 19.31 & 15.07 & 3 & 0.008 & 364.49 & 176.72 & 211.02 \\
0.10 & 0.18 & 8 & 0.467 & 0.632 & 12.18 & 10.16 & 4.48 & 3 & 0.037 & 80.79 & 89.64 & 47.22 \\
0.18 & 0.32 & 38 & 1.436 & 1.887 & 19.61 & 5.21 & 3.27 & 6 & 0.134 & 43.64 & 46.72 & 18.22 \\
0.32 & 0.57 & 28 & 4.090 & 5.406 & 4.67 & 2.72 & 0.98 & 11 & 0.436 & 24.23 & 18.75 & 7.61 \\
0.57 & 1.02 & 76 & 11.156 & 14.737 & 4.64 & 1.47 & 0.59 & 34 & 1.335 & 24.46 & 8.59 & 4.37 \\
1.02 & 1.80 & 112 & 27.505 & 36.675 & 2.55 & 0.87 & 0.29 & 36 & 2.792 & 11.89 & 4.18 & 2.15 \\
1.80 & 3.17 & 124 & 52.555 & 70.748 & 1.27 & 0.58 & 0.16 & 21 & 2.769 & 6.58 & 2.24 & 1.65 \\
3.17 & 5.60 & 114 & 54.931 & 71.458 & 1.06 & 0.48 & 0.15 & 4 & 1.217 & 2.29 & 1.73 & 1.64 \\

\hline
\end{tabular}
}
\end{center}
\end{table*}

\begin{figure*}
\centering
\includegraphics[width=0.7\textwidth]{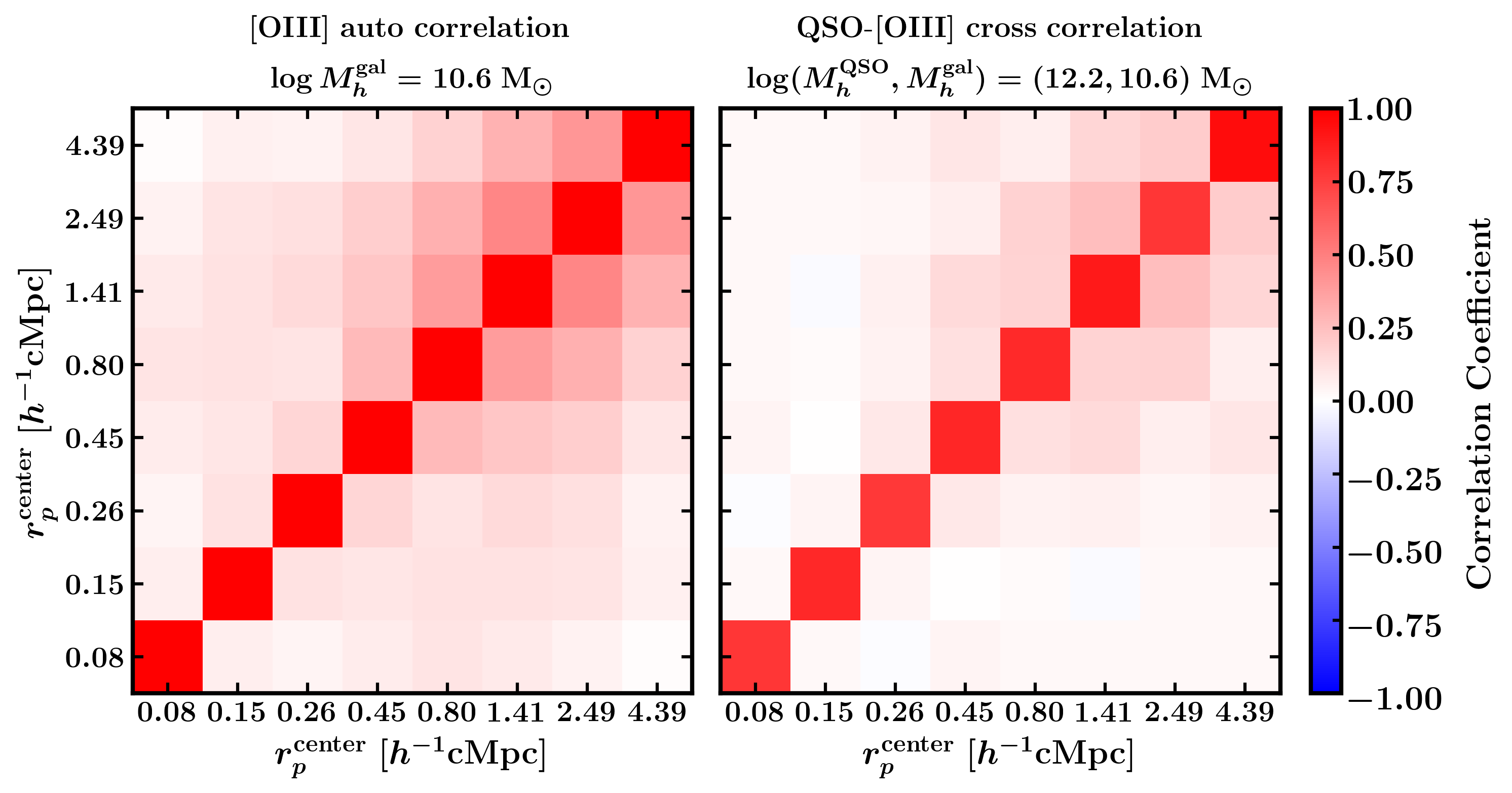}
\caption{
Correlation matrices of the volume-averaged correlation functions, normalized by their diagonal elements. 
The left panel shows the auto-correlation function of [O\,\textsc{iii}] emitters, $\chi_{\rm GG}$, and the right panel shows the quasar–[O\,\textsc{iii}] cross-correlation function, $\chi_{\rm QG}$. 
Both are computed using Eq.~\ref{eq:covariance} based on 1{,}000 mock realizations generated for the minimum-mass model with 
$\log(M_h^{\rm gal}/M_\odot)=10.6$ and $\log(M_h^{\rm QSO}/M_\odot)=12.2$, where one mock realization computes the correlation functions based on the total number counts across all 25 ASPIRE-like quasar fields.
The tick labels on each axis indicate the bin centers for the projected-radius, $r_p^{\rm center}$. 
The off-diagonal structure reflects correlated uncertainties arising from pair-count covariance and large-scale structure modes coupling across bins.
}
\label{fig:covar_sim}
\end{figure*}

\section{Covariance matrix based on mock realizations}
\label{sec:cov}
In previous studies focused on high redshift quasar clustering (e.g., \citealt{ Eilers2024, Schindler2025b}) it is commonly assumed that the uncertainty of the volume-averaged correlation function can be estimated via using the Poisson error for the pair counts, so that $\sigma _{\left \langle D_{\rm G}D_{\rm G} \right \rangle}=\sqrt{\left \langle D_{\rm G}D_{\rm G} \right \rangle}$ for the \oiii-emitter auto correlation and $\sigma _{\left \langle D_{\rm Q}D_{\rm G} \right \rangle}=\sqrt{\left \langle D_{\rm Q}D_{\rm G} \right \rangle}$ for the quasar--\oiii-emitter cross correlation function. 
However, this approach assumes that the covariance matrix is diagonal and ignores two key sources of correlated uncertainty: (1) different galaxy pairs are not statistically independent because the same galaxies contribute to multiple pairs, and (2) large-scale structure produces coherent fluctuations across the survey volume (cosmic variance). 
As a result, the Poisson approximation underestimates not only the diagonal error bars but also completely neglects the off-diagonal covariances between radial bins. Consequently, the true error budget is substantially larger and more correlated than implied by Poisson statistics alone, making simulation-based covariance estimates essential for robust clustering inferences.

In addition, with only 25 independent ASPIRE fields, a Poisson-based error estimate does not accurately capture the impact of cosmic variance (field-to-field variations), and it also ignores the correlated fluctuations between radial bins within each field; together these effects dominate the uncertainty rather than simple counting fluctuations.
Most high-redshift clustering analyses to date have assumed Poisson errors and a diagonal covariance matrix, effectively ignoring this correlated error structure. In principle, jackknife resampling can be used to estimate the covariance (e.g., \citealt{Arita2023}), but with a limited number of independent fields the jackknife does not reliably converge. The resulting covariance matrices are typically noisy and can become poorly conditioned or even singular, making jackknife resampling unsuitable for precision error estimation in our case.

The limitations of Poisson error estimates and diagonal covariance assumptions therefore necessary simulation-based covariance matrices to obtain realistic uncertainties for high-redshift clustering measurements.
Independently of getting the true error budget, simulations are also required to physically interpret the measured clustering amplitude in terms of halo masses. In previous studies, halo mass estimates are commonly obtained by fitting a power-law model to the large-scale clustering signal, converting the fitted amplitude into an effective bias, and then associating this bias with a characteristic halo mass using analytical prescriptions calibrated on linear theory or the large-scale limit of N-body simulations. This approach is not appropriate for our measurements, which probe small scales and highly biased tracers such as quasars at high redshift. At these redshifts and separations, the halo bias becomes strongly non-linear and scale dependent, and standard analytical bias models do not capture these effects. In the quasi-linear regime near the one halo to two halo transition (0.5 to 10 Mpc, for example \citealt{Zheng2005, Shuntov2025}), analytical models can significantly underpredict the clustering amplitude if these non-linearities are not treated correctly (for example \citealt{Jose2016, Shuntov2025}). Recent work has also shown that fitting a rigid power law correlation function can lead to biased halo mass estimates that reflect the imposed functional form rather than the true halo population (see Appendix Fig. C1 in \citealt{Pizzati2024b} and the discussion in \citealt{Arita2023}). In contrast, our simulation based clustering model links the observed signal to the actual spatial distribution of halos in the simulation, providing a physically grounded inference of the halo masses hosting the quasars.

This issue is especially pronounced for rare, massive halos that host quasars, where the bias is both large and strongly scale-dependent, and additional corrections arising from 
halo exclusion become necessary. To overcome these limitations, we use the FLAMINGO-10k simulation subhalo catalog \citep{Pizzati2024a, Pizzati2024b}, which adopts the ``3x2pt + all'' cosmology from Abbott et al. (2022): $\Omega_{\rm m} = 0.306$, $\Omega_{\rm b} = 0.0486$, $\sigma_8 = 0.807$, $H_0 = 68.1\, {\rm km\, s^{-1}\, Mpc^{-1}}$, $n_{\rm s} = 0.967$, with a summed neutrino mass of 0.06 eV. We compute volume-averaged correlation functions for halos above a given mass threshold using Corrfunc \citep{Sinha2020}. This simulation-based approach allows us to more accurately and self-consistently model the observed clustering signal.
Therefore, to accurately assign cosmic variance error bars to our measurements, we will need to 
generate a mock galaxy and quasar catalog in a large cosmological volume. To this end, we use the FLAMINGO-10k simulation 
(Schaller et al., in prep.; see also \citealt{Pizzati2024a, Pizzati2024b}), a dark-matter-only (DMO) run from the FLAMINGO suite of cosmological simulations \citep{Schaye2023, Kugel2023} in a $(2.8~\mathrm{cGpc})^3$ volume. It contains $10080^3$ CDM particles and $5600^3$ neutrino particles. 
The CDM particle mass is $M_{\rm dm} = 8.40 \times 10^8 ~ \msun$.
The identification of halos is performed in post-processing using the HBT-HERONS (Hierarchical Bound-Tracing) halo finder \citep{Han2012, Han2018, ForouharMoreno2025}, which tracks the formation and evolution of subhalos by following their bound particles across cosmic time. The DM halos from HBT-HERONS includes both central and satellite halos. We define subhalo mass as the peak bound mass ($M_{\rm peak}$), i.e., the maximum mass a subhalo attains over its history. HBT-HERONS directly records $M_{\rm peak}$ for each subhalo, which we adopt as our halo mass definition. Subhalo positions are defined using the most bound particle as provided by HBT-HERONS. This approach allows us to accurately represent the spatial distribution of quasars and galaxies at high redshift. A detailed description of the halo catalog from the FLAMINGO-10k simulation is provided in \citet{Pizzati2024a, Pizzati2024b}. To generate the mock quasars and \oiii-emitters, we use the simulation snapshot at $z_{\rm snap}=6.14$, which is the closest match to the median redshift of the \oiii-emitters observed in ASPIRE. We only select halos with number of dark matter particles greater than 40, which corresponds to a minimum halo mass that can be resolved in the simulation of $\log M^{\rm res}_{\rm min} = 10.5~\msun$ \citep{Pizzati2024a}.

A large number of mock realizations of \oiii-emitters around $z\sim 6.6$ quasars are generated to match the observational properties of the quasars in the ASPIRE fields. For the quasar catalog, we adopt a step-function Halo Occupation Distribution (HOD) of the form introduced in \cite{Zheng2005}, in which quasars exclusively populate halos above a minimum mass threshold $\log \mminq$. In this limit, the scatter in the halo mass threshold, $\sigma_{\log M}$, is taken to be zero, such that
\begin{equation}
\langle N_{\text{QSO}}(M_h) \rangle =
\begin{cases} 
0, & M_h < \mminq \\
1, & M_h \geq \mminq,
\end{cases}
\end{equation}
and we model \oiii-emitters in an analogous manner using a separate step-function HOD with its own minimum halo mass threshold, $\mming$. As a reference, for the $z_{\rm snap}=6.14$ snapshot, the total number of halos above a fiducial $\log \mming = 10.6~\msun$ is $2.3\times10^8$, and the total number of halos above a fiducial quasar minimum mass $\log \mminq = 12.2~\msun$ is $9.4\times10^3$.

In addition to the HOD assumption, we assign quasars and galaxies to the subhalos identified by HBT+. Each subhalo above the relevant minimum mass threshold is treated as a potential host, regardless of whether it is a central or a satellite. The positions of quasars and \oiii emitters are therefore taken to be the centers of gravitational potential of the corresponding subhalos as provided in the HBT+ catalog. No high mass cutoff is applied, so every subhalo with mass greater than the minimum mass threshold can host a quasar.
To match the observed \oiii-emitter abundance, we down-sample the halos with $M_h \ge \mming$ to reproduce the number density measured in \citep{Matthee2023}. Each selected halo is then assigned an \oiii\ luminosity drawn from the EIGER luminosity function at $z \sim 5.3$–7, under the assumption that the \cite{Matthee2023} luminosity function does not evolve strongly over this redshift range.

Note that we do not consider how \oiii\ luminosity depends on the halo mass or
the stellar mass, since our main goal is to study the impact of the minimum
halo mass on the clustering strength. A further simplification is that
centrals and satellites are treated identically: every subhalo above the
minimum mass threshold is assigned as a potential host regardless of whether
it is a central or a satellite. In practice, environmental processes acting on
satellites can alter their \oiii\ luminosity relative to centrals of the same
$M_{\rm peak}$, either suppressing or enhancing it. We show the one-halo and
two-halo contributions to the quasar--\oiii\ cross-correlation function in
Appendix~\ref{app:1h2h}. Because our clustering analysis depends primarily on
the minimum halo mass threshold rather than the detailed luminosity assignment,
and the overall error budget is dominated by cosmic variance, we expect this
simplification to have a modest impact on the inferred halo masses. A detailed
investigation of separate central and satellite treatments will be presented
in Huang et al.\ (in preparation). In practice, the \oiii\ luminosity is randomly assigned to the mass selected \oiii-emitter host halos,
and the distribution of mock $f_{\rm \oiii}$ follows the observed luminosity function with a flux limit of $f_{\rm \oiii}>2\times10^{-18}~\rm erg\,s^{-1}\,cm^{-2}$. This flux limit is below the flux limit of the actual data, and these mock sources will be folded through selection effects in the subsequent steps. 

The previous construction of the selection function (see \S \ref{sec:coverage}-\ref{sec:sensitivity}) based on each ASPIRE pointing enables us to generate realistic mock catalogs. For each mock realization, we use the quasar and \oiii-emitter catalog selected based on the minimum masses, and we randomly select 25 quasars (which is the same as the total number of the ASPIRE fields) among the selected quasar catalog. For each quasar field, we re-center the \oiii-emitters around the central quasar location to avoid the edge effect with the simulation box, taking advantage of the periodic boundary condition of the FLAMINGO-10k simulation. Then, we assign the ``observed'' sky coordinates and redshifts based on the true location of the quasars targeted in the ASPIRE survey. The sky coordinates and redshifts of the galaxies in the quasar fields are assigned based on transforming the relative position between the galaxy and the quasar in the simulation box to the observed frame. 
In this step, we include redshift space distortions such that the mock redshift ($z_{\rm mock}$) of each \oiii\ emitter reflects both its position in the simulation box and its peculiar velocity. For each quasar field, we take the observed quasar redshift $z_{\rm QSO}$ as the reference. The real space redshift of a galaxy is obtained by converting its comoving line of sight separation $r_{\parallel}$ relative to the quasar halo into a redshift offset using the cosmological comoving distance–redshift relation. We then include redshift space distortions by accounting for the galaxy peculiar velocity along the line of sight. In this way, the mock redshift encodes both the cosmological redshift offset induced by the galaxy position relative to the quasar and the additional Doppler shift from peculiar motions.


Then, we apply the spatial and spectral coverage maps field by field to each quasar field, based on the sky position and redshift of the mock \oiii-emitters. For the subset of mock \oiii-emitters that fall within the coverage map, we apply the corresponding sensitivity map to exclude emitters with fluxes below the local flux limit. We also apply the same angular mask used in the data to exclude the region around the quasar position. This procedure is identical to the random galaxy selection described in \S\ref{sec:select_random}, with the random points replaced by mock \oiii-emitters, ensuring that both samples are passed through the same selection function.

All mock realizations are generated using a single simulation snapshot at redshift $z_{\rm snap}$. The ASPIRE \oiii-emitter sample spans the redshift range $5.3 < z < 7.0$, corresponding to a comoving line of sight distance of $0.709$ cGpc. This scale is smaller than the simulation box size, allowing us to model the full redshift range using a single snapshot. In doing so, we neglect the evolution of large scale structure across this redshift interval. This approximation is reasonable given the weak evolution of the \oiii\ luminosity function over this range, although we note that any evolution in halo clustering over the light cone is not captured in our mocks.

Using the selected mock ASPIRE \oiii-emitter observation, 
we compute the correlation function with Eq. \ref{eq:auto} and \ref{eq:cross}, where the random catalog passes the same selection function as the mock \oiii-emitters. One realization contains 25 fields to assemble the ASPIRE observation of the 25 quasar fields. We repeat the steps above $N_{\rm real}=1000$ times to get 1000 correlation function measurements. 


Finally, we compute the covariance of the binned correlation function vector $\boldsymbol{\chi_a}$. 
The covariance matrix elements are computed as:
\begin{equation}
\mathcal{C}_{ij}(\chi_a)=
\frac{1}{N_{\rm real}}
\sum_{k=1}^{N_{\rm real}}
\bigl(\chi_{a,i}^{k}-\langle \chi_{a,i}\rangle\bigr)
\bigl(\chi_{a,j}^{k}-\langle \chi_{a,j}\rangle\bigr),
\label{eq:covariance}
\end{equation}
where indices $i,j=1,\dots,N_{\rm bin}$ label radial bins, and $k=1,\dots,N_{\rm real}$ label realizations. And $\chi_{a,i}^{k}$ is the $i$-th bin for the $k$-th mock realization of the volume averaged correlation function. The subscript $a\in[\rm GG, QG]$, and $N_{\rm real}=1000$ is the total number of realizations. For the mean in Eq. \ref{eq:covariance}, we choose to use the modeled halo correlation for given minimum masses computed with full halo catalog from FLAMINGO-10k simulation with mock \oiii-emitters passed through the selection function machinery. Note the normalization is $1/N_{\rm real}$ not $1/(N_{\rm real}-1)$ because our mean value is computed from the full mock catalog, rather than the $N_{\rm real}=1000$ realizations.
Since the mock realizations are noisy versions of the model, for sufficiently large $N$, the mean correlation function vector of the realizations converges to the modeled correlation function vector:
\begin{equation}
\langle{\boldsymbol{\chi_a}}\rangle
~\equiv~
\frac{1}{N_{\rm real}}\sum_{i=1}^{N_{\rm real}} \boldsymbol{\chi_a}^i
\xrightarrow{N_{\rm real} \to \infty} 
\boldsymbol{\chi_a}^{\text{model}}.
\end{equation}

In summary, the following steps outline the model dependent procedure for generating mock realizations of the correlation functions and constructing the corresponding covariance matrices for each choice of the halo mass parameters ($\mming, \mminq$):
\begin{enumerate}
    \item Select quasars and \oiii-emitters based on a step-function halo occupation distribution, with a fixed minimum halo masses ($\log \mminq$, $\log \mming$) that host quasar and \oiii-emitters.
    \item Assign the location and redshift to the quasars and galaxies, based on the ASPIRE observations. Then apply the selection function for each ASPIRE field to the mock.
    \item Compute the \oiii-emitter auto correlation functions, $\chi_{\rm GG}$, excising the \oiii-emitters in the QSO environment ($|\Delta \pi|<7 \mpch$), just like was done for the real data.
    Compute the quasar--\oiii-emitter cross correlation functions, $\chi_{\rm QG}$, with the \oiii-emitters in the quasar environments ($|\Delta \pi|<7 \mpch$).
    \item Compute the covariance matrix for $\chi_{\rm QG}$ and $\chi_{\rm GG}$ based on the correlation functions from $N_{\rm real}=1000$ realizations.
    \item Perform all of these steps for a 2d grid of minimum halo masses, ($\mming, \mminq$).
\end{enumerate}

The additional details of the mock generation based on the simulation catalog are described in Huang et al. in preparation. The left panel of Fig. \ref{fig:covar_sim} shows an example correlation matrix (covariance matrix in Eq. \ref{eq:covariance} normalized by its diagonal elements)
\begin{equation}
\mathcal{R}_{ij}(\chi_a) \equiv
\frac{\mathcal{C}_{ij}(\chi_a)}
{\sqrt{\mathcal{C}_{ii}(\chi_a),\mathcal{C}_{jj}(\chi_a)}},
\label{eq:correlation_matrix}
\end{equation}
computed using $N_{\rm real}=1000$ realizations for the
\oiii-emitter auto correlation function with \oiii\ host halo minimum mass $\log (\mming / \msun) = 10.6$. And the right panel of Fig. \ref{fig:covar_sim} shows the correlation matrix computed with ($\log (\mming / \msun), \log (\mminq / \msun)) = (10.6, 12.2)$
The positive off-diagonal elements in the correlation matrix are a natural outcome of how the auto-correlation function is measured. Since the measurement is based on counting \oiii-emitters pairs, any large-scale fluctuation in the \oiii\ number density (e.g., an overdense region) tends to boost the number of pairs at all separations. As a result, the auto-correlation function values in different bins tend to increase or decrease together, creating positive correlations between bins. In other words, since the same galaxy can contribute to multiple pair separations, we expect positive correlation in between the bins. The QSO–\oiii cross correlation shows weaker off diagonal terms because each pointing contains only one quasar, so the measurement reduces to counting \oiii emitters around that single object at different separations. A galaxy contributes to only one separation bin relative to the quasar, rather than to many pair combinations as in the auto correlation, so fluctuations at one scale do not propagate across bins as strongly. This leads to weaker off diagonal correlations in the covariance for QSO–\oiii cross correlation compared with \oiii-emitter auto correlation.

In our subsequent paper (Huang et al., in preparation), we show that the dominant source of uncertainty in estimating the halo masses of both \oiii-emitters and quasars comes from field-to-field (cosmic) variance. Relying only on Poisson counting errors underestimates the true uncertainties by  about $0.7$ dex for the quasar halo mass and about $0.5$ dex for the \oiii-emitters halo mass in the ASPIRE survey. 
Therefore, accurately accounting for field-to-field variance is critical 
for reliable halo mass measurements in clustering analysis.


\section{Inferring the correlation length and the halo mass}
\label{sec:inference}
\subsection{Correlation length}
\label{sec:corr_length}
\begin{figure}
\centering
	\includegraphics[width=\columnwidth]{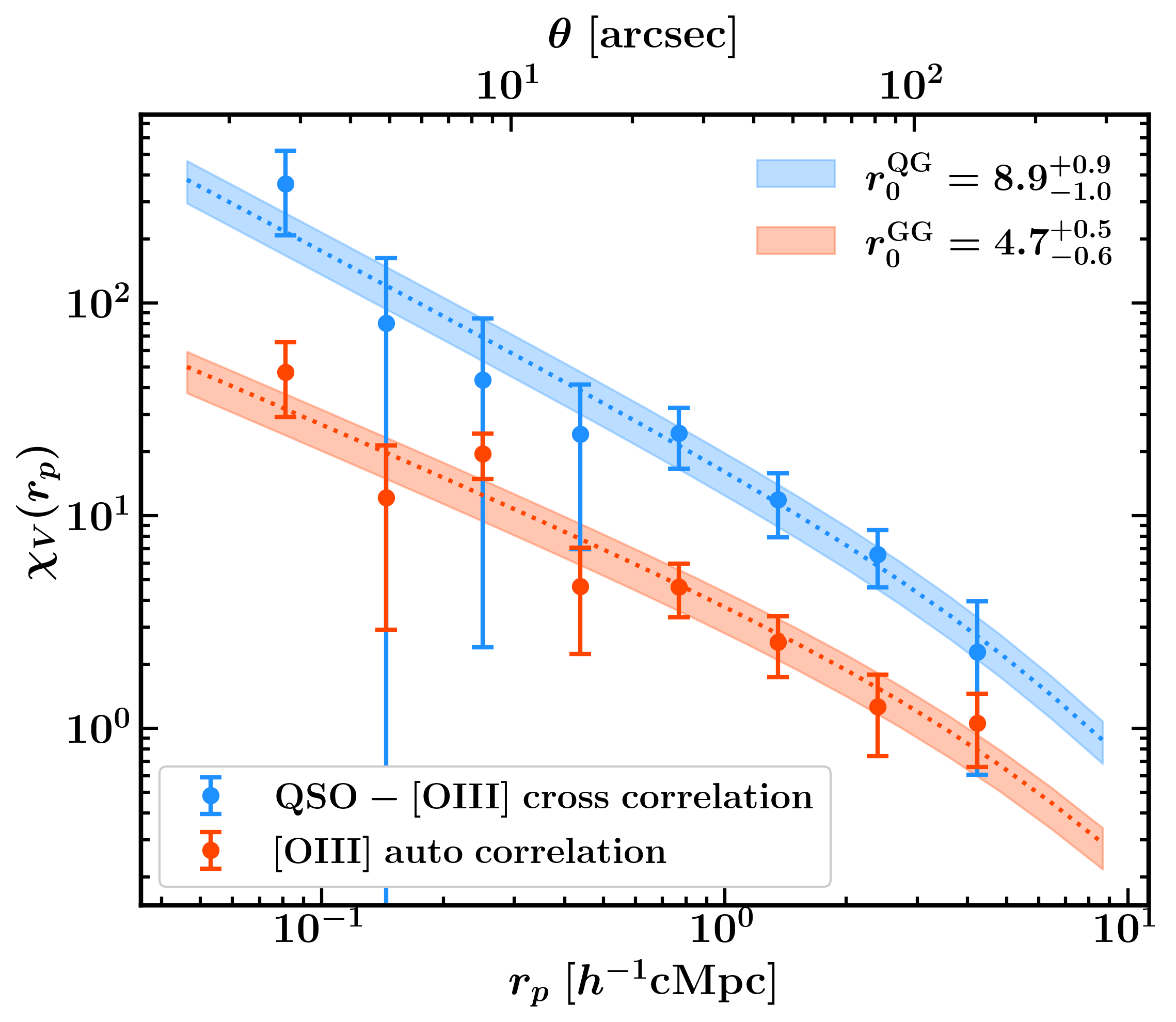}
    \caption{Power law fit to the volume averaged  \oiii-emitter auto correlation function (red) and the quasar--\oiii-emitter cross correlation function (blue). The fit assumes the powerlaw index for the 2d correlation function, $\gamma_{\rm QG}=2.0$ and $\gamma_{\rm GG}=1.8$. For each case, we sample the posterior distribution of the correlation length $r_0$, and show the median model (solid line) along with the $1\sigma$ credible interval (shaded region).
    }
    \label{fig:corrfit_r0}
\end{figure}
To determine the correlation lengths for \oiii-emitters and quasars, we employ an MCMC approach to fit the parameters $ r_{0, \rm GG} $ for the \oiii\ auto-correlation function and $r_{0, \rm QG}$ for the quasar--\oiii\ cross-correlation function.  We use a likelihood function based on the covariance matrix derived from the mock correlation functions forward modeled from the FLAMINGO-10k simulation (Schaller et al., in prep.; see also \citealt{Pizzati2024a, Pizzati2024b}),
ensuring that our parameter estimates accurately account for cosmic variance and bin-to-bin correlations.

\subsubsection{Likelihood function}
By construction, $\chi_{\rm QG}$ is computed only considering the \oiii-emitters in the quasar environment, and $\chi_{\rm GG}$ is computed with \oiii-emitters after excising the quasar environment. Therefore, we can assume statistical independence between the two correlation functions ($\chi_{\rm QG}$, $\chi_{\rm GG}$), and thus we do not consider the cross-covariance between auto- and cross-correlation functions. So the total likelihood function for the $\mathcal{L}_{\text{tot}}$ is:
\begin{equation}
\mathcal{L}_{\text{tot}}(X \mid \Theta) = \mathcal{L}_{\text{GG}} \cdot \mathcal{L}_{\text{QG}},
\label{eq:total_likelihood}
\end{equation}
where $X$ is a vector of observed correlation functions, $X = (\chi_{\text{GG}}, \chi_{\text{QG}})$ and $\Theta$ is the modeled parameters for the minimum masses of halos hosting quasars and galaxies,
$\Theta = \Theta(\mminq$ and $\mming)$. 
The individual likelihoods $\mathcal{L}_{\text{GG}}$ and $\mathcal{L}_{\text{QG}}$ are:

\begin{equation}
\mathcal{L}_{\rm GG} = \mathcal{N}\left(\chi_{\rm GG}^{\rm obs} \middle| \langle\chi_{\rm GG}\rangle(\mming), \boldsymbol{C}_{\rm GG}(\mming)\right)
\label{eq:auto_likelihood}
\end{equation}

\begin{equation}
\mathcal{L}_{\text{QG}} = \mathcal{N}(\chi_{\rm QG}(\mminq, \mming), 
\boldsymbol{C}_{\rm QG}(\mminq, \mming)).
\end{equation}
Here, $\mathcal{N}$ represents a Gaussian distribution with mean $\chi$ and covariance $\boldsymbol{C}$. Specifically, $\chi$ represents the model prediction for the correlation function, computed with all halos in FLAMINGO-10k satisfying the minimum mass threshold. And 
$\boldsymbol{C}$ represents the associated covariance matrix with given minimum masses. For each minimum mass model, 
we generate $N=1000$
realizations and compute the covariance matrix using Eq. \ref{eq:covariance}. Since the cross correlation function between quasar and \oiii-emitters depends not only on $\mminq$, but also on the tracer halo masses $\mming$, our model for cross correlation involves both halo minimum masses for quasar and \oiii-emitters. Therefore, we generate the model correlation functions and compute the covariance matrix for the quasar-\oiii-emitter cross correlation function based on a 2D grid of minimum masses, where $\log \mming=10.5-11.5~\msun$ in 0.1 dex step, and $\log \mminq=11.0-12.6~\msun$ in 0.1 dex step. Then we use the nearest grid point method to find the covariance matrix at ($\mming, \mminq$) to compute the likelihood function.

We model the correlation function as a simple power law, defined as $\xi(R,\pi)=\left(r/r_0\right)^{-\gamma}$, where the scale length $r_0$ is the correlation length and $\gamma$ is the slope. If $R$ is the projected separation perpendicular to the line of sight and $\pi$ is the separation along the line of sight, then $r=\sqrt{R^2+\pi^2}$. To compare the model with the observed measurements, we integrate $\xi(R,\pi)$ over the cylindrical ring volume (Eq. \ref{eq:volavg_corr}) to obtain the volume-averaged model prediction, which we denote as $\langle \chi_a \rangle$.

For the correlation length inference with the power law model, we adopt the full multivariate normal likelihood with a model-dependent covariance matrix,
\begin{equation}
\begin{aligned}
\log \mathcal{L}_a(r_0 \mid \chi_{a,{\rm obs}}, \boldsymbol{C}_a)
&=
-\frac{1}{2}
(\chi_{a,{\rm obs}}-\langle \chi_a \rangle)^{\top}
\boldsymbol{C}_a^{-1}
(\chi_{a,{\rm obs}}-\langle \chi_a \rangle) \\
&\quad
-\frac{1}{2}\log\det\left(2\pi\boldsymbol{C}_a\right),
\end{aligned}
\label{eq:likelihood_powerlaw}
\end{equation}
where $a \in [\rm GG, QG]$, and $\chi_{a,\rm obs}$ is the observed volume-averaged correlation function. In addition, $\boldsymbol{C}_{\rm GG}=\boldsymbol{C}(\mminq)$ and $\boldsymbol{C}_{\rm QG}=\boldsymbol{C}(\mminq, \mming)$ are the minimum halo mass-dependent covariance matrix estimated from mock realizations. The total likelihood for the power law fit is therefore given by Eq. \ref{eq:total_likelihood}. 

For the fit to the correlation length, the power law model cannot be used to generate realistic minimum halo mass dependent mocks. In particular, the fitted parameter $r_0$ does not correspond to a physical halo mass or occupation model, and therefore we cannot construct a covariance matrix $\boldsymbol{C}(r_0)$ that varies with the power law model. Hence, for the correlation length analysis, we adopt a fixed covariance matrix derived from representative fiducial halo masses. The covariance matrix is computed using fiducial minimum masses $\log(\mming/\msun,\log(\mminq/\msun))=(10.6,12.2)$. For these fiducial halo masses, the normalization term in Eq. \ref{eq:likelihood_powerlaw}, $\log\det(2\pi\boldsymbol{C}_a)$, is constant with respect to $r_0$ and can be omitted without affecting the inferred posteriors. However, because elsewhere the covariance matrix $\boldsymbol{C}_a=\boldsymbol{C}_a(\mming,\mminq)$ varies with the halo mass model, we retain the full likelihood expression to emphasize that the determinant term is model dependent and, in general, cannot be neglected.
\begin{figure}
\centering
	\includegraphics[width=\columnwidth]{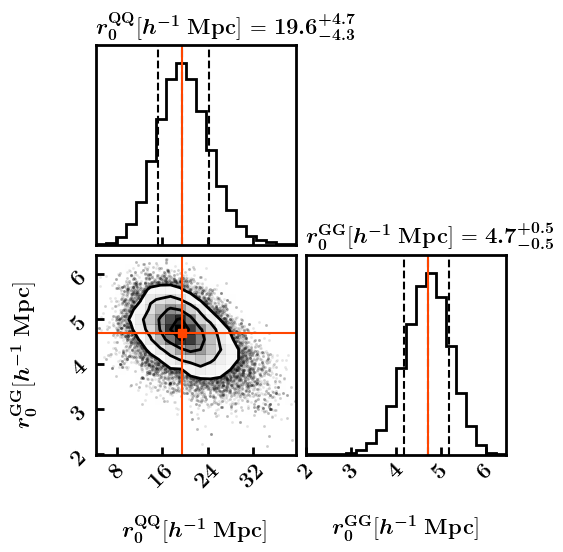}
    \caption{MCMC Posterior distribution joint power law model fit to the ASPIRE \oiii-emitter auto correlation function and quasar-\oiii-emitter cross correlation with fixed power law index $\gamma_{\rm QQ}=2$ and $\gamma_{\rm GG}=1.8$. The likelihood function is given by Eq. \ref{eq:total_likelihood}. The 16-84th percentile range in the marginalized posterior is shown with vertical dashed lines. 
    }
    \label{fig:corner_r0_fit}
\end{figure}

\begin{figure*}
    \centering
    \includegraphics[width=0.7\textwidth]{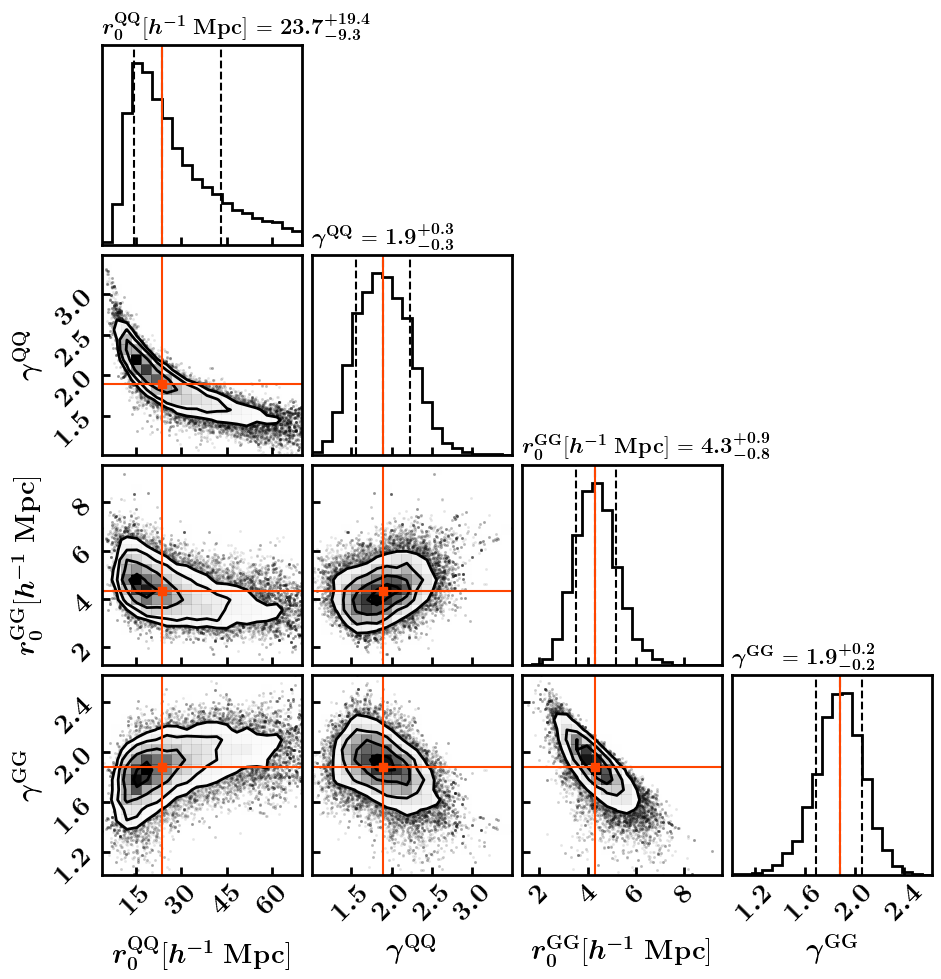}
    \caption{MCMC Posterior distribution for the joint power law model fit to the ASPIRE \oiii-emitter auto correlation function and quasar-\oiii-emitter cross correlation allowing the power law slope to vary. The 16-84th percentile range in the marginalized posterior is shown with vertical dashed lines.}
    \label{fig:corner_QQ_varygamma}
\end{figure*}

Fig. \ref{fig:corrfit_r0} shows the power-law fit to correlation function with the covariance matrix.
We fix the power-law index to $\gamma_{\rm QG}=2.0$ and $\gamma_{\rm GG}=1.8$. The fit yields $r_0^{\rm QG}=8.9^{+0.9}_{-1.0}\mpch$ for quasar--\oiii-emitter cross correlation function and $r_0^{\rm GG}=4.7^{+0.5}_{-0.6}\mpch$ for \oiii-emitter auto correlation function,
where the uncertainties are estimated as the 16-84th percentile interval around the median. 

Assuming deterministic bias, where galaxies and quasars trace the same underlying dark matter distribution, the quasar auto correlation function can be expressed as $\xi_{\rm QQ} = \xi^2_{\rm QG}/\xi_{\rm GG}$ (see e.g., \citealt{Croom2001, GarciaVergara2017, Eilers2024}). We follow \citep{Eilers2024} to jointly fit the \oiii-emitter auto correlation and the quasar-\oiii-emitter cross correlation function using the joint likelihood function in Eq. \ref{eq:total_likelihood}. We assume a power law correlation function with the fixed power law index ($\gamma_{\rm QQ}=2$) for quasar auto correlation function to be consistent with previous work (e.g., \citealt{Shen2007, Eftekharzadeh2015, Eilers2024}) for better comparison. And we keep $\gamma_{\rm GG}=1.8$ for the \oiii-emitter auto correlation function. The resulting posterior distribution for the joint fit of $r_0^{\rm QQ}$ and $r_0^{\rm GG}$ is shown in Fig. \ref{fig:corner_r0_fit}.
And our joint fit yields quasar auto correlation length $r_0^{\rm QQ}=19.6^{+4.7}_{-4.3}\mpch$.

When we allow the power-law index to vary, we obtain 
$r_0^{\rm QG} = 10.3^{+3.4}_{-2.3} \mpch$ with $\gamma_{\rm QG} = 1.8 \pm 0.3$ and 
$r_0^{\rm GG} = 4.6^{+0.9}_{-0.8} \mpch$ with $\gamma_{\rm GG} = 1.9 \pm 0.2$. 
For the joint fit to the \oiii-emitter auto correlation and the quasar-\oiii-emitter cross correlation function, we find $r_0^{\rm QQ} = 23.7^{+19.4}_{-9.3} \mpch$ and $\gamma_{\rm QQ} = 1.9 \pm 0.2$, as shown in Fig. \ref{fig:corner_QQ_varygamma}. This quasar auto correlation length is larger than the quasar auto-correlation length inferred when the power-law index $\gamma_{\rm QG}$ is fixed. This discrepancy arises because, for a fixed $r_0^{\rm QQ}$, an increase in $\gamma_{\rm QQ}$ leads to a higher model prediction for $\chi_{\rm QQ}$ at small projected separations ($r < r_0^{\rm QQ}$), where most of the bins are located. To compensate for this increase in $\gamma_{\rm QQ}$, a smaller $r_0^{\rm QG}$ is required.

\subsection{Halo Masses}
\label{sec:minmass}
In the previous section (\S \ref{sec:corr_length}), we jointly fitted the correlation functions with the power law model to get the quasar auto correlation length. However, instead of deriving halo masses from analytical models that relate the clustering length to the large scale bias, we use simulations to predict the small scale correlation functions. This is because our simulation-based halo model explicitly depends on the minimum halo masses and more accurately reproduces the observed correlation shape. We then directly fit the measured quasar–\oiii-emitter cross correlation function together with the \oiii-emitter auto correlation function to jointly infer the minimum halo masses of quasars and galaxies traced by \oiii emitters.



\begin{figure*}
    \centering
    \begin{subfigure}[t]{0.55\textwidth}
        \includegraphics[width=\linewidth]{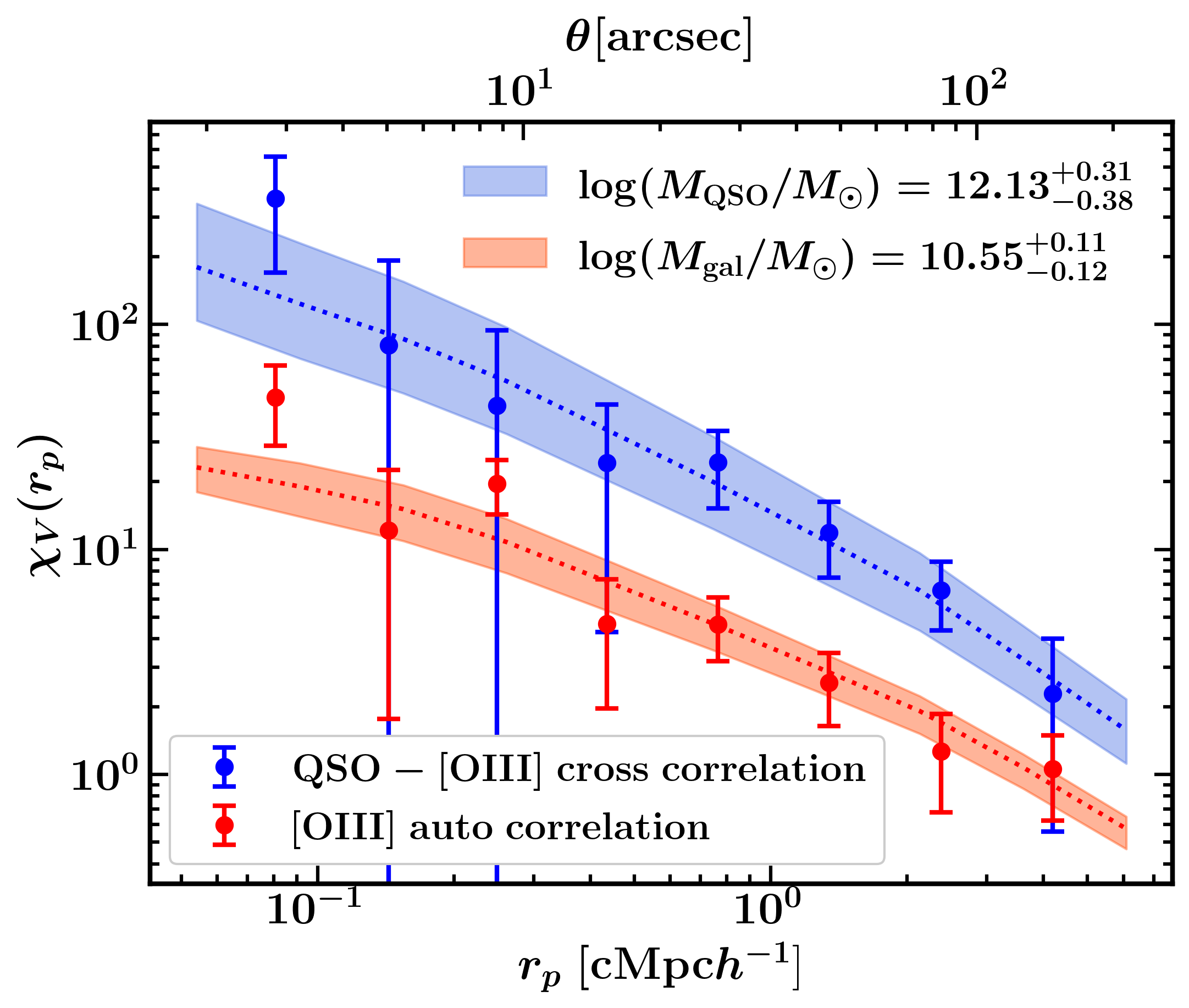}
        \label{fig:corr_fit_min}
    \end{subfigure}
    \begin{subfigure}[t]{0.4\textwidth}
        \includegraphics[width=\linewidth]{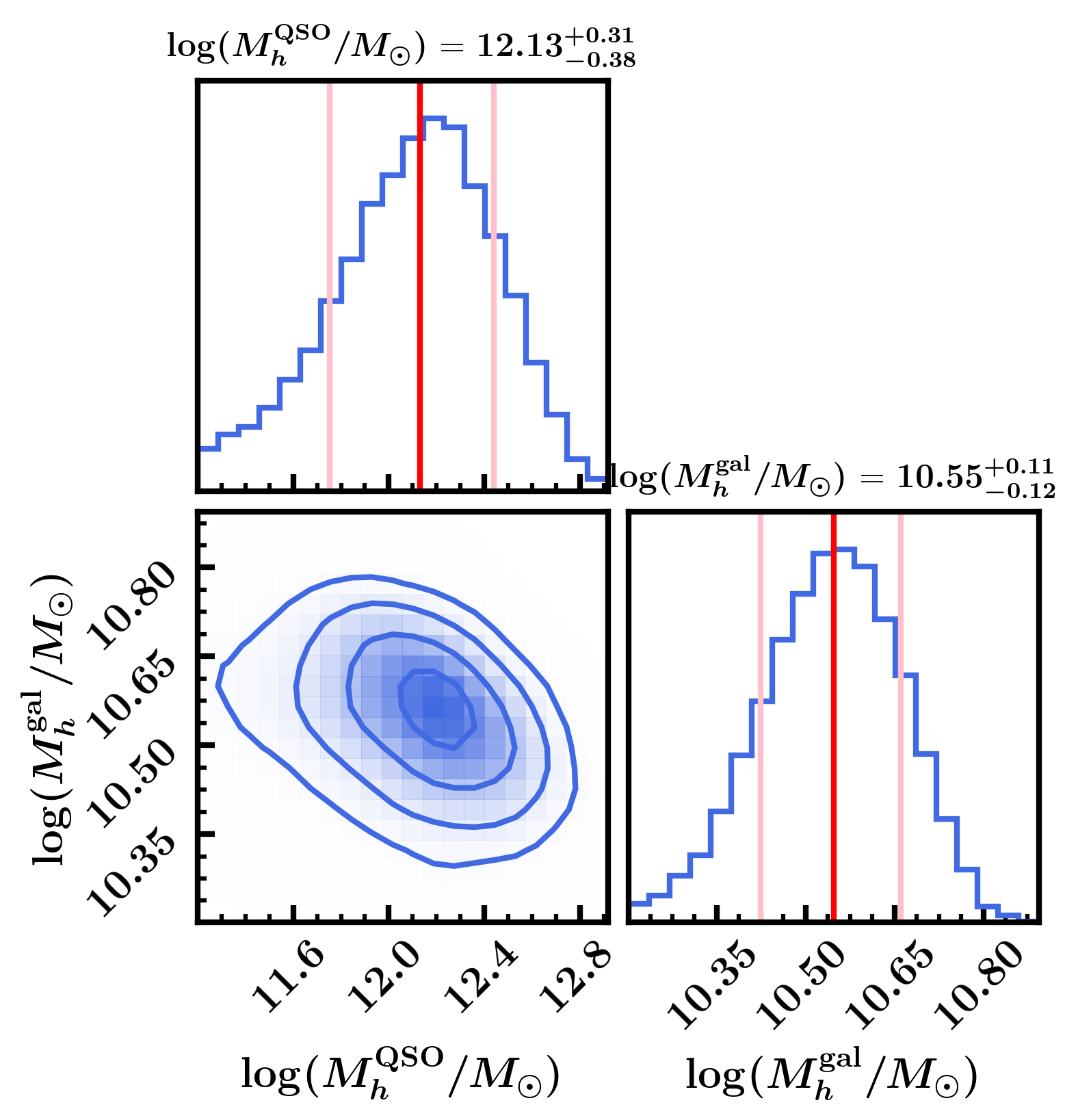}
        \label{fig:corner_minfit}
    \end{subfigure}
    \caption{Joint analysis of quasar and \oiii-emitters clustering in the ASPIRE sample. 
    \textbf{\textit{Left:}} Joint fit to the volume-averaged \oiii-emitters auto-correlation function (red) and quasar–\oiii-emitters cross-correlation function (blue). 
        The red dotted curve and shaded region show the best-fit minimum galaxy host halo mass model and its 1$\sigma$ range ($\log \mming/\rm \msun = 10.55^{+0.11}_{-0.12}$). 
        The blue dotted curve and shaded region show the fitted quasar host minimum mass model ($\log \mminq/\rm \msun = 12.13^{+0.31}_{-0.38}$).
    \textbf{\textit{Right:}} MCMC posterior distribution for the joint minimum halo mass fit using the combined \oiii-emitters auto and quasar–\oiii\ cross correlation likelihoods (blue; see Eq.~\ref{eq:total_likelihood}). 
        The 1$\sigma$ marginalized posterior ranges are marked by vertical pink lines.}
    \label{fig:aspire_joint}
\end{figure*}

The result of the MCMC fitting is shown in Fig.~\ref{fig:aspire_joint}. We jointly sample the posterior probability distribution in the two-dimensional parameter space of galaxy and quasar minimum halo masses, $\log \mming$ and $\log \mminq$. The red curve and shaded region 
represent the median model with the 16$\%$ and 84$\%$
credible interval of the \oiii-emitters auto-correlation function, corresponding to a galaxy minimum halo mass of $\log \mming/\rm \msun = 10.55^{+0.11}_{-0.12}$. The blue curve and shaded region show the corresponding result for the quasar--\oiii-emitters cross-correlation function, yielding a quasar minimum halo mass of $\log \mminq/\rm \msun = 12.13^{+0.31}_{-0.38}$. The shaded regions are constructed by sampling the joint posterior distribution of $(\mming,\mminq)$ and computing the distribution of the corresponding model correlation functions. Compared to the power law fit in Fig.~\ref{fig:corrfit_r0}, the credible intervals here are broader because we use a mass-dependent covariance matrix: as the halo mass increases, the tracer bias and clustering amplitude grow, which amplifies both the diagonal variances and off-diagonal correlations in the covariance matrix.

Because the quasar and galaxy minimum halo masses both influence the cross-correlation signal, we perform a joint fit to constrain them consistently. As indicated in Fig.~\ref{fig:aspire_joint}, the two parameters are correlated: increasing $\mming$ boosts both the \oiii\ auto-correlation function and the quasar–\oiii\ cross-correlation function, so a corresponding decrease in $\mminq$ is required to reproduce the observed cross-correlation amplitude.


\section{Quasar and Galaxy Occupation Fraction}
\label{sec:duty_cycle}
The duty cycle, or the "occupation fraction", is defined as the fraction of observed quasars or galaxies relative to the total number of dark matter halos that can host quasars or galaxies. Under the step function HOD assumption (see, e.g., \citealt{Berlind&Weinberg2002} and Section 2 in \citealt{Pizzati2024a}),
the number density of the quasar/galaxy host halos is
\begin{equation}
    n_{\rm halo}=\int^{\infty}_{0}dM\frac{dn}{dM} {\rm HOD}(M)=\int^{\infty}_{M_{\rm min}}dM\frac{dn}{dM}.
\end{equation}
Therefore, an increase in the minimum halo mass threshold $M_{\rm min}$ (which corresponds to stronger clustering) reduces the total number of halos capable of hosting a quasar or galaxy ($n_{\rm halo}$), thus lowering the inferred duty cycle, which is defined as
\begin{equation}
f_{\rm duty}=\frac{n}{n_{\rm halo}(M_h>M_{h, \rm min})},
\end{equation}
where $n$ is the number density of quasar or galaxy. This duty cycle reflects the fraction of halos above this mass threshold that are actively hosting a quasar or \oiii-emitter 
at a given time. When the duty cycle approaches unity, every halo above the minimum halo mass is assumed to host an active quasar or galaxy. Conversely, a low duty cycle implies that only a small fraction of eligible halos are active at any given time, indicating that quasar or galaxy activity is highly episodic or short-lived relative to the Hubble time. In the following analysis, we use the minimum halo mass inferred from clustering measurements to estimate the duty cycle.

\subsection{Duty cycle or \oiii-emitter}
\label{sec:o3_duty}
From the joint fit, we derived the minimum mass hosting \oiii-emitters to be $\log (\mming / \msun) = 10.55^{+0.11}_{-0.12}$. Based on a step-function HOD, the number density of the halos above the minimum mass is computed by integrating the halo mass function for masses larger than the inferred minimum mass. Instead of assuming a certain analytical form of the halo mass function, we can directly compute the number density of the halo from FLAMINGO-10k. We find
$n_{\rm halo}(M_h>\mming)=6.25^{+3.21}_{-2.21}\times 10^{-3} ~\rm cMpc^{-3}$. Integrating the \oiii\ luminosity function from \cite{Matthee2023} using Eq. \ref{eq:sample_lf_luminosity} without using the selection function, yields an observed \oiii\ number density above the flux limit of $f_{\rm min} = 2\times 10^{-18}~\rm erg\,s^{-1}\,cm^{-2}$ to be $n_{\rm \oiii}=1.54\times 10^{-4} ~\rm cMpc^{-3}$. This corresponds to a duty cycle of $f^{\rm \oiii}_{\rm duty} = \frac{n_{\rm \oiii}} {n_{\rm halo}(M_h>\mming)}\times100\% = 2.5^{+1.3}_{-0.8}\%$. 

\subsection{Quasar duty cycle}
\label{sec:qso_duty}
Similar to the \oiii-emitters duty cycle, we can infer the duty cycle for quasars using their halo minimum mass from the joint fit the for quasar-\oiii-emitter cross correlation function and the \oiii-emitter auto correlation function, 
where $\log (\mminq / \msun) = 12.13^{+0.31}_{-0.38}$. 
This yields $n_{\rm halo}(M_h>\mminq)=7.67^{+109.30}_{-7.16}\times 10^{-7} ~\rm cMpc^{-3}$. Integrating the observed $z\approx 6$ quasar UV luminosity function from \cite{Schindler2023} with $-25.24>M_{1450}>-27.38$, matching the luminosity range of the 25 ASPIRE quasars, yields the observed quasar number density $n_{\rm QSO}=2.65 ~\rm cGpc^{-3}$. 
Therefore, the quasar duty cycle is $f^{\rm QSO}_{\rm duty} = 0.3^{+4.8}_{-0.3}\%$.
However, we note that the uncertainty in the quasar duty cycle is substantial due to the steep exponential cutoff in the halo mass function. Since the number density of massive halos declines sharply for $M_h\gtrsim 10^{12} \rm \msun$, even a small increase in the inferred minimum halo mass significantly reduces the number of available quasar-host halos. This, in turn, enlarges the estimated duty cycle, making it highly sensitive to the inferred mass threshold.

\begin{figure}
    \centering
    \includegraphics[width=0.47\textwidth]{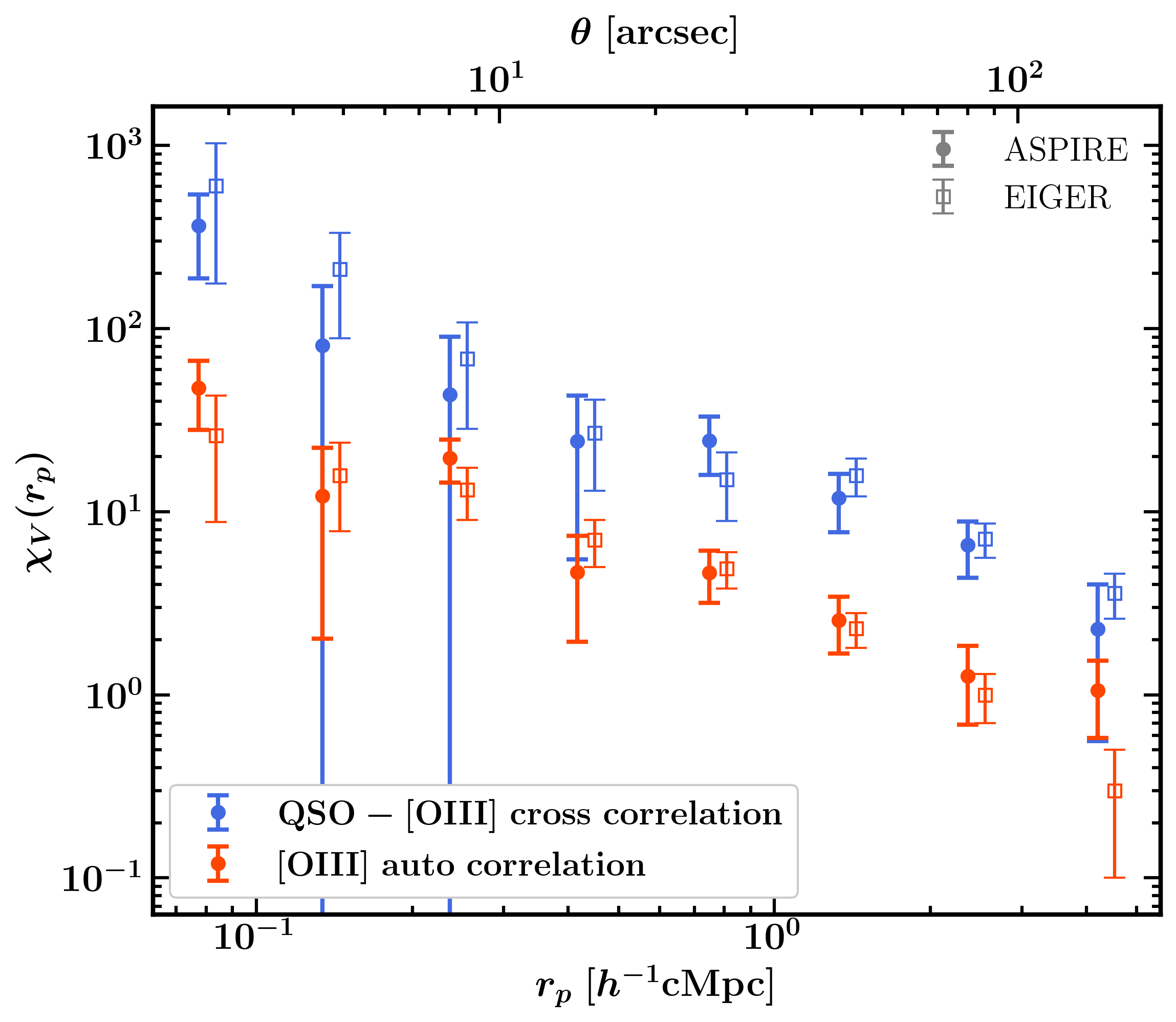}
    \caption{Comparison between EIGER (open square) and ASPIRE (filled circle) correlation functions. The quasar-\oiii-emitter cross correlation is shown in blue, and the \oiii-emitter auto correlation is shown in red. The EIGER correlation function data is taken from Table 2 in \citet{Eilers2024}. Error bars for EIGER represent Poisson uncertainties only, while ASPIRE error bars are computed as the square root of the diagonal elements in the covariace matrix shown in Fig. \ref{fig:covar_sim}. The EIGER data points are slightly shifted to the right for better visual representation. }
    \label{fig:corr_compare}
\end{figure}

\begin{figure*}
\centering
	\includegraphics[width=2.0\columnwidth]{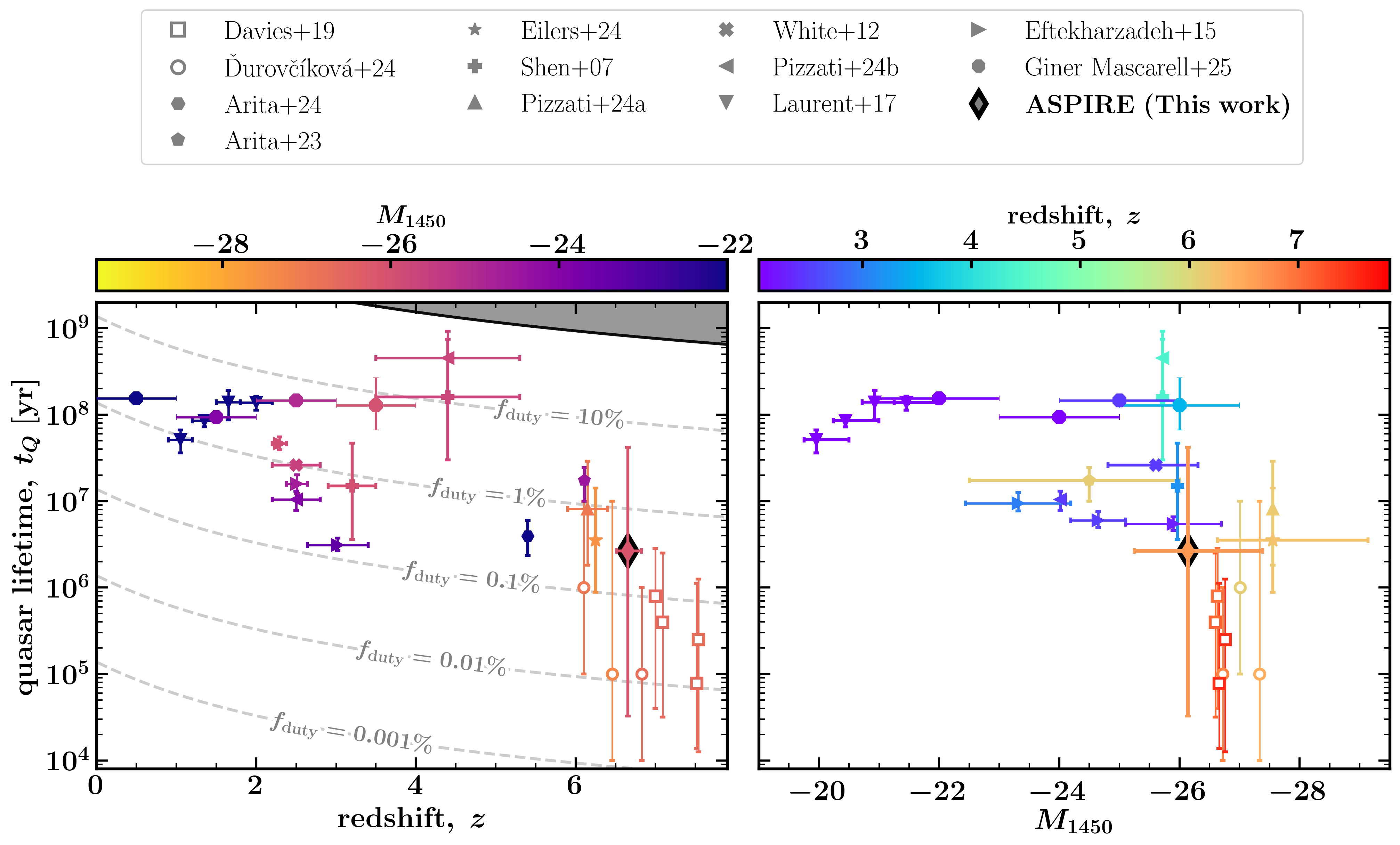}
    \caption{\textbf{\textit{Left:}} Redshift evolution of the quasar lifetime. The dashed curves show the lifetime with fixed quasar duty cycle, and the gray shaded region shows the forbidden region where $t_{\rm Q}>t_{\rm H}$ or, equivalently, $n_{\rm Q}>n_{\rm halo}$ where $n_{\rm halo}$ is the number density of quasar host halos.  The color bar shows the $M_{1450}$ magnitude of the quasar sample, with darker colors indicating lower luminosity. 
    Our result for the ASPIRE sample is marked as the diamond data point. The measurement from $\rm Ly\alpha$ damping wings for individual quasars at $z>6$ are shown as the round \citep{Durovcikova2024} and square open makers \citep{Davies2019, Wang2020, Yang2020}. For $z\gtrsim6$ quasars, we include the EIGER measurement, which uses \oiii-emitters as tracers to measure the quasar cross-correlation by putting together four quasar fields (shown as the star data point,  \citealt{Eilers2024}). We also include measurements from \citet{Pizzati2024b} and \citet{Arita2024}. For lower redshift quasars, we include the clustering measurement from \citet{Shen2007, White2012, Eftekharzadeh2015, Laurent2017, Pizzati2024a, GinerMascarell2025}. \textbf{\textit{Right:}} luminosity dependence of the quasar lifetime. The color bar shows the mean redshift of each sample. 
    }
    \label{fig:fduty_z_m1450}
\end{figure*}

\section{Comparison to previous work}
In this section, we compare our measurements of the quasar–\oiii-emitter cross-correlation and the \oiii-emitter auto-correlation in ASPIRE with previous clustering studies at $z\gtrsim6$, including both correlation function measurements and the inferred halo properties.
We compare our results with the previous work on the clustering analysis at $z\gtrsim6$, in particular, we compare our results with EIGER \citep{Eilers2024}, which measures both quasar-\oiii-emitter cross correlation function and the \oiii-emitter auto correlation, and with the
$z\sim 6$ quasar auto correlation measured by \citep{Arita2023} with the SHELLQs low luminosity quasars.

In Section \ref{sec:corr_length}, we have computed the quasar auto correlation length based on the assumption that galaxies and quasars trace the same underlying dark matter distribution (i.e., deterministic bias), which yields $r_0^{\rm QQ}=21.3^{+8.2}_{-8.7}\mpch$, only slightly lower than the EIGER correlation length $r_0^{\rm QQ}=22.0^{+3.0}_{-2.9}\mpch$ \citep{Eilers2024}. However, EIGER has a smaller uncertainty than our estimates despite having a comparable number of total \oiii-emitters as ASPIRE. The difference arises because we use a covariance matrix derived from mock realizations that accounts for cosmic variance and large scale structure, which dominates the error budget on the scales relevant for the correlation length. In contrast, the EIGER analysis estimates uncertainties using Poisson errors, which do not capture cosmic variance and therefore underestimate the true uncertainties. As both surveys are cosmic variance limited on these scales, the greater depth of EIGER does not significantly reduce the uncertainties, while the larger number of independent quasar fields in ASPIRE partially mitigates cosmic variance.
This is reflected in the quasar-\oiii-emitter cross correlation function as well (See Fig. \ref{fig:corr_compare}). 

The EIGER quasar-\oiii-emitter cross correlation is higher than ASPIRE result measured in this work in nearly all $r_p$ bins, except for one $0.57<r_p<1.02 \mpch$. 
The quasar auto correlation measured by \citep{Arita2023} yields $r_0^{\rm QQ}=23.7\pm 11 \mpch$, which is comparable to both the ASPIRE and EIGER results.

For the halo mass estimate, our joint fit yields the minimum quasar host halo mass of $\log \mminq/\rm \msun = 12.13^{+0.31}_{-0.38}$; 
for the EIGER \citep{Eilers2024} measurement, the $\log \mminq/\rm \msun = 12.43^{+0.13}_{-0.15}$, 
Although we adopt a similar modeling framework, including the same underlying simulation and halo occupation prescription for assigning quasars and \oiii-emitting galaxies to dark matter halos as \citep{Eilers2024}, the two analyses differ in how the likelihood is constructed. In this work, we use a model dependent covariance matrix derived from mock realizations, whereas the EIGER analysis adopts Poisson errors. As discussed above, accounting for cosmic variance in the covariance matrix leads to larger, but more realistic, uncertainties on the inferred halo masses.
This can also be seen from Fig. \ref{fig:corr_compare}, 
where the ASPIRE quasar-\oiii-emitter cross correlation is slightly weaker than the EIGER measurement especially at $r_p<0.5 \mpch$ (in the one-halo regime). 

There are two main contributing factors. First, EIGER targets four quasar fields, and with such a small sample, a single overdense field can disproportionately boost the stacked clustering signal and bias the inferred amplitude high. This effect reflects field-to-field variance rather than statistical uncertainty. Although both surveys contain similar numbers of \oiii-emitters and therefore have comparable Poisson noise, the impact of field-to-field variance scales roughly as $\sigma_{\rm cv} \propto 1/\sqrt{N_{\rm field}}$, making the ASPIRE measurement more representative of the average quasar environment. Second, the two surveys probe different luminosity ranges: ASPIRE reaches a lower UV luminosity limit of $M^{\rm ASPIRE}_{1450,~\rm lim}=-25.25$, whereas EIGER targets only very luminous quasars with $M^{\rm EIGER}_{1450,~\rm lim}=-26.63$. Although most studies find only a weak dependence of quasar halo mass on luminosity \citep[e.g.][]{Lidz2006, Shen2009, Krolewski2015} at $z \lesssim 3$, focusing on the bright end with only four sightlines increases the likelihood of sampling rare, highly overdense environments. Together, the more extreme luminosity selection and the small number of fields can lead to the slightly higher inferred halo mass for the EIGER sample relative to 
ASPIRE. In addition, the EIGER uncertainties are estimated using Poisson errors and do not account for cosmic variance, which likely leads to an underestimate of the true uncertainty on the inferred halo mass.

In addition, the quasar host halo mass inferred by \citet{Arita2023} is $\log \mminq = 12.7^{+0.4}_{-0.7}\rm~\msun$, 
about 0.5 dex higher than our ASPIRE estimate but within the error bars, even though the quasar auto-correlation length in their sample is comparable to ours. This difference is likely to arise from the methodology used to convert clustering into halo mass. \citet{Arita2023} 
derives the halo mass by translating the measured linear bias into a characteristic halo mass using the analytic bias–mass relation of \citet{Tinker2010}.

At a luminosity range more comparable to ASPIRE, \citet{Meng2026} measured the quasar auto-correlation function at $5.0 \leq z < 6.3$ using photometrically
selected candidates from the Legacy Survey and WISE, finding
$\log(M_{\rm h}/M_\odot) = 12.2^{+0.2}_{-0.7}$ at $z \sim 5.3$ and
$\log(M_{\rm h}/M_\odot) = 11.9^{+0.3}_{-0.7}$ at $z \sim 6.0$ for quasars
with $M_{1450} \sim -25$ to $-25.5$. Their characteristic halo mass at
$z \sim 6$ is broadly consistent with our inferred minimum halo mass of
$\log(M^{\rm QSO}_{h,\rm min}/M_\odot) = 12.13^{+0.31}_{-0.38}$. Their inferred duty cycles of $0.8^{+13.5}_{-0.7}\%$ at $z \sim 5.3$ and
$0.3^{+4.7}_{-0.3}\%$ at $z \sim 6.0$ are also consistent with our ASPIRE
estimate.

However, our measurement is based on fitting the full clustering signal to simulations using a step-function HOD to determine the minimum halo mass that reproduces the observed cross-correlation. Because the bias–mass conversion yields an effective characteristic halo mass rather than a minimum mass threshold, the resulting values lead to a higher inferred halo mass in \citet{Arita2023}. This is also discussed in Appendix C in \citet{Pizzati2024b}. As shown in Fig. C1 of \citet{Pizzati2024b}, power-law parametrizations of the quasar autocorrelation function can provide a formally better fit to the data than linear-theory-based halo models, but only by favoring very large autocorrelation lengths ($r_{0,\rm QQ} \approx 20$–$50$ cMpc). Because these large clustering amplitudes are then translated into halo masses through a bias–mass relation, this approach naturally leads to high inferred halo masses, consistent with the values reported by \citet{Arita2023}. While power-law parametrization can reproduce the observed correlation function, it does not account for scale-dependent halo bias or the one-halo to two-halo transition relevant for highly biased tracers at high redshift. In contrast, our analysis uses simulation-based halo correlation functions and an explicit halo occupation prescription, linking the inferred halo masses directly to the underlying halo population in the simulation.

We also compare our results for the tracer halo mass estimates with that inferred from \citep{Shuntov2025}, which computed the angular auto correlation function using the H$\alpha$ emitters in the FRESCO survey \citep{Oesch2023} at $4.9<z<6.7$. \cite{Shuntov2025} finds that the mean halo mass for the H$\alpha$ emitters to be $\log M_h=11.24^{+0.03}_{-0.04}~\msun$. 
We can similarly compute the mean halo mass, using the halo mass function with FLAMINGO-10k:
\begin{equation}
\langle M_h \rangle =
\frac{
\int_{M_{\rm min}}^{\infty} M_h \, \frac{dn}{dM_h} \, dM_h}{\int_{M_{\rm min}}^{\infty} \frac{dn}{dM_h} \, dM_h}.
\end{equation}
This yields a mean halo mass of $\log M^{\rm \oiii}_{h, \rm mean}=10.7~\msun$, for our minimum galaxy halo mass $\log \mming=10.55^{+0.11}_{-0.12} \msun$. This is $0.5~\rm dex$ smaller than the estimate from \cite{Shuntov2025}. However, we note that the tracer population used in our analysis is different: our ASPIRE results and EIGER \citep{Eilers2024} uses \oiii-emitters, whereas \cite{Shuntov2025} uses the H$\alpha$ emitters as tracers which may reside in halos with slightly different masses. 

\section{Discussion}
\label{sec:discussion}

One of the most central questions in astrophysics is how supermassive black holes assemble their masses in the early Universe \citep{Fan2023}. This question has been re-visited thanks to recent \textit{JWST} observations \citep{Eilers2024, Bosman2025}. 
In the standard picture, black holes grow through sustained gas accretion in their accretion disks. Under this scenario, the black hole mass can be described by an exponential growth law
\begin{equation}
    M_{\rm SMBH}(t) = M_{\rm seed} e^{(t-t_{\rm seed})/t_s},
\label{eq:BH_growth}
\end{equation}
where $M_{\rm seed}$ is the mass of the original black hole seed. Such seeds may form through direct collapse in pre-galactic halos \citep{Begelman2006}, as remnants of massive Pop III stars \citep{Madau2001}, or via runaway stellar collisions in dense young star clusters (see e.g., \citealt{Devecchi&Volonteri2009, Regan2024} 
and references therein). 
And $t_s$ is the Salpeter timescale ($t_s = 45 (\epsilon/0.1) (L/L_{\rm Edd})^{-1}~\rm Myr$, where $\epsilon$ is the radiative efficiency and $L/L_{\rm Edd}$ is the Eddington ratio). For a fixed Eddington ratio, a lower radiative efficiency will correspond to a shorter Salpeter time, resulting in a more rapid SMBH mass growth, according to Eq. \ref{eq:BH_growth}. A direct consequence of a small radiative efficiency is thus a small UV luminous duty cycle: heuristically, the duty cycle is of order a few Salpeter times divided by the Hubble time, $f_{\rm duty}\sim{\rm few}\times t_s/t_{\rm H}$, so that reducing $\epsilon$ decreases the fraction of cosmic time during which the quasar luminosity remains above the detection limit.

Another implication of a small UV luminous quasar duty cycle is that a significant fraction of SMBH growth at high redshift could occur during phases that are not selected in rest-frame UV quasar samples. The standard Salpeter time argument implies that sustained gas accretion leads to exponential mass growth, independent of whether the emission is observable in the rest-frame UV. In this context, the UV luminous duty cycle measures the fraction of time that an actively accreting SMBH is both UV bright enough to be detected in rest-frame UV selected quasar samples.
If a substantial portion of accretion occurs during obscured phases, then the UV luminous duty cycle will underestimate the total time spent in active growth, because it does not include accretion episodes whose UV emission is attenuated by dust. In this case, a short UV duty cycle does not by itself require a short total accretion lifetime.
In other words, a short duty cycle in the UV does not necessarily imply a short total black hole lifetime; rather, it may reflect that much of the accretion proceeds in phases that are not detectable in UV-selected quasar samples. Conversely, if obscured accretion is subdominant, or if the obscured and unobscured active fractions are comparable, then a short UV duty cycle would imply a short total accretion timescale.

In Fig. \ref{fig:fduty_z_m1450}, we plot our measurement for the quasar duty cycle in comparison with the previous results measured from quasar clustering \citep{Shen2007, White2012, Eftekharzadeh2015, Laurent2017, Pizzati2024a, Pizzati2024b, Arita2024, GinerMascarell2025} as well as from quasar proximity zone size measurement and $\rm Ly\alpha$ damping wings \citep{Davies2019, Wang2020, Yang2020, Durovcikova2024}.

The duty cycle is commonly approximated as $t_{\rm Q}/t_{\rm H}$, where $t_{\rm Q}$ is the quasar lifetime, and $t_{\rm H}$ is the Hubble time at the redshift we observe the quasar. This approximation is based on the assumption that each halo experiences one quasar phase of duration $t_{\rm Q}$ during the age of the Universe. In this case, the duty cycle reflects the fraction of time a typical halo is actively accreting and luminous as a quasar. Interestingly, we find that the duty cycle inferred for the ASPIRE quasars ($-25.25 > M_{1450} > -27.38$) is comparable to that of the EIGER quasar sample \citep{Eilers2024}, which focus on relatively bright quasars ($-26.63 > M_{1450} > -29.14$),
despite the ASPIRE quasars being more than $\sim$1 magnitude fainter in the UV on average. Since the duty cycle is defined as the ratio of the quasar number density to the number density of dark matter halos above a given mass threshold, we would expect that the faint quasars yield a larger duty cycle due to their larger abundance relative to the bright quasars. Specifically, raising the lower UV luminosity threshold from $M_{1450}=-25.25$ to $-26.63$ decreases the predicted quasar number density, such that $n^{\rm EIGER}_{\rm qso}/n^{\rm ASPIRE}_{\rm qso}\simeq12$, according to the quasar luminosity function from \cite{Schindler2023}. In other words, if the underlying dark matter halo population for EIGER and ASPIRE quasars are the same, the inferred quasar duty cycle would be $12\times$ higher for the ASPIRE quasars than the EIGER quasars.

This trend can also be seen for the low redshift quasar sample in Fig. \ref{fig:fduty_z_m1450}. For example, at $z\sim2$, \cite{Eftekharzadeh2015} sample (right-pointing triangles) probes a higher luminosity range, hence getting a smaller duty cycle relative to the fainter quasars at similar redshift in the \citep{Laurent2017} sample (downward-pointing triangles in blue). 
This indicates that faint quasars are active for a larger fraction of time than bright quasars. The luminosity dependence arises because the characteristic host halo mass shows little variation with quasar luminosity. 
The key point is that the clustering strength, and hence the characteristic host halo mass, is similar for bright and faint quasars, so the denominator in the duty cycle, the abundance of dark matter halos at that mass, is essentially fixed. Since faint quasars are more numerous, the numerator in the duty cycle, the space density of quasars, is larger for the faint population, which directly leads to a larger inferred duty cycle.

However, this luminosity dependence is no longer the case at high redshift ($z\gtrsim4$), as indicated in the right panel of Fig. \ref{fig:fduty_z_m1450}, in addition to the large scatter in the quasar duty cycle, there is no apparent correlation between the quasar lifetime and the UV luminosity of quasar. The lifetime from damping wing measurements \citep{Davies2019, Wang2020, Yang2020, Durovcikova2024} (open error bars in the right panel of Fig. \ref{fig:fduty_z_m1450}) do not show a significant luminosity dependence within their uncertainties. These measurements are broadly consistent with the clustering based constraints given the size of the error bars, although the two methods cannot be compared more precisely without a quantitative assessment of the uncertainties.
Taken together, these measurements suggest that the simple luminosity–duty-cycle relation that holds at lower redshift does not extend to the reionization era. In this regime, duty cycle is no longer dependent on the quasar luminosity. This is what we observe for the ASPIRE and EIGER quasars. Although the ASPIRE sample is on average $\sim$1 magnitude fainter in the UV and should therefore exhibit a substantially larger duty cycle under the low-redshift luminosity-scaling argument, the inferred duty cycles of the two samples are comparable within uncertainties. As noted above, if both samples traced the same halo population, ASPIRE would yield a much larger duty cycle. Instead, the inferred $f_{\rm duty}$ values are comparable because the clustering analysis yields a lower $M_{\rm min}$ for ASPIRE. The halo abundance $n_{\rm halo}(>M_{\rm min})$ is extremely sensitive to $M_{\rm min}$ in the exponential tail of the mass function, so a modest shift in $M_{\rm min}$ offsets the difference in $n_{\rm QSO}$.
Instead, the inferred $f_{\mathrm{duty}}$ values for ASPIRE and EIGER are similar. This outcome occurs because the clustering analysis requires a different $M_{\mathrm{min}}$ for ASPIRE. The larger space density in the numerator is compensated by a larger denominator, $n_{\mathrm{halo}}(>M_{\mathrm{min}})$, which is extremely sensitive to $M_{\mathrm{min}}$ in the exponential tail of the halo mass function. A modest shift in $M_{\mathrm{min}}$ produces an order of magnitude change in $n_{\mathrm{halo}}$, which offsets the factor of 12 difference in $n_{\mathrm{QSO}}$. As a result, both samples yield comparable duty cycles within the uncertainties.

\begin{figure}
\centering
	\includegraphics[width=\columnwidth]{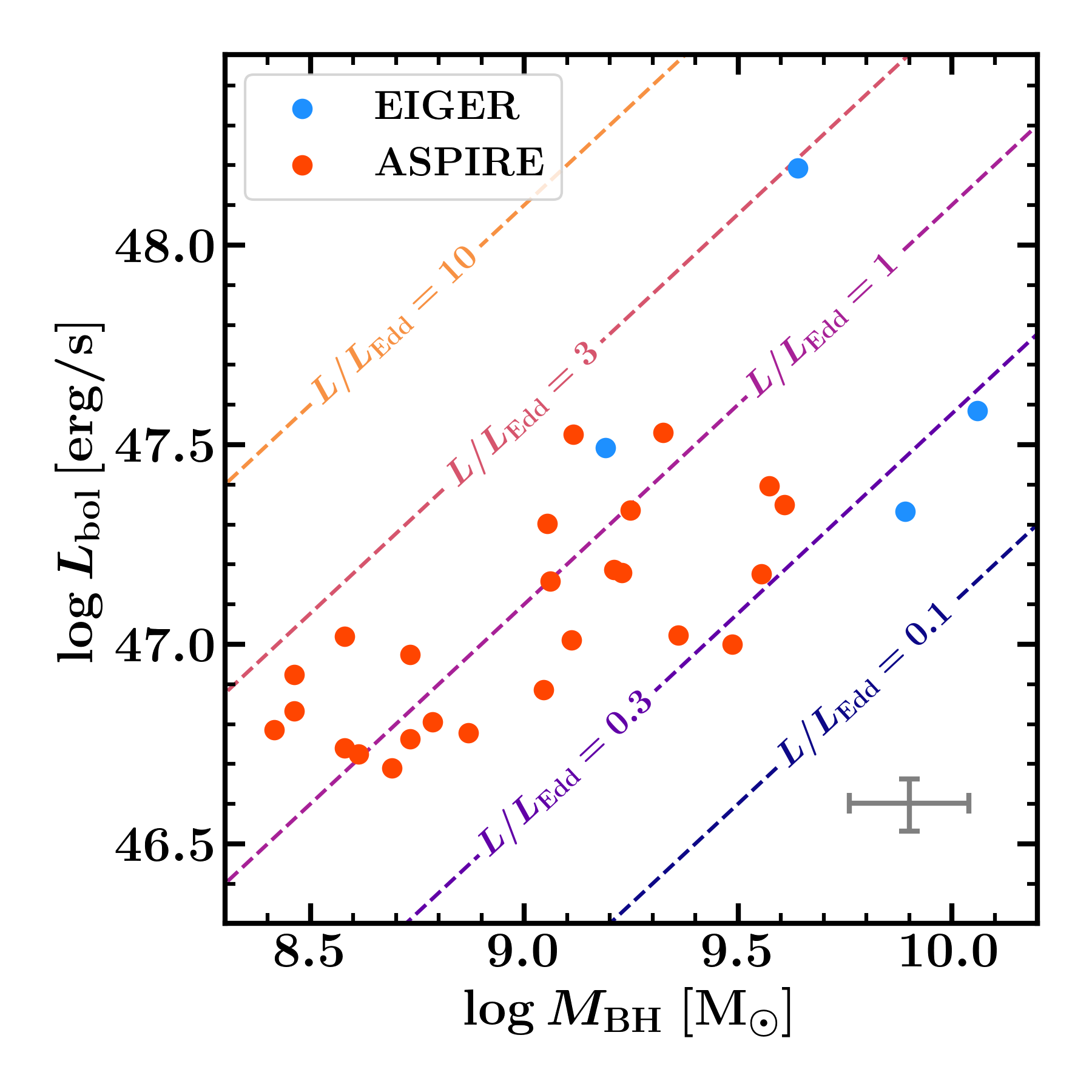}
    \caption{$L_{\rm bol}-M_{\rm BH}$ distribution of the EIGER (blue) \citep{Eilers2024} and ASPIRE (red) quasars. Diagonal lines mark the constant Eddington ratio. The typical error for $L_{\rm bol}$ and $M_{\rm BH}$ is plotted as the gray error bar.  }
    \label{fig:ledd_aspire_eiger}
\end{figure}

To understand the effect on the lower lifetime probed by ASPIRE quasar on the SMBH growth, we plot distribution on the $M_{\rm BH}-L_{\rm bol}$ plane for the ASPIRE (red) and EIGER (blue) quasars in Fig. \ref{fig:ledd_aspire_eiger}. We show the constant Eddington ratio as the diagonal dashed lines. We show that even though 
the faint luminosity limit for ASPIRE and EIGER defer by $0.6$ dex ($1.4$ in magnitude), 
the mean Eddington ratio between the two samples are similar, due to a slightly more massive SMBH for the EIGER sample. The mean Eddington ratios are $\lambda^{\rm EIGER}_{\rm Edd} = 1.22$ and $\lambda^{\rm ASPIRE}_{\rm Edd} = 1.04$. The resulting mean Salpeter time for the two samples are $t^{\rm EIGER}_{\rm S}=40~\rm Myr$ and $t^{\rm ASPIRE}_{\rm S}=48~\rm Myr$. We compare the Salpeter timescale to the measured quasar lifetime, $t^{\rm EIGER}_{\rm Q}=3.54~\rm Myr$ \citep{Eilers2024} and $t^{\rm ASPIRE}_{\rm Q}=2.64^{+39.15}_{-2.61}~\rm Myr$. This indicates that the measured quasar lifetime is only a small fraction of the Salpeter time, where $t_{\rm Q}/t_{\rm S} = 9\%$ for EIGER quasars and $5\%$ for ASPIRE. 
Therefore, although there is a large uncertainty induced by steep halo mass function at high mass end (as discussed in Sec. \ref{sec:qso_duty}), the low duty cycle measured by ASPIRE places a stronger constraint on quasar activity and consequently on black hole growth compared to EIGER.

The tighter constraint implied by ASPIRE also can be viewed from the perspective of quasar light curves, where the UV luminosity is proportional to the accretion rate ($L\propto \epsilon \dot{M}c^2$). For a given luminosity threshold set by the lower luminosity limit of the survey, the fraction of time where the light curve exceeds the threshold is the quasar duty cycle (see, e.g, \citealt{Hopkins2005} and references therein). Because ASPIRE targets a fainter luminosity limit ($M_{1450} = -26.63$) than EIGER, one would expect ASPIRE to produce a larger duty cycle if both samples were drawn from quasars with the same underlying accretion history: a lower luminosity threshold should include more of each quasar’s light curve above the limit.

However, we observe the opposite: ASPIRE quasars have a smaller duty cycle than EIGER. This can only be explained if the ASPIRE quasar light curves spend even less time above the (lower) luminosity threshold, meaning that their UV luminosities, and therefore their accretion rates, remain lower for most of their lifetimes than the EIGER quasars. As a result, the time-averaged accretion rate during UV-bright phases is more limited, which in turn reduces the total mass that the black hole can grow through this observable mode of activity.

When these faint quasars also have short UV-luminous lifetimes (e.g., $t_{\rm Q} \sim$ a few Myr), they can only grow by a small amount during the cosmic epoch. For example, even at the Eddington limit, a 4 Myr episode increases the black hole mass by less than 10\%. Since most black holes fall into this faint category, their short observable lifetimes imply that the bulk of their mass assembly must occur in phases that do not produce long lived UV bright emission.
This growth could happen during heavily obscured phases, during radiatively inefficient accretion, or during short periods of super-Eddington accretion.

In comparison, the bright quasars are rare and likely correspond to short episodes of very high luminosity. Their short duty cycle does not imply limited growth, because a substantial fraction of their mass assembly may have occurred earlier, when the black hole was accreting more rapidly than it is at the redshift of observation. For the bulk of the population traced by the fainter ASPIRE quasars, however, the inferred UV luminous lifetime combined with the observed luminosity is insufficient to account for the black hole mass under standard assumptions. The Soltan argument \citep{Soltan1982} links the total accreted mass onto the SMBHs to the total radiation emitted by quasars, assuming that the SMBH mass is grown via accretion:
\begin{equation}
    \rho_{\bullet, \rm acc}^{\rm QSO}(z) = \int_{z}^{\infty} \frac{dt}{dz} dz \int_{0}^{\infty} \frac{(1 - \epsilon)L_{\rm bol}}{\epsilon c^2} \Phi(L, z) dL.
\label{eq:Soltan_argument}
\end{equation}
Where $\rho_{\bullet, \rm acc}^{\rm QSO}(z)$ is the density of total mass accreted onto all SMBHs at redshift z, which is equivalent to the SMBH mass density at redshift z. And $L_{\rm bol}$ is the bolometric luminosity of the accreting quasar, and $\Phi(L_{\rm bol},z)$ is the (UV luminous) quasar luminosity function. At $z\sim0$, the Soltan argument yields a radiative efficiency $\epsilon\sim10\%$. While the Soltan argument has been used at low redshift to estimate the average radiative efficiency by comparing the integrated quasar luminosity to the observed SMBH mass density, this approach is less robust at high redshift due to the lack of direct constraints on the total SMBH mass density. Instead, we can consider the standard picture of black hole growth. In this framework, a small duty cycle implies that the black hole is actively accreting for only a small fraction of time, which reduces the effective growth rate. This presents a challenge at high redshift, where the available cosmic time is limited. The low duty cycle measured by ASPIRE is therefore in tension with the need to grow to $\sim 10^9~\msun$ black holes by $z \sim 6$. 

In particular, assuming a radiative efficiency $\epsilon \sim 0.1$, the total energy emitted over the inferred lifetime, $L_{\rm obs} t_{\rm Q} / \epsilon$, falls well short of $M_{\rm BH} c^2$. This mismatch implies that either the effective radiative efficiency during most of the growth was lower than the canonical value, the black hole experienced earlier phases of much higher luminosity that are not represented by the current population, or the total accretion timescale is significantly longer than the UV luminous timescale inferred from clustering and proximity zone measurements. Because faint quasars dominate the number density, they provide the strongest constraints on these scenarios and indicate that the present day luminosity and inferred UV luminous lifetime alone cannot account for the bulk of supermassive black hole growth.

Based on Eq. \ref{eq:Soltan_argument}, the two possibilities for getting a large black hole mass density are:
\begin{itemize}
    \item a large population of SMBHs grow in an obscured phase. This implies that the observed UV quasar luminosity function, $\Phi(L_{\rm bol},z)$, only accounts for \textit{a small fraction of} the full population of growing SMBH;
    \item the radiative efficiency, $\epsilon$, is small.
\end{itemize}

In the first scenario, a large obscured fraction ($f_{\rm obsc.} \equiv n_{\rm obsc.}/n_{\rm unobsc.} \gg 1$, where $n_{\rm obsc.}$ and $n_{\rm unobsc.}$ are the number densities of obscured and unobscured quasars, respectively) would imply that most SMBH growth occurs while the quasar is heavily obscured. In this case, the black hole could accrete a large amount of mass without appearing as a UV–bright quasar, even though the total radiative output ($\propto \epsilon \dot{M} c^2$) is the same before dust reprocessing. Consequently, the UV luminosity function would no longer provide a reliable census of the total emission associated with SMBH growth, because a substantial fraction of the radiative energy would be absorbed and re-emitted in the infrared.

While this level of obscuration may help explain the abundance of faint \textit{JWST} AGN, it is difficult to accommodate the much higher number density of the ``little red dots'' without invoking an unrealistically large population of actively accreting SMBHs. Left panel in Fig. 1 of \cite{Pizzati2025} show that the obscured:unobscured ratio would reach 2300:1 implied by the LRD and quasar luminosity function at $z\sim7$. As a result, a purely obscured growth scenario remains in tension with the halo-based duty cycle constraints derived from clustering. Moreover, if LRDs represent quasars in their obscured phase, then they should occupy essentially the same dark matter halos as UV-luminous quasars and therefore exhibit comparable clustering strength. Although some studies suggest that obscured AGN may reside in slightly lower-mass halos than unobscured AGN \citep[e.g.,][]{Allevato2014}, the inferred differences are modest; both populations are thought to trace roughly the same underlying massive halo population. However, this prediction is not reflected in the recent clustering results of LRDs. Although there is some evidence that some LRDs may reside in overdense regions \citep{Schindler2025a}, most LRDs have weaker clustering than that of ASPIRE quasars \citep{Arita2024, Lin2025}, indicating that LRDs do not trace the same halo population as UV-bright quasars.

In the second scenario, if the radiative efficiency, $\epsilon$, is small, then only a small fraction of the accreted mass is converted into radiation during the accretion process onto the SMBH. This leads to a higher inferred mass accretion rate for a given observed luminosity, effectively increasing the total mass density of SMBHs. Specifically, since the luminosity integral in Eq.~\ref{eq:Soltan_argument} is proportional to $1/\epsilon$, a small $\epsilon$ results in a significantly larger inferred SMBH mass density \citep{Volonteri2006, Trakhtenbrot2017, Davies2019}. 
Moreover, a small radiative efficiency is also consistent with a small duty cycle. If quasar luminosity varies stochastically over time (i.e., follows a quasar light curve), a lower $\epsilon$ implies that the mean bolometric luminosity of the quasar is lower. As a result, the fraction of time during which the SMBH appears as a UV-luminous quasar is reduced, meaning that we are less likely to observe it in this active phase. Although a small $\epsilon$ can explain the rapid SMBH growth at high redshift with apparent small UV luminous quasar duty cycle, we do not expect $\epsilon$ to be small at low redshift, as this will over produce the SMBH mass density that we observe at $z=0$, violating the Soltan argument.

\section{Conclusion}
\label{sec:conclusion}

In this work, we present clustering analysis of 25 quasar fields from the ASPIRE program using \textit{JWST}/NIRCam Wide Field Slitless Spectroscopy (WFSS). We jointly analyze the auto-correlation function of \oiii-emitting galaxies and the cross-correlation function between quasars and \oiii-emitters at $z \sim 6$, with realistic selection functions derived from data and synthetic mocks built from the \texttt{FLAMINGO-10k} cosmological simulation. Our key results are summarized below:

\begin{enumerate}
    \item We measure a correlation length of $r_0^{\rm GG} = 4.7 ^{+0.5}_{-0.6}~\mpch$ with fixed power law slope of $\gamma_{\rm GG}=1.8$ for the \oiii-emitters auto-correlation function and $r_0^{\rm QG} = 8.9^{+0.9}_{-1.0}~\mpch$ for $\gamma_{\rm QG}=2.0$ for the quasar--\oiii-emitters cross-correlation function. 
    Assuming deterministic bias, this yields quasar auto corelation length, $r_0^{\rm QQ}=19.7^{+4.7}_{-4.3}\mpch$.
    
    \item By jointly fitting both correlation functions, we infer the characteristic host halo masses as $\log (\mming / \msun) = 10.55^{+0.11}_{-0.12}$ for [OIII] emitters and $\log (\mminq / \msun) = 12.13^{+0.31}_{-0.38}$ for quasars.
    
    \item Using the inferred minimum halo masses and the observed number densities, we estimate the duty cycle to be $2.5^{+1.3}_{-0.8}\%$ for \oiii-emitters and $0.3^{+4.8}_{-0.3}\%$ for quasars.
\end{enumerate}

To properly account for uncertainties in our clustering measurements, we construct covariance matrices using 1000 mock realizations per mass model from the FLAMINGO-10k simulation, incorporating the observational selection functions of ASPIRE. As we show in our companion paper (Huang et al., in prep.), cosmic variance dominates over Poisson noise, and the bin-to-bin correlations in the covariance matrix arise from large-scale density fluctuations. These mock-based covariance estimates provide a more realistic characterization of the measurement uncertainties than approaches assuming diagonal or Poisson-only errors. Methods that neglect cosmic variance and bin-to-bin correlations significantly underestimate the true uncertainties in the inferred halo masses. As a result, our simulation-based, model-dependent inference yields more robust halo mass constraints.

The inferred low duty cycle of high-redshift quasars suggests that their UV-luminous phases are both rare and short-lived. Comparing the measured quasar lifetimes to the Salpeter timescale shows that the UV-bright phase accounts for only a small fraction of an e-folding time, with $t_{\rm Q}/t_{\rm S}\approx 9\%$ for EIGER quasars and $t_{\rm Q}/t_{\rm S}\approx 5\%$ for ASPIRE. Although ASPIRE quasars have
lower SMBH masses than the EIGER sample, their Eddington ratios are comparable (see Fig.~\ref{fig:Lbol_MBH}), resulting in similar Salpeter times. The smaller ratio inferred for ASPIRE implies a tighter limit on the amount of black hole growth that can occur during the observed UV-luminous phase. As a result, the ASPIRE duty cycle places a stronger constraint on quasar activity and on the contribution of unobscured accretion to early SMBH mass assembly than EIGER.

Taken together, our clustering analysis provides an independent constraint on the minimum halo masses of quasar and galaxy hosts, as well as the quasar duty cycle. 
The measured duty cycle of quasars provides implications on the early SMBH growth: the observed high-redshift quasars suggest rapid SMBH assembly, yet standard accretion models struggle to explain the required mass buildup within cosmic time constraints. Two primary scenarios emerge to reconcile this discrepancy: (i) a significant fraction of SMBH growth occurs in an obscured phase, where the observed UV luminosity function underestimates the total accretion rate, and (ii) the radiative efficiency is smaller than the canonical $\sim 10\%$, allowing for more efficient mass accumulation. Both scenarios have observational implications. 



Finally, our results demonstrate that joint clustering analysis of galaxies and quasars, calibrated with realistic simulations and observational selection effects, provides a powerful route to constraining halo masses and duty cycles. A companion study (Huang et al., in prep.) will extend this framework by modeling the field-to-field variation in galaxy counts within the quasar environment using counts-in-cells statistics. This approach not only measures the mean overdensity across sightlines, as in the correlation function analysis, but also captures the full probability distribution of overdensities. Because this distribution is sensitive to the underlying halo mass, the counts-in-cells statistics provide an additional and complementary constraint on quasar environments and host halo masses.

Our results imply that the UV-luminous phase contributes only a small fraction of the mass growth required to assemble the observed high-redshift SMBHs, raising the question of where the remaining growth occurs. Future multi-wavelength observations provide a direct way to test this. If infrared, radio, and X-ray surveys reveal a large population of heavily obscured quasars at $z\gtrsim6$, this would support a scenario in which most SMBH growth occurs in obscured phases that are missed by rest-UV samples. Conversely, if the obscured fraction is found to be modest, then the short UV-bright lifetimes inferred here would instead favor low radiative efficiencies, allowing efficient mass buildup despite limited observable accretion time. In this way, improved constraints on obscured AGN fractions and quasar lifetimes will directly determine whether early SMBH growth is dominated by hidden accretion or by intrinsically inefficient radiation.

In addition, future clustering measurements with larger quasar samples, particularly those reaching fainter luminosities and extending to higher redshifts, will further improve constraints on halo occupation and duty cycle estimates. Slitless spectroscopic surveys with JWST enable ASPIRE-like clustering analyses to be carried out across a range of redshifts, allowing the evolution of quasar host halo masses and duty cycles to be directly probed. Wide-area slitless spectroscopic surveys with Euclid and the Nancy Grace Roman Space Telescope will provide complementary samples at higher redshift and larger volumes, together leading to a more complete understanding of the connection between quasars and their dark matter environments.



\section*{Acknowledgements}
This work is based on observations made with the NASA/ESA/CSA James Webb Space Telescope. The data were obtained from the Mikulski Archive for Space Telescopes at the Space Telescope Science Institute, which is operated by the Association of Universities for Research in Astronomy, Inc., under NASA contract NAS 5-03127 for JWST. These observations are associated with program \#2078. Support for program \#2078 was provided by NASA through a grant from the Space Telescope Science Institute, which is operated by the Association of Universities for Research in Astronomy, Inc., under NASA contract NAS 5-03127. The specific observations analyzed can be accessed via \doi[10.17909/vt74-kd84]{https://doi.org/10.17909/vt74-kd84}.

We acknowledge helpful conversations with the ENIGMA group at UC Santa Barbara and Leiden University. JH is grateful to the discussions with Shane Bechtel, Daming Yang, Ben Wang, and Xiaojing Lin. JH and JFH acknowledges support from the National Science Foundation under Grant No. 2307180.
JFH also acknowledges support for program XXX provided by NASA through a grant from the Space Telescope Science Institute, which is operated by the Association of Universities for Research in Astronomy, Inc., under NASA contract NAS 5-03127.
FW acknowledges support from NSF award AST-2513040. CM acknowledges support from Fondecyt Iniciacion grant 11240336 and the ANID BASAL project FB210003.

\section*{Data Availability}

The data underlying the findings of this study, together with the analysis code used to produce the results, are available from the corresponding author upon request.



\bibliographystyle{mnras}
\bibliography{aspire_citations} 



\appendix

\section{One-halo and two-halo decomposition of the quasar--[O\,\textsc{iii}]
cross-correlation function}
\label{app:1h2h}

\begin{figure}
    \centering
    \includegraphics[width=\columnwidth]{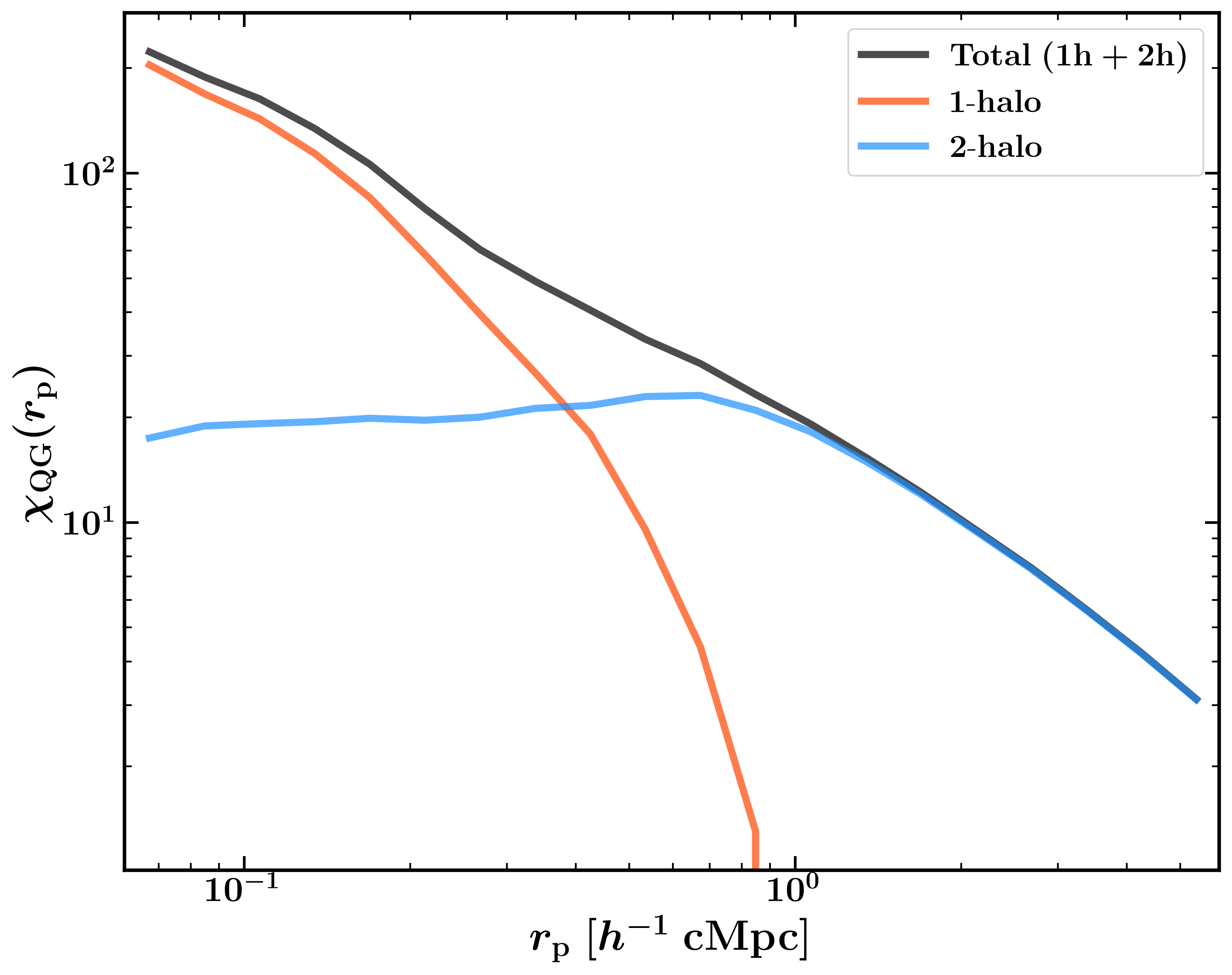}
    \caption{One-halo and two-halo decomposition of the volume-averaged
    quasar--[O\,\textsc{iii}]-emitter cross-correlation function,
    $\chi_{\rm QG}(r_p)$, computed from the FLAMINGO-10k simulation for the
    fiducial minimum mass model with
    $\log(M^{\rm [O\,III]}_{h,\rm min}/M_\odot) = 10.6$ and
    $\log(M^{\rm QSO}_{h,\rm min}/M_\odot) = 12.2$. The one-halo term
    (orange) counts [O\,\textsc{iii}]-emitting satellite subhaloes within the
    same host halo as the central quasar, while the two-halo term (blue)
    counts [O\,\textsc{iii}]-emitters in distinct host halos. The total
    (black) is the sum of both contributions.}
    \label{fig:1h2h}
\end{figure}

We decompose the modelled quasar--[O\,\textsc{iii}]-emitter cross-correlation
function into one-halo and two-halo contributions for the fiducial minimum
mass model with $\log(M^{\rm [O\,III]}_{h,\rm min}/M_\odot) = 10.6$ and
$\log(M^{\rm QSO}_{h,\rm min}/M_\odot) = 12.2$. Since quasars are placed in
central subhaloes, the one-halo term counts [O\,\textsc{iii}]-emitting
satellite subhaloes residing within the same host halo as the quasar, while
the two-halo term counts [O\,\textsc{iii}]-emitters in separate central subhalos.
As shown in Fig.~\ref{fig:1h2h}, the one-halo term dominates at
$r_{\rm p} \lesssim 0.3\;h^{-1}\,\mathrm{cMpc}$, while the two-halo term dominates
at $r_{\rm p} \gtrsim 0.5\;h^{-1}\,\mathrm{cMpc}$.

\section{Example Injected Mock \oiii-emitters extracted using \texttt{unfold\_jwst}}

\begin{figure*}
    \centering
    \includegraphics[width=0.7\textwidth]{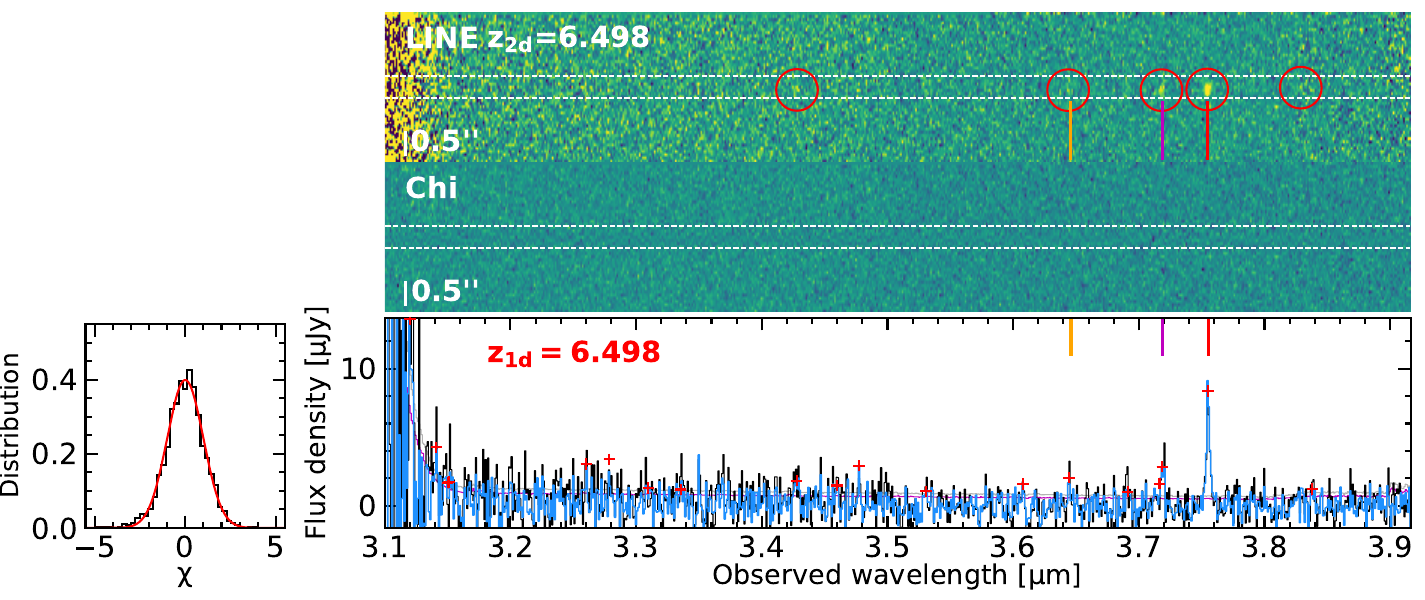}
    
    \includegraphics[width=0.7\textwidth]
    {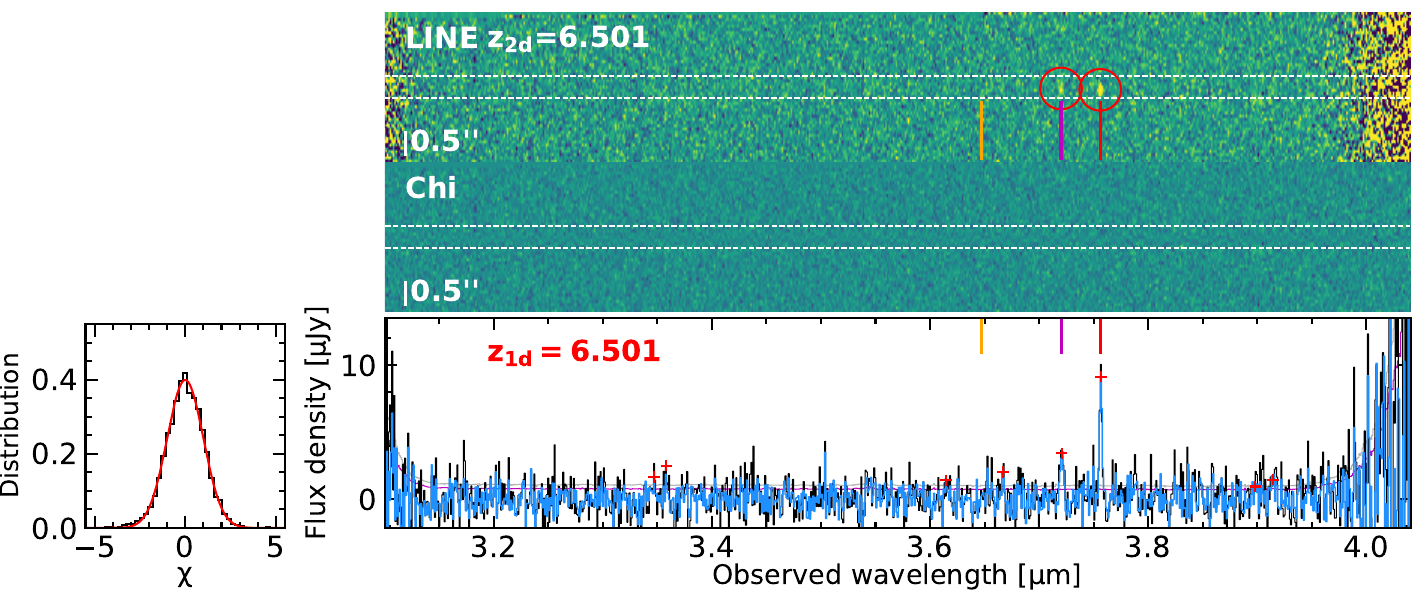}
    \caption{Example of two injected mock \oiii-emitters with $z_{\rm mock}=6.50$ in the quasar field J0109-3047. The mock \oiii-emitters are reduced and extracted with \texttt{unfold\_jwst}.}
    \label{fig:appendix_mock_o3}
\end{figure*}


\begin{table*}
\centering
\begin{tabular}{l c c c c c c c c c}
\hline
\textbf{Field} & \multicolumn{8}{c}{$r_p ~ [\mpch]$} & $N_{\rm \oiii}$ \\
 & 0.06-0.10 & 0.10-0.18 & 0.18-0.32 & 0.32-0.57 & 0.57-1.02 & 1.02-1.80 & 1.80-3.17 & 3.17-5.60 &  \\
\hline
\hline
J0109-3047  & 1  & 0  & 1  & 0  & 0  & 0  & 2  & 0  & 4  \\
J0218+0007  & 0  & 0  & 0  & 0  & 6  & 6  & 0  & 0  & 12  \\
J0224-4711  & 0  & 0  & 0  & 0  & 1  & 0  & 0  & 1  & 2  \\
J0226+0302  & 0  & 0  & 3  & 1  & 8  & 7  & 0  & 1  & 20  \\
J0229-0808  & 0  & 0  & 0  & 0  & 2  & 2  & 0  & 0  & 4  \\
J0244-5008  & 0  & 1  & 0  & 3  & 4  & 1  & 0  & 0  & 9  \\
J0305-3150  & 1  & 0  & 0  & 0  & 2  & 3  & 6  & 0  & 12  \\
J0430-1445  & 0  & 0  & 0  & 0  & 2  & 0  & 0  & 0  & 2  \\
J0525-2406  & 0  & 0  & 0  & 0  & 0  & 0  & 0  & 0  & 0  \\
J0706+2921  & 0  & 0  & 0  & 0  & 0  & 0  & 2  & 1  & 3  \\
J0910+1656  & 1  & 0  & 2  & 1  & 0  & 7  & 2  & 0  & 13  \\
J0910-0414  & 0  & 0  & 0  & 1  & 0  & 0  & 1  & 0  & 2  \\
J0921+0007  & 0  & 0  & 0  & 0  & 1  & 0  & 1  & 0  & 2  \\
J0923+0402  & 0  & 1  & 0  & 0  & 1  & 2  & 0  & 0  & 4  \\
J0923+0753  & 0  & 0  & 0  & 0  & 1  & 0  & 0  & 0  & 1  \\
J1048-0109  & 0  & 0  & 0  & 1  & 1  & 4  & 2  & 0  & 8  \\
J1058+2930  & 0  & 0  & 0  & 0  & 0  & 0  & 1  & 0  & 1  \\
J1104+2134  & 0  & 0  & 0  & 0  & 0  & 0  & 0  & 0  & 0  \\
J1110-1329  & 0  & 0  & 0  & 0  & 0  & 1  & 0  & 0  & 1  \\
J1129+1846  & 0  & 0  & 0  & 2  & 4  & 1  & 0  & 0  & 7  \\
J1526-2050  & 0  & 0  & 0  & 0  & 0  & 2  & 3  & 0  & 5  \\
J2002-3013  & 0  & 1  & 0  & 0  & 1  & 0  & 1  & 0  & 3  \\
J2102-1458  & 0  & 0  & 0  & 2  & 0  & 0  & 0  & 0  & 2  \\
J2132+1217  & 0  & 0  & 0  & 0  & 0  & 0  & 0  & 1  & 1  \\
\hline
Total & 3 & 3 & 6 & 11 & 34 & 36 & 21 & 4 & 118 \\
\hline
\end{tabular}
\caption{Quasar field name, observed \oiii-emitters counts in each bin range, and total \oiii-emitters counts ($N_{\rm \oiii}$) for each field.}
\label{tab:counts}
\end{table*}

\section {Corner plot for the individulal power law fit}
\begin{figure*}
    \centering
    \includegraphics[width=0.47\textwidth]{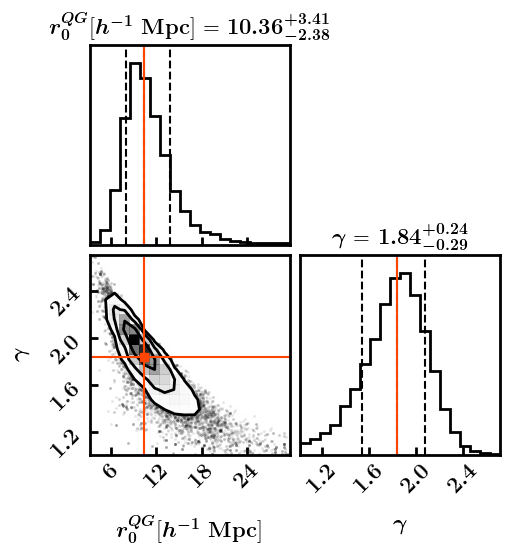}
    \caption{MCMC Posterior distribution for the joint power law model fit to the ASPIRE \oiii-emitter auto correlation function and quasar-\oiii-emitter cross correlation allowing the power law slope to vary.}
    \label{fig:appendix_corner_QG}
\end{figure*}

\begin{figure*}
    \centering
    \includegraphics[width=0.47\textwidth]{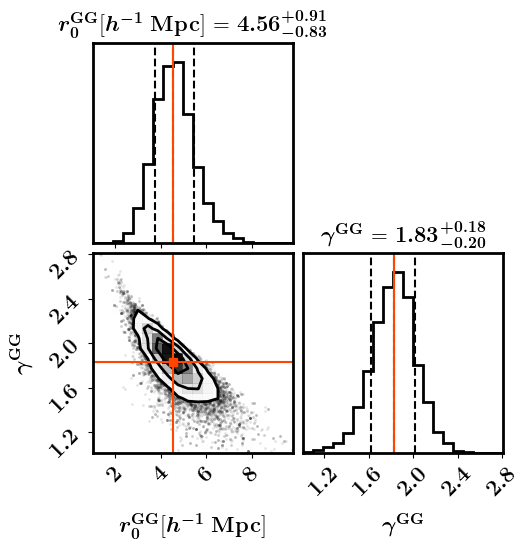}
    \caption{Same as Fig. \ref{fig:appendix_corner_QG}, but for the ASPIRE \oiii-emitter auto correlation function.}
    \label{fig:appendix_corner_GG}
\end{figure*}

We show the MCMC Posterior distribution for the power law model fit to the ASPIRE quasar-\oiii-emitter cross correlation (Fig. \ref{fig:appendix_corner_QG}) and the \oiii-emitter auto correlation function (Fig. \ref{fig:appendix_corner_GG}).

\label{sec:appendix_corner}

\section{QG pair count for each ASPIRE quasar field}
We list the quasar--\oiii-emitters pair count in cylindrical shells for each of the 25 ASPIRE quasar fields in Tab. \ref{tab:counts}. 


\bsp	
\label{lastpage}
\end{document}

%% file: authors_list.tex
\author[J. Huang et al.]{
Jiamu Huang$^{1}$\thanks{E-mail: jiamu\_huang@ucsb.edu}\orcid{0000-0002-5721-0709},
Joseph Hennawi$^{1,2}$\orcid{0000-0002-7054-4332},
Elia Pizzati$^{2,3}$\orcid{0000-0002-9712-0038},
Feige Wang$^{4}$\orcid{0000-0002-7633-431X},
Jinyi Yang$^{4}$\orcid{0000-0001-5287-4242},
\newauthor
~Jaclyn B. Champagne$^{5}$\orcid{0000-0002-6184-9097},
Xiaohui Fan$^{5}$\orcid{},
Eduardo Ba\~nados$^{7}$\orcid{},
Xiangyu Jin$^{4}$\orcid{},
Koki Kakiichi$^{8}$\orcid{},
\newauthor
~Romain A. Meyer$^{9}$\orcid{},
Fengwu Sun$^{11}$\orcid{},
Yunjing Wu$^{12}$\orcid{},
Haowen Zhang$^{10}$\orcid{},
Chiara Mazzucchelli$^{7}$\orcid{0000-0002-5941-5214},
\newauthor
~Anna-Christina Eilers$^{14}$\orcid{0000-0003-2895-6218},
Maria Pudoka$^{5}$\orcid{0000-0003-4924-5941},
Huanian Zhang$^{16}$\orcid{},
Jan-Torge Schindler$^{15}$\orcid{0000-0002-4544-8242},
\newauthor
~Matthieu Schaller$^{2,6}$\orcid{},
Joop Schaye$^{2}$\orcid{},
~Ben Snyder$^{1}$\orcid{},
Yi Kang$^{1}$\orcid{},
Silvia Onorato$^{2}$\orcid{}
\\
$^{1}$ Department of Physics, University of California, Santa Barbara, CA 93106, USA\\
$^{2}$ Leiden Observatory, Leiden University, 2300 RA Leiden, The Netherlands\\
$^{3}$ Department of Astronomy, Harvard University, 60 Garden Street, Cambridge, MA 02138, USA\\
$^{4}$ Department of Astronomy, University of Michigan, 1085 S. University Ave., Ann Arbor, MI 48109, USA\\
$^{5}$ Steward Observatory, University of Arizona, 933 N. Cherry Ave., Tucson, AZ 85721, USA\\
$^{6}$ Lorentz Institute for Theoretical Physics, Leiden University, PO Box 9506, NL-2300 RA Leiden, The Netherlands\\
$^{7}$ Max-Planck-Institut f\"ur Astronomie, K\"onigstuhl 17, D-69117 Heidelberg, Germany\\
$^{8}$ Cosmic Dawn Center (DAWN), Niels Bohr Institute, University of Copenhagen, Jagtvej 128, DK-2200 Copenhagen N, Denmark\\
$^{9}$ Department of Astronomy, University of Geneva, Chemin Pegasi 51, 1290 Versoix, Switzerland\\
$^{10}$ Canadian Institute for Theoretical Astrophysics, 60 St George St, Toronto, ON M5S 3H8, Canada\\
$^{11}$ Center for Astrophysics | Harvard \& Smithsonian, 60 Garden St., Cambridge, MA 02138, USA\\
$^{12}$ Department of Astronomy, Tsinghua University, Beijing 100084, People's Republic of China\\
$^{13}$ Instituto de Estudios Astrof\'{\i}sicos, Facultad de Ingenier\'{\i}a y Ciencias, Universidad Diego Portales, Avenida Ejercito Libertador 441, Santiago, Chile\\
$^{14}$ Department of Physics and MIT Kavli Institute for Astrophysics and Space Research, Massachusetts Institute of Technology, Cambridge, MA 02139, USA\\
$^{15}$ Hamburger Sternwarte, Universit\"at Hamburg, Gojenbergsweg 112, D-21029 Hamburg, Germany\\
$^{16}$ Department of Astronomy, Huazhong University of Science and Technology, Wuhan, Hubei 430074, People's Republic of China
}

%% file: aspire_citations.bib
@ARTICLE{Eilers2024,
       author = {{Eilers}, Anna-Christina and {Mackenzie}, Ruari and {Pizzati}, Elia and {Matthee}, Jorryt and {Hennawi}, Joseph F. and {Zhang}, Haowen and {Bordoloi}, Rongmon and {Kashino}, Daichi and {Lilly}, Simon J. and {Naidu}, Rohan P. and {Simcoe}, Robert A. and {Yue}, Minghao and {Frenk}, Carlos S. and {Helly}, John C. and {Schaller}, Matthieu and {Schaye}, Joop},
        title = "{EIGER VI. The Correlation Function, Host Halo Mass and Duty Cycle of Luminous Quasars at $zrsim6$}",
      journal = {arXiv e-prints},
         year = 2024,
        month = mar,
          eid = {arXiv:2403.07986},
        pages = {arXiv:2403.07986},
          doi = {10.48550/arXiv.2403.07986},
archivePrefix = {arXiv},
       eprint = {2403.07986},
 primaryClass = {astro-ph.GA},
       adsurl = {https://ui.adsabs.harvard.edu/abs/2024arXiv240307986E}
}

@ARTICLE{Wang2023,
       author = {{Wang}, Feige and {Yang}, Jinyi and {Hennawi}, Joseph F. and {Fan}, Xiaohui and {Sun}, Fengwu and {Champagne}, Jaclyn B. and {Costa}, Tiago and {Habouzit}, Melanie and {Endsley}, Ryan and {Li}, Zihao and {Lin}, Xiaojing and {Meyer}, Romain A. and {Schindler}, Jan-Torge and {Wu}, Yunjing and {Ba{\~n}ados}, Eduardo and {Barth}, Aaron J. and {Bhowmick}, Aklant K. and {Bieri}, Rebekka and {Blecha}, Laura and {Bosman}, Sarah and {Cai}, Zheng and {Colina}, Luis and {Connor}, Thomas and {Davies}, Frederick B. and {Decarli}, Roberto and {De Rosa}, Gisella and {Drake}, Alyssa B. and {Egami}, Eiichi and {Eilers}, Anna-Christina and {Evans}, Analis E. and {Farina}, Emanuele Paolo and {Haiman}, Zoltan and {Jiang}, Linhua and {Jin}, Xiangyu and {Jun}, Hyunsung D. and {Kakiichi}, Koki and {Khusanova}, Yana and {Kulkarni}, Girish and {Li}, Mingyu and {Liu}, Weizhe and {Loiacono}, Federica and {Lupi}, Alessandro and {Mazzucchelli}, Chiara and {Onoue}, Masafusa and {Pudoka}, Maria A. and {Rojas-Ruiz}, Sof{\'\i}a and {Shen}, Yue and {Strauss}, Michael A. and {Tee}, Wei Leong and {Trakhtenbrot}, Benny and {Trebitsch}, Maxime and {Venemans}, Bram and {Volonteri}, Marta and {Walter}, Fabian and {Xie}, Zhang-Liang and {Yue}, Minghao and {Zhang}, Haowen and {Zhang}, Huanian and {Zou}, Siwei},
        title = "{A SPectroscopic Survey of Biased Halos in the Reionization Era (ASPIRE): JWST Reveals a Filamentary Structure around a z = 6.61 Quasar}",
      journal = {\apjl},
         year = 2023,
        month = jul,
       volume = {951},
       number = {1},
          eid = {L4},
        pages = {L4},
          doi = {10.3847/2041-8213/accd6f},
archivePrefix = {arXiv},
       eprint = {2304.09894},
 primaryClass = {astro-ph.GA},
       adsurl = {https://ui.adsabs.harvard.edu/abs/2023ApJ...951L...4W}
}

@ARTICLE{Matthee2023,
       author = {{Matthee}, Jorryt and {Mackenzie}, Ruari and {Simcoe}, Robert A. and {Kashino}, Daichi and {Lilly}, Simon J. and {Bordoloi}, Rongmon and {Eilers}, Anna-Christina},
        title = "{EIGER. II. First Spectroscopic Characterization of the Young Stars and Ionized Gas Associated with Strong H{\ensuremath{\beta}} and [O III] Line Emission in Galaxies at z = 5-7 with JWST}",
      journal = {\apj},
         year = 2023,
        month = jun,
       volume = {950},
       number = {1},
          eid = {67},
        pages = {67},
          doi = {10.3847/1538-4357/acc846},
archivePrefix = {arXiv},
       eprint = {2211.08255},
 primaryClass = {astro-ph.GA},
       adsurl = {https://ui.adsabs.harvard.edu/abs/2023ApJ...950...67M}
}

@MISC{Sun2024jwst,
       author = {{Sun}, Fengwu and {Boyer}, Martha L. and {Egami}, Eiichi and {Pirzkal}, Norbert and {Rieke}, Marcia J. and {Sun}, Jiayi},
        title = "{A Novel Wavelength Calibration of NIRCam WFSS with a Nearby Star-Forming Galaxy}",
 howpublished = {JWST Proposal. Cycle 3, ID. \#4924},
         year = 2024,
        month = feb,
        pages = {4924},
       adsurl = {https://ui.adsabs.harvard.edu/abs/2024jwst.prop.4924S}
}

@ARTICLE{Schaye2023,
       author = {{Schaye}, Joop and {Kugel}, Roi and {Schaller}, Matthieu and {Helly}, John C. and {Braspenning}, Joey and {Elbers}, Willem and {McCarthy}, Ian G. and {van Daalen}, Marcel P. and {Vandenbroucke}, Bert and {Frenk}, Carlos S. and {Kwan}, Juliana and {Salcido}, Jaime and {Bah{\'e}}, Yannick M. and {Borrow}, Josh and {Chaikin}, Evgenii and {Hahn}, Oliver and {Hu{\v{s}}ko}, Filip and {Jenkins}, Adrian and {Lacey}, Cedric G. and {Nobels}, Folkert S.~J.},
        title = "{The FLAMINGO project: cosmological hydrodynamical simulations for large-scale structure and galaxy cluster surveys}",
      journal = {\mnras},
         year = 2023,
        month = dec,
       volume = {526},
       number = {4},
        pages = {4978-5020},
          doi = {10.1093/mnras/stad2419},
archivePrefix = {arXiv},
       eprint = {2306.04024},
 primaryClass = {astro-ph.CO},
       adsurl = {https://ui.adsabs.harvard.edu/abs/2023MNRAS.526.4978S}
}

@ARTICLE{Kugel2023,
       author = {{Kugel}, Roi and {Schaye}, Joop and {Schaller}, Matthieu and {Helly}, John C. and {Braspenning}, Joey and {Elbers}, Willem and {Frenk}, Carlos S. and {McCarthy}, Ian G. and {Kwan}, Juliana and {Salcido}, Jaime and {van Daalen}, Marcel P. and {Vandenbroucke}, Bert and {Bah{\'e}}, Yannick M. and {Borrow}, Josh and {Chaikin}, Evgenii and {Hu{\v{s}}ko}, Filip and {Jenkins}, Adrian and {Lacey}, Cedric G. and {Nobels}, Folkert S.~J. and {Vernon}, Ian},
        title = "{FLAMINGO: calibrating large cosmological hydrodynamical simulations with machine learning}",
      journal = {\mnras},
         year = 2023,
        month = dec,
       volume = {526},
       number = {4},
        pages = {6103-6127},
          doi = {10.1093/mnras/stad2540},
archivePrefix = {arXiv},
       eprint = {2306.05492},
 primaryClass = {astro-ph.CO},
       adsurl = {https://ui.adsabs.harvard.edu/abs/2023MNRAS.526.6103K}
}

@ARTICLE{GarciaVergara2017,
       author = {{Garc{\'\i}a-Vergara}, Cristina and {Hennawi}, Joseph F. and {Barrientos}, L. Felipe and {Rix}, Hans-Walter},
        title = "{Strong Clustering of Lyman Break Galaxies around Luminous Quasars at Z {\ensuremath{\sim}} 4}",
      journal = {\apj},
         year = 2017,
        month = oct,
       volume = {848},
       number = {1},
          eid = {7},
        pages = {7},
          doi = {10.3847/1538-4357/aa8b69},
archivePrefix = {arXiv},
       eprint = {1701.01114},
 primaryClass = {astro-ph.GA},
       adsurl = {https://ui.adsabs.harvard.edu/abs/2017ApJ...848....7G}
}

@ARTICLE{LZ1993,
       author = {{Landy}, Stephen D. and {Szalay}, Alexander S.},
        title = "{Bias and Variance of Angular Correlation Functions}",
      journal = {\apj},
         year = 1993,
        month = jul,
       volume = {412},
        pages = {64},
          doi = {10.1086/172900},
       adsurl = {https://ui.adsabs.harvard.edu/abs/1993ApJ...412...64L}
}

@ARTICLE{Pizzati2024a,
       author = {{Pizzati}, Elia and {Hennawi}, Joseph F. and {Schaye}, Joop and {Schaller}, Matthieu},
        title = "{Revisiting the extreme clustering of z {\ensuremath{\approx}} 4 quasars with large volume cosmological simulations}",
      journal = {\mnras},
         year = 2024,
        month = mar,
       volume = {528},
       number = {3},
        pages = {4466-4489},
          doi = {10.1093/mnras/stae329},
archivePrefix = {arXiv},
       eprint = {2311.17181},
 primaryClass = {astro-ph.GA},
       adsurl = {https://ui.adsabs.harvard.edu/abs/2024MNRAS.528.4466P}
}

@ARTICLE{Pizzati2024b,
       author = {{Pizzati}, Elia and {Hennawi}, Joseph F. and {Schaye}, Joop and {Schaller}, Matthieu and {Eilers}, Anna-Christina and {Wang}, Feige and {Frenk}, Carlos S. and {Elbers}, Willem and {Helly}, John C. and {Mackenzie}, Ruari and {Matthee}, Jorryt and {Bordoloi}, Rongmon and {Kashino}, Daichi and {Naidu}, Rohan P. and {Yue}, Minghao},
        title = "{A unified model for the clustering of quasars and galaxies at z {\ensuremath{\approx}} 6}",
      journal = {\mnras},
         year = 2024,
        month = nov,
       volume = {534},
       number = {4},
        pages = {3155-3175},
          doi = {10.1093/mnras/stae2307},
archivePrefix = {arXiv},
       eprint = {2403.12140},
 primaryClass = {astro-ph.GA},
       adsurl = {https://ui.adsabs.harvard.edu/abs/2024MNRAS.534.3155P}
}

@ARTICLE{Soltan1982,
       author = {{Soltan}, A.},
        title = "{Masses of quasars.}",
      journal = {\mnras},
         year = 1982,
        month = jul,
       volume = {200},
        pages = {115-122},
          doi = {10.1093/mnras/200.1.115},
       adsurl = {https://ui.adsabs.harvard.edu/abs/1982MNRAS.200..115S}
}

@ARTICLE{Fan2023,
       author = {{Fan}, Xiaohui and {Ba{\~n}ados}, Eduardo and {Simcoe}, Robert A.},
        title = "{Quasars and the Intergalactic Medium at Cosmic Dawn}",
      journal = {\araa},
         year = 2023,
        month = aug,
       volume = {61},
        pages = {373-426},
          doi = {10.1146/annurev-astro-052920-102455},
archivePrefix = {arXiv},
       eprint = {2212.06907},
 primaryClass = {astro-ph.GA},
       adsurl = {https://ui.adsabs.harvard.edu/abs/2023ARA&A..61..373F}
}

@ARTICLE{Efstathiou1988,
       author = {{Efstathiou}, G. and {Rees}, M.~J.},
        title = "{High-redshift quasars in the Cold Dark Matter cosmogony}",
      journal = {\mnras},
         year = 1988,
        month = feb,
       volume = {230},
        pages = {5p-11p},
          doi = {10.1093/mnras/230.1.5P},
       adsurl = {https://ui.adsabs.harvard.edu/abs/1988MNRAS.230P...5E}
}

@ARTICLE{Kaiser1984,
       author = {{Kaiser}, N.},
        title = "{On the spatial correlations of Abell clusters.}",
      journal = {\apjl},
         year = 1984,
        month = sep,
       volume = {284},
        pages = {L9-L12},
          doi = {10.1086/184341},
       adsurl = {https://ui.adsabs.harvard.edu/abs/1984ApJ...284L...9K}
}

@ARTICLE{Bogdan2024,
       author = {{Bogd{\'a}n}, {\'A}kos and {Goulding}, Andy D. and {Natarajan}, Priyamvada and {Kov{\'a}cs}, Orsolya E. and {Tremblay}, Grant R. and {Chadayammuri}, Urmila and {Volonteri}, Marta and {Kraft}, Ralph P. and {Forman}, William R. and {Jones}, Christine and {Churazov}, Eugene and {Zhuravleva}, Irina},
        title = "{Evidence for heavy-seed origin of early supermassive black holes from a z {\ensuremath{\approx}} 10 X-ray quasar}",
      journal = {Nature Astronomy},
         year = 2024,
        month = jan,
       volume = {8},
        pages = {126-133},
          doi = {10.1038/s41550-023-02111-9},
archivePrefix = {arXiv},
       eprint = {2305.15458},
 primaryClass = {astro-ph.GA},
       adsurl = {https://ui.adsabs.harvard.edu/abs/2024NatAs...8..126B}
}

@ARTICLE{Begelman2006,
       author = {{Begelman}, Mitchell C. and {Volonteri}, Marta and {Rees}, Martin J.},
        title = "{Formation of supermassive black holes by direct collapse in pre-galactic haloes}",
      journal = {\mnras},
         year = 2006,
        month = jul,
       volume = {370},
       number = {1},
        pages = {289-298},
          doi = {10.1111/j.1365-2966.2006.10467.x},
archivePrefix = {arXiv},
       eprint = {astro-ph/0602363},
 primaryClass = {astro-ph},
       adsurl = {https://ui.adsabs.harvard.edu/abs/2006MNRAS.370..289B}
}

@ARTICLE{Regan2024,
       author = {{Regan}, John and {Volonteri}, Marta},
        title = "{Massive Black Hole Seeds}",
      journal = {The Open Journal of Astrophysics},
         year = 2024,
        month = sep,
       volume = {7},
          eid = {72},
        pages = {72},
          doi = {10.33232/001c.123239},
archivePrefix = {arXiv},
       eprint = {2405.17975},
 primaryClass = {astro-ph.GA},
       adsurl = {https://ui.adsabs.harvard.edu/abs/2024OJAp....7E..72R}
}

@ARTICLE{Madau2001,
       author = {{Madau}, Piero and {Rees}, Martin J.},
        title = "{Massive Black Holes as Population III Remnants}",
      journal = {\apjl},
         year = 2001,
        month = apr,
       volume = {551},
       number = {1},
        pages = {L27-L30},
          doi = {10.1086/319848},
archivePrefix = {arXiv},
       eprint = {astro-ph/0101223},
 primaryClass = {astro-ph},
       adsurl = {https://ui.adsabs.harvard.edu/abs/2001ApJ...551L..27M}
}

@ARTICLE{Kashino2023,
       author = {{Kashino}, Daichi and {Lilly}, Simon J. and {Matthee}, Jorryt and {Eilers}, Anna-Christina and {Mackenzie}, Ruari and {Bordoloi}, Rongmon and {Simcoe}, Robert A.},
        title = "{EIGER. I. A Large Sample of [O III]-emitting Galaxies at 5.3 < z < 6.9 and Direct Evidence for Local Reionization by Galaxies}",
      journal = {\apj},
         year = 2023,
        month = jun,
       volume = {950},
       number = {1},
          eid = {66},
        pages = {66},
          doi = {10.3847/1538-4357/acc588},
archivePrefix = {arXiv},
       eprint = {2211.08254},
 primaryClass = {astro-ph.GA},
       adsurl = {https://ui.adsabs.harvard.edu/abs/2023ApJ...950...66K}
}

@ARTICLE{Yang2005,
       author = {{Yang}, Xiaohu and {Mo}, H.~J. and {van den Bosch}, Frank C. and {Jing}, Y.~P.},
        title = "{The two-point correlation of galaxy groups: probing the clustering of dark matter haloes}",
      journal = {\mnras},
         year = 2005,
        month = feb,
       volume = {357},
       number = {2},
        pages = {608-618},
          doi = {10.1111/j.1365-2966.2005.08667.x},
archivePrefix = {arXiv},
       eprint = {astro-ph/0406593},
 primaryClass = {astro-ph},
       adsurl = {https://ui.adsabs.harvard.edu/abs/2005MNRAS.357..608Y}
}

@ARTICLE{Arita2024,
       author = {{Arita}, Junya and {Kashikawa}, Nobunari and {Onoue}, Masafusa and {Yoshioka}, Takehiro and {Takeda}, Yoshihiro and {Hoshi}, Hiroki and {Shimizu}, Shunta},
        title = "{The nature of low-luminosity AGNs discovered by JWST based on clustering analysis: Progenitors of low-z quasars?}",
      journal = {\mnras},
         year = 2024,
        month = dec,
          doi = {10.1093/mnras/stae2765},
archivePrefix = {arXiv},
       eprint = {2410.08707},
 primaryClass = {astro-ph.GA},
       adsurl = {https://ui.adsabs.harvard.edu/abs/2024MNRAS.tmp.2641A}
}

@ARTICLE{MoWhite2002,
       author = {{Mo}, H.~J. and {White}, S.~D.~M.},
        title = "{The abundance and clustering of dark haloes in the standard {\ensuremath{\Lambda}}CDM cosmogony}",
      journal = {\mnras},
         year = 2002,
        month = oct,
       volume = {336},
       number = {1},
        pages = {112-118},
          doi = {10.1046/j.1365-8711.2002.05723.x},
archivePrefix = {arXiv},
       eprint = {astro-ph/0202393},
 primaryClass = {astro-ph},
       adsurl = {https://ui.adsabs.harvard.edu/abs/2002MNRAS.336..112M}
}

@ARTICLE{Hennawi2006,
       author = {{Hennawi}, Joseph F. and {Strauss}, Michael A. and {Oguri}, Masamune and {Inada}, Naohisa and {Richards}, Gordon T. and {Pindor}, Bartosz and {Schneider}, Donald P. and {Becker}, Robert H. and {Gregg}, Michael D. and {Hall}, Patrick B. and {Johnston}, David E. and {Fan}, Xiaohui and {Burles}, Scott and {Schlegel}, David J. and {Gunn}, James E. and {Lupton}, Robert H. and {Bahcall}, Neta A. and {Brunner}, Robert J. and {Brinkmann}, Jon},
        title = "{Binary Quasars in the Sloan Digital Sky Survey: Evidence for Excess Clustering on Small Scales}",
      journal = {\aj},
         year = 2006,
        month = jan,
       volume = {131},
       number = {1},
        pages = {1-23},
          doi = {10.1086/498235},
archivePrefix = {arXiv},
       eprint = {astro-ph/0504535},
 primaryClass = {astro-ph},
       adsurl = {https://ui.adsabs.harvard.edu/abs/2006AJ....131....1H}
}

@ARTICLE{Salpeter1964,
       author = {{Salpeter}, E.~E.},
        title = "{Accretion of Interstellar Matter by Massive Objects.}",
      journal = {\apj},
         year = 1964,
        month = aug,
       volume = {140},
        pages = {796-800},
          doi = {10.1086/147973},
       adsurl = {https://ui.adsabs.harvard.edu/abs/1964ApJ...140..796S}
}

@ARTICLE{Wang2021,
       author = {{Wang}, Feige and {Yang}, Jinyi and {Fan}, Xiaohui and {Hennawi}, Joseph F. and {Barth}, Aaron J. and {Banados}, Eduardo and {Bian}, Fuyan and {Boutsia}, Konstantina and {Connor}, Thomas and {Davies}, Frederick B. and {Decarli}, Roberto and {Eilers}, Anna-Christina and {Farina}, Emanuele Paolo and {Green}, Richard and {Jiang}, Linhua and {Li}, Jiang-Tao and {Mazzucchelli}, Chiara and {Nanni}, Riccardo and {Schindler}, Jan-Torge and {Venemans}, Bram and {Walter}, Fabian and {Wu}, Xue-Bing and {Yue}, Minghao},
        title = "{A Luminous Quasar at Redshift 7.642}",
      journal = {\apjl},
         year = 2021,
        month = jan,
       volume = {907},
       number = {1},
          eid = {L1},
        pages = {L1},
          doi = {10.3847/2041-8213/abd8c6},
archivePrefix = {arXiv},
       eprint = {2101.03179},
 primaryClass = {astro-ph.GA},
       adsurl = {https://ui.adsabs.harvard.edu/abs/2021ApJ...907L...1W}
}

@ARTICLE{Matsuoka2019,
       author = {{Matsuoka}, Yoshiki and {Onoue}, Masafusa and {Kashikawa}, Nobunari and {Strauss}, Michael A. and {Iwasawa}, Kazushi and {Lee}, Chien-Hsiu and {Imanishi}, Masatoshi and {Nagao}, Tohru and {Akiyama}, Masayuki and {Asami}, Naoko and {Bosch}, James and {Furusawa}, Hisanori and {Goto}, Tomotsugu and {Gunn}, James E. and {Harikane}, Yuichi and {Ikeda}, Hiroyuki and {Izumi}, Takuma and {Kawaguchi}, Toshihiro and {Kato}, Nanako and {Kikuta}, Satoshi and {Kohno}, Kotaro and {Komiyama}, Yutaka and {Koyama}, Shuhei and {Lupton}, Robert H. and {Minezaki}, Takeo and {Miyazaki}, Satoshi and {Murayama}, Hitoshi and {Niida}, Mana and {Nishizawa}, Atsushi J. and {Noboriguchi}, Akatoki and {Oguri}, Masamune and {Ono}, Yoshiaki and {Ouchi}, Masami and {Price}, Paul A. and {Sameshima}, Hiroaki and {Schulze}, Andreas and {Shirakata}, Hikari and {Silverman}, John D. and {Sugiyama}, Naoshi and {Tait}, Philip J. and {Takada}, Masahiro and {Takata}, Tadafumi and {Tanaka}, Masayuki and {Tang}, Ji-Jia and {Toba}, Yoshiki and {Utsumi}, Yousuke and {Wang}, Shiang-Yu and {Yamashita}, Takuji},
        title = "{Discovery of the First Low-luminosity Quasar at z > 7}",
      journal = {\apjl},
         year = 2019,
        month = feb,
       volume = {872},
       number = {1},
          eid = {L2},
        pages = {L2},
          doi = {10.3847/2041-8213/ab0216},
archivePrefix = {arXiv},
       eprint = {1901.10487},
 primaryClass = {astro-ph.GA},
       adsurl = {https://ui.adsabs.harvard.edu/abs/2019ApJ...872L...2M}
}

@ARTICLE{Yang2020,
       author = {{Yang}, Jinyi and {Wang}, Feige and {Fan}, Xiaohui and {Hennawi}, Joseph F. and {Davies}, Frederick B. and {Yue}, Minghao and {Banados}, Eduardo and {Wu}, Xue-Bing and {Venemans}, Bram and {Barth}, Aaron J. and {Bian}, Fuyan and {Boutsia}, Konstantina and {Decarli}, Roberto and {Farina}, Emanuele Paolo and {Green}, Richard and {Jiang}, Linhua and {Li}, Jiang-Tao and {Mazzucchelli}, Chiara and {Walter}, Fabian},
        title = "{P{\={o}}niu{\={a}}'ena: A Luminous z = 7.5 Quasar Hosting a 1.5 Billion Solar Mass Black Hole}",
      journal = {\apjl},
         year = 2020,
        month = jul,
       volume = {897},
       number = {1},
          eid = {L14},
        pages = {L14},
          doi = {10.3847/2041-8213/ab9c26},
archivePrefix = {arXiv},
       eprint = {2006.13452},
 primaryClass = {astro-ph.GA},
       adsurl = {https://ui.adsabs.harvard.edu/abs/2020ApJ...897L..14Y}
}

@ARTICLE{Tanaka2009,
       author = {{Tanaka}, Takamitsu and {Haiman}, Zolt{\'a}n},
        title = "{The Assembly of Supermassive Black Holes at High Redshifts}",
      journal = {\apj},
         year = 2009,
        month = may,
       volume = {696},
       number = {2},
        pages = {1798-1822},
          doi = {10.1088/0004-637X/696/2/1798},
archivePrefix = {arXiv},
       eprint = {0807.4702},
 primaryClass = {astro-ph},
       adsurl = {https://ui.adsabs.harvard.edu/abs/2009ApJ...696.1798T}
}

@ARTICLE{Cole1989,
       author = {{Cole}, Shaun and {Kaiser}, Nick},
        title = "{Biased clustering in the cold dark matter cosmogony.}",
      journal = {\mnras},
         year = 1989,
        month = apr,
       volume = {237},
        pages = {1127-1146},
          doi = {10.1093/mnras/237.4.1127},
       adsurl = {https://ui.adsabs.harvard.edu/abs/1989MNRAS.237.1127C}
}

@ARTICLE{Martini2001,
       author = {{Martini}, Paul and {Weinberg}, David H.},
        title = "{Quasar Clustering and the Lifetime of Quasars}",
      journal = {\apj},
         year = 2001,
        month = jan,
       volume = {547},
       number = {1},
        pages = {12-26},
          doi = {10.1086/318331},
archivePrefix = {arXiv},
       eprint = {astro-ph/0002384},
 primaryClass = {astro-ph},
       adsurl = {https://ui.adsabs.harvard.edu/abs/2001ApJ...547...12M}
}

@ARTICLE{Osmer1981,
       author = {{Osmer}, P.~S.},
        title = "{The three-dimensional distribution of quasars in the CTIO surveys}",
      journal = {\apj},
         year = 1981,
        month = aug,
       volume = {247},
        pages = {762-773},
          doi = {10.1086/159087},
       adsurl = {https://ui.adsabs.harvard.edu/abs/1981ApJ...247..762O}
}

@ARTICLE{Shen2007,
       author = {{Shen}, Yue and {Strauss}, Michael A. and {Oguri}, Masamune and {Hennawi}, Joseph F. and {Fan}, Xiaohui and {Richards}, Gordon T. and {Hall}, Patrick B. and {Gunn}, James E. and {Schneider}, Donald P. and {Szalay}, Alexander S. and {Thakar}, Anirudda R. and {Vanden Berk}, Daniel E. and {Anderson}, Scott F. and {Bahcall}, Neta A. and {Connolly}, Andrew J. and {Knapp}, Gillian R.},
        title = "{Clustering of High-Redshift (z >= 2.9) Quasars from the Sloan Digital Sky Survey}",
      journal = {\aj},
         year = 2007,
        month = may,
       volume = {133},
       number = {5},
        pages = {2222-2241},
          doi = {10.1086/513517},
archivePrefix = {arXiv},
       eprint = {astro-ph/0702214},
 primaryClass = {astro-ph},
       adsurl = {https://ui.adsabs.harvard.edu/abs/2007AJ....133.2222S}
}

@ARTICLE{Porciani2004,
       author = {{Porciani}, Cristiano and {Magliocchetti}, Manuela and {Norberg}, Peder},
        title = "{Cosmic evolution of quasar clustering: implications for the host haloes}",
      journal = {\mnras},
         year = 2004,
        month = dec,
       volume = {355},
       number = {3},
        pages = {1010-1030},
          doi = {10.1111/j.1365-2966.2004.08408.x},
archivePrefix = {arXiv},
       eprint = {astro-ph/0406036},
 primaryClass = {astro-ph},
       adsurl = {https://ui.adsabs.harvard.edu/abs/2004MNRAS.355.1010P}
}

@ARTICLE{Croom2005,
       author = {{Croom}, Scott M. and {Boyle}, B.~J. and {Shanks}, T. and {Smith}, R.~J. and {Miller}, L. and {Outram}, P.~J. and {Loaring}, N.~S. and {Hoyle}, F. and {da {\^A}ngela}, J.},
        title = "{The 2dF QSO Redshift Survey - XIV. Structure and evolution from the two-point correlation function}",
      journal = {\mnras},
         year = 2005,
        month = jan,
       volume = {356},
       number = {2},
        pages = {415-438},
          doi = {10.1111/j.1365-2966.2004.08379.x},
archivePrefix = {arXiv},
       eprint = {astro-ph/0409314},
 primaryClass = {astro-ph},
       adsurl = {https://ui.adsabs.harvard.edu/abs/2005MNRAS.356..415C}
}

@ARTICLE{Coil2007,
       author = {{Coil}, Alison L. and {Hennawi}, Joseph F. and {Newman}, Jeffrey A. and {Cooper}, Michael C. and {Davis}, Marc},
        title = "{The DEEP2 Galaxy Redshift Survey: Clustering of Quasars and Galaxies at z = 1}",
      journal = {\apj},
         year = 2007,
        month = jan,
       volume = {654},
       number = {1},
        pages = {115-124},
          doi = {10.1086/509099},
archivePrefix = {arXiv},
       eprint = {astro-ph/0607454},
 primaryClass = {astro-ph},
       adsurl = {https://ui.adsabs.harvard.edu/abs/2007ApJ...654..115C}
}

@ARTICLE{Wang2019,
       author = {{Wang}, Feige and {Yang}, Jinyi and {Fan}, Xiaohui and {Wu}, Xue-Bing and {Yue}, Minghao and {Li}, Jiang-Tao and {Bian}, Fuyan and {Jiang}, Linhua and {Ba{\~n}ados}, Eduardo and {Schindler}, Jan-Torge and {Findlay}, Joseph R. and {Davies}, Frederick B. and {Decarli}, Roberto and {Farina}, Emanuele P. and {Green}, Richard and {Hennawi}, Joseph F. and {Huang}, Yun-Hsin and {Mazzuccheli}, Chiara and {McGreer}, Ian D. and {Venemans}, Bram and {Walter}, Fabian and {Dye}, Simon and {Lyke}, Brad W. and {Myers}, Adam D. and {Nunez}, Evan Haze},
        title = "{Exploring Reionization-era Quasars. III. Discovery of 16 Quasars at 6.4 {\ensuremath{\lesssim}} z {\ensuremath{\lesssim}} 6.9 with DESI Legacy Imaging Surveys and the UKIRT Hemisphere Survey and Quasar Luminosity Function at z {\ensuremath{\sim}} 6.7}",
      journal = {\apj},
         year = 2019,
        month = oct,
       volume = {884},
       number = {1},
          eid = {30},
        pages = {30},
          doi = {10.3847/1538-4357/ab2be5},
archivePrefix = {arXiv},
       eprint = {1810.11926},
 primaryClass = {astro-ph.GA},
       adsurl = {https://ui.adsabs.harvard.edu/abs/2019ApJ...884...30W}
}

@ARTICLE{Schindler2023,
       author = {{Schindler}, Jan-Torge and {Ba{\~n}ados}, Eduardo and {Connor}, Thomas and {Decarli}, Roberto and {Fan}, Xiaohui and {Farina}, Emanuele Paolo and {Mazzucchelli}, Chiara and {Nanni}, Riccardo and {Rix}, Hans-Walter and {Stern}, Daniel and {Venemans}, Bram P. and {Walter}, Fabian},
        title = "{The Pan-STARRS1 z > 5.6 Quasar Survey. III. The z {\ensuremath{\approx}} 6 Quasar Luminosity Function}",
      journal = {\apj},
         year = 2023,
        month = jan,
       volume = {943},
       number = {1},
          eid = {67},
        pages = {67},
          doi = {10.3847/1538-4357/aca7ca},
archivePrefix = {arXiv},
       eprint = {2212.04179},
 primaryClass = {astro-ph.GA},
       adsurl = {https://ui.adsabs.harvard.edu/abs/2023ApJ...943...67S}
}

@ARTICLE{Bunker2023,
       author = {{Bunker}, Andrew J. and {Saxena}, Aayush and {Cameron}, Alex J. and {Willott}, Chris J. and {Curtis-Lake}, Emma and {Jakobsen}, Peter and {Carniani}, Stefano and {Smit}, Renske and {Maiolino}, Roberto and {Witstok}, Joris and {Curti}, Mirko and {D'Eugenio}, Francesco and {Jones}, Gareth C. and {Ferruit}, Pierre and {Arribas}, Santiago and {Charlot}, Stephane and {Chevallard}, Jacopo and {Giardino}, Giovanna and {de Graaff}, Anna and {Looser}, Tobias J. and {L{\"u}tzgendorf}, Nora and {Maseda}, Michael V. and {Rawle}, Tim and {Rix}, Hans-Walter and {Del Pino}, Bruno Rodr{\'\i}guez and {Alberts}, Stacey and {Egami}, Eiichi and {Eisenstein}, Daniel J. and {Endsley}, Ryan and {Hainline}, Kevin and {Hausen}, Ryan and {Johnson}, Benjamin D. and {Rieke}, George and {Rieke}, Marcia and {Robertson}, Brant E. and {Shivaei}, Irene and {Stark}, Daniel P. and {Sun}, Fengwu and {Tacchella}, Sandro and {Tang}, Mengtao and {Williams}, Christina C. and {Willmer}, Christopher N.~A. and {Baker}, William M. and {Baum}, Stefi and {Bhatawdekar}, Rachana and {Bowler}, Rebecca and {Boyett}, Kristan and {Chen}, Zuyi and {Circosta}, Chiara and {Helton}, Jakob M. and {Ji}, Zhiyuan and {Kumari}, Nimisha and {Lyu}, Jianwei and {Nelson}, Erica and {Parlanti}, Eleonora and {Perna}, Michele and {Sandles}, Lester and {Scholtz}, Jan and {Suess}, Katherine A. and {Topping}, Michael W. and {{\"U}bler}, Hannah and {Wallace}, Imaan E.~B. and {Whitler}, Lily},
        title = "{JADES NIRSpec Spectroscopy of GN-z11: Lyman-{\ensuremath{\alpha}} emission and possible enhanced nitrogen abundance in a z = 10.60 luminous galaxy}",
      journal = {\aap},
         year = 2023,
        month = sep,
       volume = {677},
          eid = {A88},
        pages = {A88},
          doi = {10.1051/0004-6361/202346159},
archivePrefix = {arXiv},
       eprint = {2302.07256},
 primaryClass = {astro-ph.GA},
       adsurl = {https://ui.adsabs.harvard.edu/abs/2023A&A...677A..88B}
}

@ARTICLE{Mazzucchelli2017,
       author = {{Mazzucchelli}, C. and {Ba{\~n}ados}, E. and {Venemans}, B.~P. and {Decarli}, R. and {Farina}, E.~P. and {Walter}, F. and {Eilers}, A. -C. and {Rix}, H. -W. and {Simcoe}, R. and {Stern}, D. and {Fan}, X. and {Schlafly}, E. and {De Rosa}, G. and {Hennawi}, J. and {Chambers}, K.~C. and {Greiner}, J. and {Burgett}, W. and {Draper}, P.~W. and {Kaiser}, N. and {Kudritzki}, R. -P. and {Magnier}, E. and {Metcalfe}, N. and {Waters}, C. and {Wainscoat}, R.~J.},
        title = "{Physical Properties of 15 Quasars at z {\ensuremath{\gtrsim}} 6.5}",
      journal = {\apj},
         year = 2017,
        month = nov,
       volume = {849},
       number = {2},
          eid = {91},
        pages = {91},
          doi = {10.3847/1538-4357/aa9185},
archivePrefix = {arXiv},
       eprint = {1710.01251},
 primaryClass = {astro-ph.GA},
       adsurl = {https://ui.adsabs.harvard.edu/abs/2017ApJ...849...91M}
}

@ARTICLE{Banados2018,
       author = {{Ba{\~n}ados}, Eduardo and {Venemans}, Bram P. and {Mazzucchelli}, Chiara and {Farina}, Emanuele P. and {Walter}, Fabian and {Wang}, Feige and {Decarli}, Roberto and {Stern}, Daniel and {Fan}, Xiaohui and {Davies}, Frederick B. and {Hennawi}, Joseph F. and {Simcoe}, Robert A. and {Turner}, Monica L. and {Rix}, Hans-Walter and {Yang}, Jinyi and {Kelson}, Daniel D. and {Rudie}, Gwen C. and {Winters}, Jan Martin},
        title = "{An 800-million-solar-mass black hole in a significantly neutral Universe at a redshift of 7.5}",
      journal = {\nat},
         year = 2018,
        month = jan,
       volume = {553},
       number = {7689},
        pages = {473-476},
          doi = {10.1038/nature25180},
archivePrefix = {arXiv},
       eprint = {1712.01860},
 primaryClass = {astro-ph.GA},
       adsurl = {https://ui.adsabs.harvard.edu/abs/2018Natur.553..473B}
}

@INPROCEEDINGS{Shankar2010,
       author = {{Shankar}, Francesco},
        title = "{Merger-Induced Quasars, Their Light Curves, and Their Host Halos}",
    booktitle = {Co-Evolution of Central Black Holes and Galaxies},
         year = 2010,
       editor = {{Peterson}, Bradley M. and {Somerville}, Rachel S. and {Storchi-Bergmann}, Thaisa},
       series = {IAU Symposium},
       volume = {267},
        month = may,
        pages = {248-253},
          doi = {10.1017/S1743921310006356},
       adsurl = {https://ui.adsabs.harvard.edu/abs/2010IAUS..267..248S}
}

@ARTICLE{Conroy2013,
       author = {{Conroy}, Charlie and {White}, Martin},
        title = "{A Simple Model for Quasar Demographics}",
      journal = {\apj},
         year = 2013,
        month = jan,
       volume = {762},
       number = {2},
          eid = {70},
        pages = {70},
          doi = {10.1088/0004-637X/762/2/70},
archivePrefix = {arXiv},
       eprint = {1208.3198},
 primaryClass = {astro-ph.CO},
       adsurl = {https://ui.adsabs.harvard.edu/abs/2013ApJ...762...70C}
}

@ARTICLE{WMC2008,
       author = {{White}, Martin and {Martini}, Paul and {Cohn}, J.~D.},
        title = "{Constraints on the correlation between QSO luminosity and host halo mass from high-redshift quasar clustering}",
      journal = {\mnras},
         year = 2008,
        month = nov,
       volume = {390},
       number = {3},
        pages = {1179-1184},
          doi = {10.1111/j.1365-2966.2008.13817.x},
archivePrefix = {arXiv},
       eprint = {0711.4109},
 primaryClass = {astro-ph},
       adsurl = {https://ui.adsabs.harvard.edu/abs/2008MNRAS.390.1179W}
}

@ARTICLE{HaimanHui2001,
       author = {{Haiman}, Zolt{\'a}n and {Hui}, Lam},
        title = "{Constraining the Lifetime of Quasars from Their Spatial Clustering}",
      journal = {\apj},
         year = 2001,
        month = jan,
       volume = {547},
       number = {1},
        pages = {27-38},
          doi = {10.1086/318330},
archivePrefix = {arXiv},
       eprint = {astro-ph/0002190},
 primaryClass = {astro-ph},
       adsurl = {https://ui.adsabs.harvard.edu/abs/2001ApJ...547...27H}
}

@ARTICLE{Kim2009,
       author = {{Kim}, Soyoung and {Stiavelli}, Massimo and {Trenti}, M. and {Pavlovsky}, C.~M. and {Djorgovski}, S.~G. and {Scarlata}, C. and {Stern}, D. and {Mahabal}, A. and {Thompson}, D. and {Dickinson}, M. and {Panagia}, N. and {Meylan}, G.},
        title = "{The Environments of High-Redshift Quasi-Stellar Objects}",
      journal = {\apj},
         year = 2009,
        month = apr,
       volume = {695},
       number = {2},
        pages = {809-817},
          doi = {10.1088/0004-637X/695/2/809},
archivePrefix = {arXiv},
       eprint = {0805.1412},
 primaryClass = {astro-ph},
       adsurl = {https://ui.adsabs.harvard.edu/abs/2009ApJ...695..809K}
}

@ARTICLE{Morselli2014,
       author = {{Morselli}, L. and {Mignoli}, M. and {Gilli}, R. and {Vignali}, C. and {Comastri}, A. and {Sani}, E. and {Cappelluti}, N. and {Zamorani}, G. and {Brusa}, M. and {Gallozzi}, S. and {Vanzella}, E.},
        title = "{Primordial environment of super massive black holes: large-scale galaxy overdensities around z \raisebox{-0.5ex}\textasciitilde 6 quasars with LBT}",
      journal = {\aap},
         year = 2014,
        month = aug,
       volume = {568},
          eid = {A1},
        pages = {A1},
          doi = {10.1051/0004-6361/201423853},
archivePrefix = {arXiv},
       eprint = {1406.3961},
 primaryClass = {astro-ph.GA},
       adsurl = {https://ui.adsabs.harvard.edu/abs/2014A&A...568A...1M}
}

@ARTICLE{Banados2013,
       author = {{Ba{\~n}ados}, Eduardo and {Venemans}, Bram and {Walter}, Fabian and {Kurk}, Jaron and {Overzier}, Roderik and {Ouchi}, Masami},
        title = "{The Galaxy Environment of a QSO at z \raisebox{-0.5ex}\textasciitilde 5.7}",
      journal = {\apj},
         year = 2013,
        month = aug,
       volume = {773},
       number = {2},
          eid = {178},
        pages = {178},
          doi = {10.1088/0004-637X/773/2/178},
archivePrefix = {arXiv},
       eprint = {1306.6642},
 primaryClass = {astro-ph.CO},
       adsurl = {https://ui.adsabs.harvard.edu/abs/2013ApJ...773..178B}
}

@ARTICLE{Simpson2014,
       author = {{Simpson}, Chris and {Mortlock}, Daniel and {Warren}, Stephen and {Cantalupo}, Sebastiano and {Hewett}, Paul and {McLure}, Ross and {McMahon}, Richard and {Venemans}, Bram},
        title = "{No excess of bright galaxies around the redshift 7.1 quasar ULAS J1120+0641}",
      journal = {\mnras},
         year = 2014,
        month = aug,
       volume = {442},
       number = {4},
        pages = {3454-3461},
          doi = {10.1093/mnras/stu1116},
archivePrefix = {arXiv},
       eprint = {1406.0851},
 primaryClass = {astro-ph.GA},
       adsurl = {https://ui.adsabs.harvard.edu/abs/2014MNRAS.442.3454S}
}

@ARTICLE{Overzier2016,
       author = {{Overzier}, Roderik A.},
        title = "{The realm of the galaxy protoclusters. A review}",
      journal = {\aapr},
         year = 2016,
        month = nov,
       volume = {24},
       number = {1},
          eid = {14},
        pages = {14},
          doi = {10.1007/s00159-016-0100-3},
archivePrefix = {arXiv},
       eprint = {1610.05201},
 primaryClass = {astro-ph.GA},
       adsurl = {https://ui.adsabs.harvard.edu/abs/2016A&ARv..24...14O}
}

@ARTICLE{Sinha2020,
    author = {{Sinha}, Manodeep and {Garrison}, Lehman H.},
    title = "{CORRFUNC - a suite of blazing fast correlation functions on
    the CPU}",
    journal = {\mnras},
    year = "2020",
    month = "Jan",
    volume = {491},
    number = {2},
    pages = {3022-3041},
    doi = {10.1093/mnras/stz3157},
    adsurl =
    {https://ui.adsabs.harvard.edu/abs/2020MNRAS.491.3022S}
}

@ARTICLE{Zheng2005,
       author = {{Zheng}, Zheng and {Berlind}, Andreas A. and {Weinberg}, David H. and {Benson}, Andrew J. and {Baugh}, Carlton M. and {Cole}, Shaun and {Dav{\'e}}, Romeel and {Frenk}, Carlos S. and {Katz}, Neal and {Lacey}, Cedric G.},
        title = "{Theoretical Models of the Halo Occupation Distribution: Separating Central and Satellite Galaxies}",
      journal = {\apj},
         year = 2005,
        month = nov,
       volume = {633},
       number = {2},
        pages = {791-809},
          doi = {10.1086/466510},
archivePrefix = {arXiv},
       eprint = {astro-ph/0408564},
 primaryClass = {astro-ph},
       adsurl = {https://ui.adsabs.harvard.edu/abs/2005ApJ...633..791Z}
}

@ARTICLE{Robertson2010,
       author = {{Robertson}, Brant E.},
        title = "{A Method for Measuring the Bias of High-redshift Galaxies from Cosmic Variance}",
      journal = {\apjl},
         year = 2010,
        month = jun,
       volume = {716},
       number = {2},
        pages = {L229-L234},
          doi = {10.1088/2041-8205/716/2/L229},
archivePrefix = {arXiv},
       eprint = {1005.4927},
 primaryClass = {astro-ph.CO},
       adsurl = {https://ui.adsabs.harvard.edu/abs/2010ApJ...716L.229R}
}

@software{photutils,
  author       = {Larry Bradley and
                  Brigitta Sipőcz and
                  Thomas Robitaille and
                  Erik Tollerud and
                  Zé Vinícius and
                  Christoph Deil and
                  Kyle Barbary and
                  Tom J Wilson and
                  Ivo Busko and
                  Axel Donath and
                  Hans Moritz Günther and
                  Mihai Cara and
                  P. L. Lim and
                  Sebastian Meßlinger and
                  Simon Conseil and
                  Azalee Bostroem and
                  Michael Droettboom and
                  E. M. Bray and
                  Lars Andersen Bratholm and
                  Geert Barentsen and
                  Matt Craig and
                  Shivangee Rathi and
                  Sergio Pascual and
                  Gabriel Perren and
                  Iskren Y. Georgiev and
                  Miguel de Val-Borro and
                  Wolfgang Kerzendorf and
                  Yoonsoo P. Bach and
                  Bruno Quint and
                  Harrison Souchereau},
  title        = {astropy/photutils: 1.5.0},
  month        = jul,
  year         = 2022,
  publisher    = {Zenodo},
  version      = {1.5.0},
  doi          = {10.5281/zenodo.6825092},
  url          = {https://doi.org/10.5281/zenodo.6825092}
}

@ARTICLE{Pizzati2025,
       author = {{Pizzati}, Elia and {Hennawi}, Joseph F. and {Schaye}, Joop and {Eilers}, Anna-Christina and {Huang}, Jiamu and {Schindler}, Jan-Torge and {Wang}, Feige},
        title = "{'Little red dots' cannot reside in the same dark matter haloes as comparably luminous unobscured quasars}",
      journal = {\mnras},
         year = 2025,
        month = jun,
       volume = {539},
       number = {4},
        pages = {2910-2925},
          doi = {10.1093/mnras/staf660},
       adsurl = {https://ui.adsabs.harvard.edu/abs/2025MNRAS.539.2910P}
}

@ARTICLE{Trakhtenbrot2017,
       author = {{Trakhtenbrot}, Benny and {Volonteri}, Marta and {Natarajan}, Priyamvada},
        title = "{On the Accretion Rates and Radiative Efficiencies of the Highest-redshift Quasars}",
      journal = {\apjl},
         year = 2017,
        month = feb,
       volume = {836},
       number = {1},
          eid = {L1},
        pages = {L1},
          doi = {10.3847/2041-8213/836/1/L1},
archivePrefix = {arXiv},
       eprint = {1611.00772},
 primaryClass = {astro-ph.GA},
       adsurl = {https://ui.adsabs.harvard.edu/abs/2017ApJ...836L...1T}
}

@ARTICLE{Davies2019,
       author = {{Davies}, Frederick B. and {Hennawi}, Joseph F. and {Eilers}, Anna-Christina},
        title = "{Evidence for Low Radiative Efficiency or Highly Obscured Growth of z > 7 Quasars}",
      journal = {\apjl},
         year = 2019,
        month = oct,
       volume = {884},
       number = {1},
          eid = {L19},
        pages = {L19},
          doi = {10.3847/2041-8213/ab42e3},
archivePrefix = {arXiv},
       eprint = {1906.10130},
 primaryClass = {astro-ph.GA},
       adsurl = {https://ui.adsabs.harvard.edu/abs/2019ApJ...884L..19D}
}

@ARTICLE{Volonteri2006,
       author = {{Volonteri}, Marta and {Rees}, Martin J.},
        title = "{Quasars at z=6: The Survival of the Fittest}",
      journal = {\apj},
         year = 2006,
        month = oct,
       volume = {650},
       number = {2},
        pages = {669-678},
          doi = {10.1086/507444},
archivePrefix = {arXiv},
       eprint = {astro-ph/0607093},
 primaryClass = {astro-ph},
       adsurl = {https://ui.adsabs.harvard.edu/abs/2006ApJ...650..669V}
}

@ARTICLE{Davis1983,
       author = {{Davis}, M. and {Peebles}, P.~J.~E.},
        title = "{A survey of galaxy redshifts. V. The two-point position and velocity correlations.}",
      journal = {\apj},
         year = 1983,
        month = apr,
       volume = {267},
        pages = {465-482},
          doi = {10.1086/160884},
       adsurl = {https://ui.adsabs.harvard.edu/abs/1983ApJ...267..465D}
}

@ARTICLE{Durovcikova2024,
       author = {{{\v{D}}urov{\v{c}}{\'\i}kov{\'a}}, Dominika and {Eilers}, Anna-Christina and {Chen}, Huanqing and {Satyavolu}, Sindhu and {Kulkarni}, Girish and {Simcoe}, Robert A. and {Keating}, Laura C. and {Haehnelt}, Martin G. and {Ba{\~n}ados}, Eduardo},
        title = "{Chronicling the Reionization History at 6 {\ensuremath{\lesssim}} z {\ensuremath{\lesssim}} 7 with Emergent Quasar Damping Wings}",
      journal = {\apj},
         year = 2024,
        month = jul,
       volume = {969},
       number = {2},
          eid = {162},
        pages = {162},
          doi = {10.3847/1538-4357/ad4888},
archivePrefix = {arXiv},
       eprint = {2401.10328},
 primaryClass = {astro-ph.CO},
       adsurl = {https://ui.adsabs.harvard.edu/abs/2024ApJ...969..162D}
}

@ARTICLE{Wang2020,
       author = {{Wang}, Feige and {Davies}, Frederick B. and {Yang}, Jinyi and {Hennawi}, Joseph F. and {Fan}, Xiaohui and {Barth}, Aaron J. and {Jiang}, Linhua and {Wu}, Xue-Bing and {Mudd}, Dale M. and {Ba{\~n}ados}, Eduardo and {Bian}, Fuyan and {Decarli}, Roberto and {Eilers}, Anna-Christina and {Farina}, Emanuele Paolo and {Venemans}, Bram and {Walter}, Fabian and {Yue}, Minghao},
        title = "{A Significantly Neutral Intergalactic Medium Around the Luminous z = 7 Quasar J0252-0503}",
      journal = {\apj},
         year = 2020,
        month = jun,
       volume = {896},
       number = {1},
          eid = {23},
        pages = {23},
          doi = {10.3847/1538-4357/ab8c45},
archivePrefix = {arXiv},
       eprint = {2004.10877},
 primaryClass = {astro-ph.GA},
       adsurl = {https://ui.adsabs.harvard.edu/abs/2020ApJ...896...23W}
}

@ARTICLE{White2012,
       author = {{White}, Martin and {Myers}, Adam D. and {Ross}, Nicholas P. and {Schlegel}, David J. and {Hennawi}, Joseph F. and {Shen}, Yue and {McGreer}, Ian and {Strauss}, Michael A. and {Bolton}, Adam S. and {Bovy}, Jo and {Fan}, X. and {Miralda-Escude}, Jordi and {Palanque-Delabrouille}, N. and {Paris}, I. and {Petitjean}, P. and {Schneider}, D.~P. and {Viel}, M. and {Weinberg}, David H. and {Yeche}, Ch. and {Zehavi}, I. and {Pan}, K. and {Snedden}, S. and {Bizyaev}, D. and {Brewington}, H. and {Brinkmann}, J. and {Malanushenko}, V. and {Malanushenko}, E. and {Oravetz}, D. and {Simmons}, A. and {Sheldon}, A. and {Weaver}, Benjamin A.},
        title = "{The clustering of intermediate-redshift quasars as measured by the Baryon Oscillation Spectroscopic Survey}",
      journal = {\mnras},
         year = 2012,
        month = aug,
       volume = {424},
       number = {2},
        pages = {933-950},
          doi = {10.1111/j.1365-2966.2012.21251.x},
archivePrefix = {arXiv},
       eprint = {1203.5306},
 primaryClass = {astro-ph.CO},
       adsurl = {https://ui.adsabs.harvard.edu/abs/2012MNRAS.424..933W}
}

@ARTICLE{Eftekharzadeh2015,
       author = {{Eftekharzadeh}, Sarah and {Myers}, Adam D. and {White}, Martin and {Weinberg}, David H. and {Schneider}, Donald P. and {Shen}, Yue and {Font-Ribera}, Andreu and {Ross}, Nicholas P. and {Paris}, Isabelle and {Streblyanska}, Alina},
        title = "{Clustering of intermediate redshift quasars using the final SDSS III-BOSS sample}",
      journal = {\mnras},
         year = 2015,
        month = nov,
       volume = {453},
       number = {3},
        pages = {2779-2798},
          doi = {10.1093/mnras/stv1763},
archivePrefix = {arXiv},
       eprint = {1507.08380},
 primaryClass = {astro-ph.CO},
       adsurl = {https://ui.adsabs.harvard.edu/abs/2015MNRAS.453.2779E}
}

@ARTICLE{Laurent2017,
       author = {{Laurent}, Pierre and {Eftekharzadeh}, Sarah and {Le Goff}, Jean-Marc and {Myers}, Adam and {Burtin}, Etienne and {White}, Martin and {Ross}, Ashley J. and {Tinker}, Jeremy and {Tojeiro}, Rita and {Bautista}, Julian and {Brinkmann}, Jonathan and {Comparat}, Johan and {Dawson}, Kyle and {du Mas des Bourboux}, H{\'e}lion and {Kneib}, Jean-Paul and {McGreer}, Ian D. and {Palanque-Delabrouille}, Nathalie and {Percival}, Will J. and {Prada}, Francisco and {Rossi}, Graziano and {Schneider}, Donald P. and {Weinberg}, David and {Y{\`e}che}, Christophe and {Zarrouk}, Pauline and {Zhao}, Gong-Bo},
        title = "{Clustering of quasars in SDSS-IV eBOSS: study of potential systematics and bias determination}",
      journal = {\jcap},
         year = 2017,
        month = jul,
       volume = {2017},
       number = {7},
          eid = {017},
        pages = {017},
          doi = {10.1088/1475-7516/2017/07/017},
archivePrefix = {arXiv},
       eprint = {1705.04718},
 primaryClass = {astro-ph.CO},
       adsurl = {https://ui.adsabs.harvard.edu/abs/2017JCAP...07..017L}
}

@ARTICLE{Jose2016,
       author = {{Jose}, Charles and {Lacey}, Cedric G. and {Baugh}, Carlton M.},
        title = "{The clustering of dark matter haloes: scale-dependent bias on quasi-linear scales}",
      journal = {\mnras},
         year = 2016,
        month = nov,
       volume = {463},
       number = {1},
        pages = {270-281},
          doi = {10.1093/mnras/stw1702},
archivePrefix = {arXiv},
       eprint = {1509.06715},
 primaryClass = {astro-ph.CO},
       adsurl = {https://ui.adsabs.harvard.edu/abs/2016MNRAS.463..270J}
}

@ARTICLE{Shuntov2025,
       author = {{Shuntov}, Marko and {Oesch}, Pascal A. and {Toft}, Sune and {Meyer}, Romain A. and {Covelo-Paz}, Alba and {Paquereau}, Louise and {Bouwens}, Rychard and {Brammer}, Gabriel and {Gelli}, Viola and {Giovinazzo}, Emma and {Herard-Demanche}, Thomas and {Illingworth}, Garth D. and {Mason}, Charlotte and {Naidu}, Rohan P. and {Weibel}, Andrea and {Xiao}, Mengyuan},
        title = "{Constraints on the early Universe star formation efficiency from galaxy clustering and halo modeling of H{\ensuremath{\alpha}} and [O III] emitters}",
      journal = {\aap},
         year = 2025,
        month = jul,
       volume = {699},
          eid = {A231},
        pages = {A231},
          doi = {10.1051/0004-6361/202554618},
archivePrefix = {arXiv},
       eprint = {2503.14280},
 primaryClass = {astro-ph.GA},
       adsurl = {https://ui.adsabs.harvard.edu/abs/2025A&A...699A.231S}
}

@ARTICLE{Pudoka2024,
       author = {{Pudoka}, Maria and {Wang}, Feige and {Fan}, Xiaohui and {Yang}, Jinyi and {Champagne}, Jaclyn and {Jones}, Victoria and {Bian}, Fuyan and {Cai}, Zheng and {Jiang}, Linhua and {Liu}, Dezi and {Wu}, Xue-Bing},
        title = "{Large-scale Overdensity of Lyman Break Galaxies around the z = 6.3 Ultraluminous Quasar J0100 + 2802}",
      journal = {\apj},
         year = 2024,
        month = jun,
       volume = {968},
       number = {2},
          eid = {118},
        pages = {118},
          doi = {10.3847/1538-4357/ad488a},
archivePrefix = {arXiv},
       eprint = {2405.03781},
 primaryClass = {astro-ph.GA},
       adsurl = {https://ui.adsabs.harvard.edu/abs/2024ApJ...968..118P}
}

@ARTICLE{Han2012,
       author = {{Han}, Jiaxin and {Jing}, Y.~P. and {Wang}, Huiyuan and {Wang}, Wenting},
        title = "{Resolving subhaloes' lives with the Hierarchical Bound-Tracing algorithm}",
      journal = {\mnras},
         year = 2012,
        month = dec,
       volume = {427},
       number = {3},
        pages = {2437-2449},
          doi = {10.1111/j.1365-2966.2012.22111.x},
archivePrefix = {arXiv},
       eprint = {1103.2099},
 primaryClass = {astro-ph.CO},
       adsurl = {https://ui.adsabs.harvard.edu/abs/2012MNRAS.427.2437H}
}

@ARTICLE{Han2018,
       author = {{Han}, Jiaxin and {Cole}, Shaun and {Frenk}, Carlos S. and {Benitez-Llambay}, Alejandro and {Helly}, John},
        title = "{HBT+: an improved code for finding subhaloes and building merger trees in cosmological simulations}",
      journal = {\mnras},
         year = 2018,
        month = feb,
       volume = {474},
       number = {1},
        pages = {604-617},
          doi = {10.1093/mnras/stx2792},
archivePrefix = {arXiv},
       eprint = {1708.03646},
 primaryClass = {astro-ph.CO},
       adsurl = {https://ui.adsabs.harvard.edu/abs/2018MNRAS.474..604H}
}

@ARTICLE{Croom2001,
       author = {{Croom}, Scott M. and {Shanks}, T. and {Boyle}, B.~J. and {Smith}, R.~J. and {Miller}, L. and {Loaring}, N.~S. and {Hoyle}, F.},
        title = "{The 2dF QSO Redshift Survey - II. Structure and evolution at high redshift}",
      journal = {\mnras},
         year = 2001,
        month = aug,
       volume = {325},
       number = {2},
        pages = {483-496},
          doi = {10.1046/j.1365-8711.2001.04389.x},
archivePrefix = {arXiv},
       eprint = {astro-ph/0012375},
 primaryClass = {astro-ph},
       adsurl = {https://ui.adsabs.harvard.edu/abs/2001MNRAS.325..483C}
}

@ARTICLE{Findlay2025,
       author = {{Findlay}, N. and {Nadathur}, S. and {Percival}, W.~J. and {de Mattia}, A. and {Zarrouk}, P. and {Gil-Mar{\'\i}n}, H. and {Alves}, O. and {Mena-Fern{\'a}ndez}, J. and {Garcia-Quintero}, C. and {Rocher}, A. and {Ahlen}, S. and {Bianchi}, D. and {Brooks}, D. and {Claybaugh}, T. and {Cole}, S. and {de la Macorra}, A. and {Dey}, A. and {Doel}, P. and {Fanning}, K. and {Font-Ribera}, A. and {Forero-Romero}, J.~E. and {Gazta{\~n}aga}, E. and {Gutierrez}, G. and {Hahn}, C. and {Honscheid}, K. and {Howlett}, C. and {Juneau}, S. and {Levi}, M.~E. and {Meisner}, A. and {Miquel}, R. and {Moustakas}, J. and {Palanque-Delabrouille}, N. and {P{\'e}rez-R{\`a}fols}, I. and {Rossi}, G. and {Sanchez}, E. and {Schlegel}, D. and {Schubnell}, M. and {Seo}, H. and {Sprayberry}, D. and {Tarl{\'e}}, G. and {Vargas-Maga{\~n}a}, M. and {Weaver}, B.~A.},
        title = "{Exploring HOD-dependent systematics for the DESI 2024 Full-Shape galaxy clustering analysis}",
      journal = {\jcap},
         year = 2025,
        month = sep,
       volume = {2025},
       number = {9},
          eid = {007},
        pages = {007},
          doi = {10.1088/1475-7516/2025/09/007},
archivePrefix = {arXiv},
       eprint = {2411.12023},
 primaryClass = {astro-ph.CO},
       adsurl = {https://ui.adsabs.harvard.edu/abs/2025JCAP...09..007F}
}

@ARTICLE{Berlind&Weinberg2002,
       author = {{Berlind}, Andreas A. and {Weinberg}, David H.},
        title = "{The Halo Occupation Distribution: Toward an Empirical Determination of the Relation between Galaxies and Mass}",
      journal = {\apj},
         year = 2002,
        month = aug,
       volume = {575},
       number = {2},
        pages = {587-616},
          doi = {10.1086/341469},
archivePrefix = {arXiv},
       eprint = {astro-ph/0109001},
 primaryClass = {astro-ph},
       adsurl = {https://ui.adsabs.harvard.edu/abs/2002ApJ...575..587B}
}

@ARTICLE{Arita2023,
       author = {{Arita}, Junya and {Kashikawa}, Nobunari and {Matsuoka}, Yoshiki and {He}, Wanqiu and {Ito}, Kei and {Liang}, Yongming and {Ishimoto}, Rikako and {Yoshioka}, Takehiro and {Takeda}, Yoshihiro and {Iwasawa}, Kazushi and {Onoue}, Masafusa and {Toba}, Yoshiki and {Imanishi}, Masatoshi},
        title = "{Subaru High-z Exploration of Low-luminosity Quasars (SHELLQs). XVIII. The Dark Matter Halo Mass of Quasars at z   6}",
      journal = {\apj},
         year = 2023,
        month = sep,
       volume = {954},
       number = {2},
          eid = {210},
        pages = {210},
          doi = {10.3847/1538-4357/ace43a},
archivePrefix = {arXiv},
       eprint = {2307.02531},
 primaryClass = {astro-ph.GA},
       adsurl = {https://ui.adsabs.harvard.edu/abs/2023ApJ...954..210A}
}

@ARTICLE{Oesch2023,
       author = {{Oesch}, P.~A. and {Brammer}, G. and {Naidu}, R.~P. and {Bouwens}, R.~J. and {Chisholm}, J. and {Illingworth}, G.~D. and {Matthee}, J. and {Nelson}, E. and {Qin}, Y. and {Reddy}, N. and {Shapley}, A. and {Shivaei}, I. and {van Dokkum}, P. and {Weibel}, A. and {Whitaker}, K. and {Wuyts}, S. and {Covelo-Paz}, A. and {Endsley}, R. and {Fudamoto}, Y. and {Giovinazzo}, E. and {Herard-Demanche}, T. and {Kerutt}, J. and {Kramarenko}, I. and {Labbe}, I. and {Leonova}, E. and {Lin}, J. and {Magee}, D. and {Marchesini}, D. and {Maseda}, M. and {Mason}, C. and {Matharu}, J. and {Meyer}, R.~A. and {Neufeld}, C. and {Prieto Lyon}, G. and {Schaerer}, D. and {Sharma}, R. and {Shuntov}, M. and {Smit}, R. and {Stefanon}, M. and {Wyithe}, J.~S.~B. and {Xiao}, M.},
        title = "{The JWST FRESCO survey: legacy NIRCam/grism spectroscopy and imaging in the two GOODS fields}",
      journal = {\mnras},
         year = 2023,
        month = oct,
       volume = {525},
       number = {2},
        pages = {2864-2874},
          doi = {10.1093/mnras/stad2411},
archivePrefix = {arXiv},
       eprint = {2304.02026},
 primaryClass = {astro-ph.GA},
       adsurl = {https://ui.adsabs.harvard.edu/abs/2023MNRAS.525.2864O}
}

@ARTICLE{Bosman2025,
       author = {{Bosman}, Sarah E.~I. and {{\'A}lvarez-M{\'a}rquez}, Javier and {Davies}, Frederick B. and {Protu{\v{s}}ov{\'a}}, Klaudia and {Hennawi}, Joseph F. and {Yang}, Jinyi and {Spina}, Benedetta and {Colina}, Luis and {Fan}, Xiaohui and {{\"O}stlin}, G{\"o}ran and {Walter}, Fabian and {Wang}, Feige and {Ward}, Martin and {Alonso Herrero}, Almudena and {Barth}, Aaron J. and {Belladitta}, Silvia and {Boogaard}, Leindert and {Caputi}, Karina I. and {Connor}, Thomas and {{\v{D}}urov{\v{c}}{\'\i}kov{\'a}}, Dominika and {Eilers}, Anna-Christina and {Crespo G{\'o}mez}, Alejandro and {Hjorth}, Jens and {Jun}, Hyunsung D. and {Langeroodi}, Danial and {Liu}, Weizhe and {Lupi}, Alessandro and {Mazzucchelli}, Chiara and {Pye}, John P. and {Rinaldi}, Pierluigi and {van der Werf}, Paul and {Volonteri}, Marta},
        title = "{A close look at the black hole masses and hot dusty toruses of the first quasars with MIRI-MRS}",
      journal = {arXiv e-prints},
         year = 2025,
        month = nov,
          eid = {arXiv:2511.02902},
        pages = {arXiv:2511.02902},
          doi = {10.48550/arXiv.2511.02902},
archivePrefix = {arXiv},
       eprint = {2511.02902},
 primaryClass = {astro-ph.GA},
       adsurl = {https://ui.adsabs.harvard.edu/abs/2025arXiv251102902B}
}

@ARTICLE{Lin2025,
       author = {{Lin}, Xiaojing and {Fan}, Xiaohui and {Sun}, Fengwu and {Zhang}, Junyu and {Egami}, Eiichi and {Helton}, Jakob M. and {Wang}, Feige and {Zhang}, Haowen and {Bunker}, Andrew J. and {Cai}, Zheng and {Ji}, Zhiyuan and {Jin}, Xiangyu and {Maiolino}, Roberto and {Pudoka}, Maria Anne and {Rinaldi}, Pierluigi and {Robertson}, Brant and {Tacchella}, Sandro and {Tee}, Wei Leong and {Sun}, Yang and {Willmer}, Christopher N.~A. and {Willott}, Chris and {Zhu}, Yongda},
        title = "{The Large-scale Environments of Low-luminosity AGNs at $3.9 < z < 6$ and Implications for Their Host Dark Matter Halos from a Complete NIRCam Grism Redshift Survey}",
      journal = {arXiv e-prints},
         year = 2025,
        month = may,
          eid = {arXiv:2505.02896},
        pages = {arXiv:2505.02896},
          doi = {10.48550/arXiv.2505.02896},
archivePrefix = {arXiv},
       eprint = {2505.02896},
 primaryClass = {astro-ph.GA},
       adsurl = {https://ui.adsabs.harvard.edu/abs/2025arXiv250502896L}
}

@ARTICLE{Allevato2014,
       author = {{Allevato}, V. and {Finoguenov}, A. and {Civano}, F. and {Cappelluti}, N. and {Shankar}, F. and {Miyaji}, T. and {Hasinger}, G. and {Gilli}, R. and {Zamorani}, G. and {Lanzuisi}, G. and {Salvato}, M. and {Elvis}, M. and {Comastri}, A. and {Silverman}, J.},
        title = "{Clustering of Moderate Luminosity X-Ray-selected Type 1 and Type 2 AGNS at Z \raisebox{-0.5ex}\textasciitilde 3}",
      journal = {\apj},
         year = 2014,
        month = nov,
       volume = {796},
       number = {1},
          eid = {4},
        pages = {4},
          doi = {10.1088/0004-637X/796/1/4},
archivePrefix = {arXiv},
       eprint = {1409.7693},
 primaryClass = {astro-ph.GA},
       adsurl = {https://ui.adsabs.harvard.edu/abs/2014ApJ...796....4A}
}

@ARTICLE{Shen2009,
       author = {{Shen}, Yue and {Strauss}, Michael A. and {Ross}, Nicholas P. and {Hall}, Patrick B. and {Lin}, Yen-Ting and {Richards}, Gordon T. and {Schneider}, Donald P. and {Weinberg}, David H. and {Connolly}, Andrew J. and {Fan}, Xiaohui and {Hennawi}, Joseph F. and {Shankar}, Francesco and {Vanden Berk}, Daniel E. and {Bahcall}, Neta A. and {Brunner}, Robert J.},
        title = "{Quasar Clustering from SDSS DR5: Dependences on Physical Properties}",
      journal = {\apj},
         year = 2009,
        month = jun,
       volume = {697},
       number = {2},
        pages = {1656-1673},
          doi = {10.1088/0004-637X/697/2/1656},
archivePrefix = {arXiv},
       eprint = {0810.4144},
 primaryClass = {astro-ph},
       adsurl = {https://ui.adsabs.harvard.edu/abs/2009ApJ...697.1656S}
}

@ARTICLE{Krolewski2015,
       author = {{Krolewski}, Alex G. and {Eisenstein}, Daniel J.},
        title = "{Measuring the Luminosity and Virial Black Hole Mass Dependence of Quasar-Galaxy Clustering At z {\ensuremath{\sim}} 0.8}",
      journal = {\apj},
         year = 2015,
        month = apr,
       volume = {803},
       number = {1},
          eid = {4},
        pages = {4},
          doi = {10.1088/0004-637X/803/1/4},
archivePrefix = {arXiv},
       eprint = {1501.03898},
 primaryClass = {astro-ph.GA},
       adsurl = {https://ui.adsabs.harvard.edu/abs/2015ApJ...803....4K}
}

@ARTICLE{Lidz2006,
       author = {{Lidz}, Adam and {Hopkins}, Philip F. and {Cox}, Thomas J. and {Hernquist}, Lars and {Robertson}, Brant},
        title = "{The Luminosity Dependence of Quasar Clustering}",
      journal = {\apj},
         year = 2006,
        month = apr,
       volume = {641},
       number = {1},
        pages = {41-49},
          doi = {10.1086/500444},
archivePrefix = {arXiv},
       eprint = {astro-ph/0507361},
 primaryClass = {astro-ph},
       adsurl = {https://ui.adsabs.harvard.edu/abs/2006ApJ...641...41L}
}

@ARTICLE{Mortlock2011,
       author = {{Mortlock}, Daniel J. and {Warren}, Stephen J. and {Venemans}, Bram P. and {Patel}, Mitesh and {Hewett}, Paul C. and {McMahon}, Richard G. and {Simpson}, Chris and {Theuns}, Tom and {Gonz{\'a}les-Solares}, Eduardo A. and {Adamson}, Andy and {Dye}, Simon and {Hambly}, Nigel C. and {Hirst}, Paul and {Irwin}, Mike J. and {Kuiper}, Ernst and {Lawrence}, Andy and {R{\"o}ttgering}, Huub J.~A.},
        title = "{A luminous quasar at a redshift of z = 7.085}",
      journal = {\nat},
         year = 2011,
        month = jun,
       volume = {474},
       number = {7353},
        pages = {616-619},
          doi = {10.1038/nature10159},
archivePrefix = {arXiv},
       eprint = {1106.6088},
 primaryClass = {astro-ph.CO},
       adsurl = {https://ui.adsabs.harvard.edu/abs/2011Natur.474..616M}
}

@ARTICLE{Hopkins2005,
       author = {{Hopkins}, Philip F. and {Hernquist}, Lars and {Cox}, Thomas J. and {Di Matteo}, Tiziana and {Robertson}, Brant and {Springel}, Volker},
        title = "{Luminosity-dependent Quasar Lifetimes: A New Interpretation of the Quasar Luminosity Function}",
      journal = {\apj},
         year = 2005,
        month = sep,
       volume = {630},
       number = {2},
        pages = {716-720},
          doi = {10.1086/432463},
archivePrefix = {arXiv},
       eprint = {astro-ph/0504252},
 primaryClass = {astro-ph},
       adsurl = {https://ui.adsabs.harvard.edu/abs/2005ApJ...630..716H}
}

@ARTICLE{Champagne2025,
       author = {{Champagne}, Jaclyn B. and {Wang}, Feige and {Zhang}, Haowen and {Yang}, Jinyi and {Fan}, Xiaohui and {Hennawi}, Joseph F. and {Sun}, Fengwu and {Ba{\~n}ados}, Eduardo and {Bosman}, Sarah E.~I. and {Costa}, Tiago and {Eilers}, Anna-Christina and {Endsley}, Ryan and {Jin}, Xiangyu and {Jun}, Hyunsung D. and {Li}, Mingyu and {Lin}, Xiaojing and {Liu}, Weizhe and {Loiacono}, Federica and {Lupi}, Alessandro and {Mazzucchelli}, Chiara and {Pudoka}, Maria and {Protu{\v{s}}ov{\`a}}, Klaudia and {Rojas-Ruiz}, Sof{\'\i}a and {Tee}, Wei Leong and {Trebitsch}, Maxime and {Venemans}, Bram P. and {Zhuang}, Ming-Yang and {Zou}, Siwei},
        title = "{A Quasar-anchored Protocluster at z = 6.6 in the ASPIRE Survey. I. Properties of [O III] Emitters in a 10 Mpc Overdensity Structure}",
      journal = {\apj},
         year = 2025,
        month = mar,
       volume = {981},
       number = {2},
          eid = {113},
        pages = {113},
          doi = {10.3847/1538-4357/adb1bd},
archivePrefix = {arXiv},
       eprint = {2410.03826},
 primaryClass = {astro-ph.GA},
       adsurl = {https://ui.adsabs.harvard.edu/abs/2025ApJ...981..113C}
}

@ARTICLE{Lin2025_lf,
       author = {{Lin}, Xiaojing and {Egami}, Eiichi and {Sun}, Fengwu and {Zhang}, Haowen and {Fan}, Xiaohui and {Helton}, Jakob M. and {Wang}, Feige and {Bunker}, Andrew J. and {Cai}, Zheng and {Eisenstein}, Daniel J. and {Jaffe}, Daniel T. and {Ji}, Zhiyuan and {Jin}, Xiangyu and {Pudoka}, Maria Anne and {Tacchella}, Sandro and {Tee}, Wei Leong and {Rinaldi}, Pierluigi and {Robertson}, Brant and {Sun}, Yang and {Willmer}, Christopher N.~A. and {Willott}, Chris and {Zhang}, Junyu and {Zhu}, Yongda},
        title = "{The Luminosity Function and Clustering of H$α$ Emitting Galaxies at $z\approx4-6$ from a Complete NIRCam Grism Redshift Survey}",
      journal = {arXiv e-prints},
         year = 2025,
        month = apr,
          eid = {arXiv:2504.08028},
        pages = {arXiv:2504.08028},
          doi = {10.48550/arXiv.2504.08028},
archivePrefix = {arXiv},
       eprint = {2504.08028},
 primaryClass = {astro-ph.GA},
       adsurl = {https://ui.adsabs.harvard.edu/abs/2025arXiv250408028L}
}

@ARTICLE{Schindler2025a,
       author = {{Schindler}, Jan-Torge and {Hennawi}, Joseph F. and {Davies}, Frederick B. and {Bosman}, Sarah E.~I. and {Endsley}, Ryan and {Wang}, Feige and {Yang}, Jinyi and {Barth}, Aaron J. and {Eilers}, Anna-Christina and {Fan}, Xiaohui and {Kakiichi}, Koki and {Maseda}, Michael and {Pizzati}, Elia and {Nanni}, Riccardo},
        title = "{A little red dot at z = 7.3 within a large galaxy overdensity}",
      journal = {Nature Astronomy},
         year = 2025,
        month = sep,
          doi = {10.1038/s41550-025-02660-1},
archivePrefix = {arXiv},
       eprint = {2411.11534},
 primaryClass = {astro-ph.GA},
       adsurl = {https://ui.adsabs.harvard.edu/abs/2025NatAs.tmp..191S}
}

@ARTICLE{Willott2005,
       author = {{Willott}, Chris J. and {Percival}, Will J. and {McLure}, Ross J. and {Crampton}, David and {Hutchings}, John B. and {Jarvis}, Matt J. and {Sawicki}, Marcin and {Simard}, Luc},
        title = "{Imaging of SDSS z > 6 Quasar Fields: Gravitational Lensing, Companion Galaxies, and the Host Dark Matter Halos}",
      journal = {\apj},
         year = 2005,
        month = jun,
       volume = {626},
       number = {2},
        pages = {657-665},
          doi = {10.1086/430168},
archivePrefix = {arXiv},
       eprint = {astro-ph/0503202},
 primaryClass = {astro-ph},
       adsurl = {https://ui.adsabs.harvard.edu/abs/2005ApJ...626..657W}
}

@ARTICLE{Zheng2006,
       author = {{Zheng}, W. and {Overzier}, R.~A. and {Bouwens}, R.~J. and {White}, R.~L. and {Ford}, H.~C. and {Ben{\'\i}tez}, N. and {Blakeslee}, J.~P. and {Bradley}, L.~D. and {Jee}, M.~J. and {Martel}, A.~R. and {Mei}, S. and {Zirm}, A.~W. and {Illingworth}, G.~D. and {Clampin}, M. and {Hartig}, G.~F. and {Ardila}, D.~R. and {Bartko}, F. and {Broadhurst}, T.~J. and {Brown}, R.~A. and {Burrows}, C.~J. and {Cheng}, E.~S. and {Cross}, N.~J.~G. and {Demarco}, R. and {Feldman}, P.~D. and {Franx}, M. and {Golimowski}, D.~A. and {Goto}, T. and {Gronwall}, C. and {Holden}, B. and {Homeier}, N. and {Infante}, L. and {Kimble}, R.~A. and {Krist}, J.~E. and {Lesser}, M.~P. and {Menanteau}, F. and {Meurer}, G.~R. and {Miley}, G.~K. and {Motta}, V. and {Postman}, M. and {Rosati}, P. and {Sirianni}, M. and {Sparks}, W.~B. and {Tran}, H.~D. and {Tsvetanov}, Z.~I.},
        title = "{An Overdensity of Galaxies near the Most Distant Radio-loud Quasar}",
      journal = {\apj},
         year = 2006,
        month = apr,
       volume = {640},
       number = {2},
        pages = {574-578},
          doi = {10.1086/500167},
archivePrefix = {arXiv},
       eprint = {astro-ph/0511734},
 primaryClass = {astro-ph},
       adsurl = {https://ui.adsabs.harvard.edu/abs/2006ApJ...640..574Z}
}

@ARTICLE{Bosman2020,
       author = {{Bosman}, Sarah E.~I. and {Kakiichi}, Koki and {Meyer}, Romain A. and {Gronke}, Max and {Laporte}, Nicolas and {Ellis}, Richard S.},
        title = "{Three Ly{\ensuremath{\alpha}} Emitting Galaxies within a Quasar Proximity Zone at z {\ensuremath{\sim}} 5.8}",
      journal = {\apj},
         year = 2020,
        month = jun,
       volume = {896},
       number = {1},
          eid = {49},
        pages = {49},
          doi = {10.3847/1538-4357/ab85cd},
archivePrefix = {arXiv},
       eprint = {1912.11486},
 primaryClass = {astro-ph.GA},
       adsurl = {https://ui.adsabs.harvard.edu/abs/2020ApJ...896...49B}
}

@ARTICLE{Mignoli2020,
       author = {{Mignoli}, Marco and {Gilli}, Roberto and {Decarli}, Roberto and {Vanzella}, Eros and {Balmaverde}, Barbara and {Cappelluti}, Nico and {Cassar{\`a}}, Letizia P. and {Comastri}, Andrea and {Cusano}, Felice and {Iwasawa}, Kazushi and {Marchesi}, Stefano and {Prandoni}, Isabella and {Vignali}, Cristian and {Vito}, Fabio and {Zamorani}, Giovanni and {Chiaberge}, Marco and {Norman}, Colin},
        title = "{Web of the giant: Spectroscopic confirmation of a large-scale structure around the z = 6.31 quasar SDSS J1030+0524}",
      journal = {\aap},
         year = 2020,
        month = oct,
       volume = {642},
          eid = {L1},
        pages = {L1},
          doi = {10.1051/0004-6361/202039045},
archivePrefix = {arXiv},
       eprint = {2009.00024},
 primaryClass = {astro-ph.GA},
       adsurl = {https://ui.adsabs.harvard.edu/abs/2020A&A...642L...1M}
}

@ARTICLE{Meyer2022,
       author = {{Meyer}, Romain A. and {Decarli}, Roberto and {Walter}, Fabian and {Li}, Qiong and {Wang}, Ran and {Mazzucchelli}, Chiara and {Ba{\~n}ados}, Eduardo and {Farina}, Emanuele P. and {Venemans}, Bram},
        title = "{Constraining Galaxy Overdensities around Three z 6.5 Quasars with ALMA and MUSE}",
      journal = {\apj},
         year = 2022,
        month = mar,
       volume = {927},
       number = {2},
          eid = {141},
        pages = {141},
          doi = {10.3847/1538-4357/ac4f67},
archivePrefix = {arXiv},
       eprint = {2201.09720},
 primaryClass = {astro-ph.GA},
       adsurl = {https://ui.adsabs.harvard.edu/abs/2022ApJ...927..141M}
}

@ARTICLE{Garcia-Vergara2022,
       author = {{Garc{\'\i}a-Vergara}, Cristina and {Rybak}, Matus and {Hodge}, Jacqueline and {Hennawi}, Joseph F. and {Decarli}, Roberto and {Gonz{\'a}lez-L{\'o}pez}, Jorge and {Arrigoni-Battaia}, Fabrizio and {Aravena}, Manuel and {Farina}, Emanuele P.},
        title = "{ALMA Reveals a Large Overdensity and Strong Clustering of Galaxies in Quasar Environments at z 4}",
      journal = {\apj},
         year = 2022,
        month = mar,
       volume = {927},
       number = {1},
          eid = {65},
        pages = {65},
          doi = {10.3847/1538-4357/ac469d},
archivePrefix = {arXiv},
       eprint = {2109.09754},
 primaryClass = {astro-ph.GA},
       adsurl = {https://ui.adsabs.harvard.edu/abs/2022ApJ...927...65G}
}

@ARTICLE{Tinker2010,
       author = {{Tinker}, Jeremy L. and {Robertson}, Brant E. and {Kravtsov}, Andrey V. and {Klypin}, Anatoly and {Warren}, Michael S. and {Yepes}, Gustavo and {Gottl{\"o}ber}, Stefan},
        title = "{The Large-scale Bias of Dark Matter Halos: Numerical Calibration and Model Tests}",
      journal = {\apj},
         year = 2010,
        month = dec,
       volume = {724},
       number = {2},
        pages = {878-886},
          doi = {10.1088/0004-637X/724/2/878},
archivePrefix = {arXiv},
       eprint = {1001.3162},
 primaryClass = {astro-ph.CO},
       adsurl = {https://ui.adsabs.harvard.edu/abs/2010ApJ...724..878T}
}

@ARTICLE{Schindler2025b,
       author = {{Schindler}, Jan-Torge and {Hennawi}, Joseph F. and {Davies}, Frederick B. and {Bosman}, Sarah E.~I. and {Wang}, Feige and {Yang}, Jinyi and {Eilers}, Anna-Christina and {Fan}, Xiaohui and {Kakiichi}, Koki and {Pizzati}, Elia and {Nanni}, Riccardo},
        title = "{A first look at quasar-galaxy clustering at $z\simeq7.3$}",
      journal = {arXiv e-prints},
         year = 2025,
        month = oct,
          eid = {arXiv:2510.08455},
        pages = {arXiv:2510.08455},
          doi = {10.48550/arXiv.2510.08455},
archivePrefix = {arXiv},
       eprint = {2510.08455},
 primaryClass = {astro-ph.GA},
       adsurl = {https://ui.adsabs.harvard.edu/abs/2025arXiv251008455S}
}

@ARTICLE{Devecchi&Volonteri2009,
       author = {{Devecchi}, B. and {Volonteri}, M.},
        title = "{Formation of the First Nuclear Clusters and Massive Black Holes at High Redshift}",
      journal = {\apj},
         year = 2009,
        month = mar,
       volume = {694},
       number = {1},
        pages = {302-313},
          doi = {10.1088/0004-637X/694/1/302},
archivePrefix = {arXiv},
       eprint = {0810.1057},
 primaryClass = {astro-ph},
       adsurl = {https://ui.adsabs.harvard.edu/abs/2009ApJ...694..302D}
}

@ARTICLE{Lambert2024,
       author = {{Lambert}, Trystan S. and {Assef}, R.~J. and {Mazzucchelli}, C. and {Ba{\~n}ados}, E. and {Aravena}, M. and {Barrientos}, F. and {Gonz{\'a}lez-L{\'o}pez}, J. and {Hu}, W. and {Infante}, L. and {Malhotra}, S. and {Moya-Sierralta}, C. and {Rhoads}, J. and {Valdes}, F. and {Wang}, J. and {Wold}, I.~G.~B. and {Zheng}, Z.},
        title = "{A lack of Lyman {\ensuremath{\alpha}} emitters within 5 Mpc of a luminous quasar in an overdensity at z = 6.9: Potential evidence of negative quasar feedback at protocluster scales}",
      journal = {\aap},
         year = 2024,
        month = sep,
       volume = {689},
          eid = {A331},
        pages = {A331},
          doi = {10.1051/0004-6361/202449566},
       adsurl = {https://ui.adsabs.harvard.edu/abs/2024A&A...689A.331L}
}

@ARTICLE{ForouharMoreno2025,
       author = {{Forouhar Moreno}, Victor J. and {Helly}, John and {McGibbon}, Robert and {Schaye}, Joop and {Schaller}, Matthieu and {Han}, Jiaxin and {Kugel}, Roi and {Bah{\'e}}, Yannick M.},
        title = "{Assessing subhalo finders in cosmological hydrodynamical simulations}",
      journal = {\mnras},
         year = 2025,
        month = oct,
       volume = {543},
       number = {2},
        pages = {1339-1372},
          doi = {10.1093/mnras/staf1478},
archivePrefix = {arXiv},
       eprint = {2502.06932},
 primaryClass = {astro-ph.CO},
       adsurl = {https://ui.adsabs.harvard.edu/abs/2025MNRAS.543.1339F}
}

@ARTICLE{GarciaVergara2019,
       author = {{Garc{\'\i}a-Vergara}, Cristina and {Hennawi}, Joseph F. and {Barrientos}, L. Felipe and {Arrigoni Battaia}, Fabrizio},
        title = "{Clustering of Ly{\ensuremath{\alpha}} Emitters around Quasars at z {\ensuremath{\sim}} 4}",
      journal = {\apj},
         year = 2019,
        month = dec,
       volume = {886},
       number = {2},
          eid = {79},
        pages = {79},
          doi = {10.3847/1538-4357/ab4d52},
archivePrefix = {arXiv},
       eprint = {1904.05894},
 primaryClass = {astro-ph.GA},
       adsurl = {https://ui.adsabs.harvard.edu/abs/2019ApJ...886...79G}
}

@ARTICLE{Meng2026,
       author = {{Meng}, Hao and {Zhang}, Huanian and {Ye}, Guangping},
        title = "{Probing The Dark Matter Halo of High-redshift Quasar from Wide-Field Clustering Analysis}",
      journal = {arXiv e-prints},
         year = 2026,
        month = feb,
          eid = {arXiv:2602.02778},
        pages = {arXiv:2602.02778},
archivePrefix = {arXiv},
       eprint = {2602.02778},
 primaryClass = {astro-ph.GA},
}

@ARTICLE{GinerMascarell2025,
       author = {{Giner Mascarell}, Mariona and {Eilers}, Anna-Christina and {Storey-Fisher}, Kate},
        title = "{Quasar clustering and duty cycle measurements at $0\leq z\leq 4$ with the Gaia-unWISE Catalog}",
      journal = {arXiv e-prints},
         year = 2025,
        month = nov,
          eid = {arXiv:2511.17413},
        pages = {arXiv:2511.17413},
          doi = {10.48550/arXiv.2511.17413},
archivePrefix = {arXiv},
       eprint = {2511.17413},
 primaryClass = {astro-ph.GA},
       adsurl = {https://ui.adsabs.harvard.edu/abs/2025arXiv251117413G}
}

@ARTICLE{Baados2025,
       author = {{Ba{\~n}ados}, Eduardo and {Momjian}, Emmanuel and {Connor}, Thomas and {Belladitta}, Silvia and {Decarli}, Roberto and {Mazzucchelli}, Chiara and {Venemans}, Bram P. and {Walter}, Fabian and {Wang}, Feige and {Xie}, Zhang-Liang and {Barth}, Aaron J. and {Eilers}, Anna-Christina and {Fan}, Xiaohui and {Khusanova}, Yana and {Schindler}, Jan-Torge and {Stern}, Daniel and {Yang}, Jinyi and {Andika}, Irham Taufik and {Carilli}, Christopher L. and {Farina}, Emanuele P. and {Fabian}, Andrew and {Hennawi}, Joseph F. and {Pensabene}, Antonio and {Rojas-Ruiz}, Sof{\'\i}a},
        title = "{A blazar in the epoch of reionization}",
      journal = {Nature Astronomy},
         year = 2025,
        month = feb,
       volume = {9},
        pages = {293-301},
          doi = {10.1038/s41550-024-02431-4},
archivePrefix = {arXiv},
       eprint = {2407.07236},
 primaryClass = {astro-ph.GA},
       adsurl = {https://ui.adsabs.harvard.edu/abs/2025NatAs...9..293B}
}
